\documentclass[11pt]{article}
\usepackage{authblk}
\usepackage{multirow}
\usepackage{xcolor}
\usepackage{setspace}
\usepackage{amsmath}
\usepackage{mathtools}
\usepackage{bm}
\usepackage{amsthm}
\usepackage{amssymb}
\usepackage{algorithm}
\usepackage{algpseudocode}
\usepackage{graphicx,psfrag,epsf}
\usepackage{enumerate}
\usepackage{subcaption}
\usepackage{caption}
\usepackage{natbib}
\usepackage{tikz}
\usepackage{url}  
\usepackage{appendix}
\usepackage{titlesec}
\usepackage{etoolbox}
\usepackage{makecell}
\makeatletter
\def\@seccntformat#1{\@ifundefined{#1@cntformat}%
   {\csname the#1\endcsname\quad}
   {\csname #1@cntformat\endcsname}}
\apptocmd\appendix{%
    \newcommand\section@cntformat{\appendixname\ }
    \addtocontents{toc}{\bigskip\noindent\textbf{Appendix Material}\par}
    {}{}}
\makeatother

\newcommand{\blind}{1}

\newtheorem{prop}{Proposition}
\newtheorem{theo}{Theorem}
\newtheorem{coro}{Corollary}
\newtheorem{remark}{Remark}
\newtheorem{lemma}{Lemma}

\DeclarePairedDelimiter\floor{\lfloor}{\rfloor}

\begin{document}

\def\spacingset#1{\renewcommand{\baselinestretch}%
{#1}\small\normalsize} \spacingset{1}


\if1\blind
{
  \title{\bf A Class of Dependent Random Distributions Based on Atom Skipping} 
  \author[1]{Dehua Bi}  
  \author[1]{Yuan Ji}
  \affil[1]{Department of Public Health Sciences, The University of Chicago, IL}
  \maketitle
} \fi

\if0\blind
{
  \bigskip
  \bigskip
  \bigskip
  \begin{center}
    {\LARGE\bf A Class of Dependent Random Distributions Based on Atom Skipping}
\end{center}
  \medskip
} \fi

\bigskip
\begin{abstract}
\noindent We propose the Plaid Atoms Model (PAM), a novel Bayesian nonparametric model for grouped data. Founded on an idea of `atom skipping', PAM is part of a well-established category of models that generate dependent random distributions and clusters across multiple groups. Atom skipping referrs to stochastically assigning 0 weights to atoms in an infinite mixture. Deploying atom skipping across groups, PAM produces a dependent clustering pattern with overlapping and non-overlapping clusters across groups. As a result, interpretable posterior inference is possible such as reporting the posterior probability of a cluster being exclusive to a single group or shared among a subset of groups. We discuss the theoretical properties of the proposed and related models. Minor extensions of the proposed model for multivariate or count data are presented. Simulation studies and applications using real-world datasets illustrate the performance of the new models with comparison to  existing models.
\end{abstract}

\noindent%
{\it Keywords:}  Atom skipping; Clustering; Dependent clustering; Dirichlet process; MCMC; Slice sampler; Stick-breaking process. 
\vfill

\newpage
\spacingset{1.45} 
\section{Introduction}\label{sec:intro}

Clustering, or unsupervised learning, is a primary tool for data analysis and scientific exploration. Representative clustering methods include algorithmic approaches like K-Means \citep{macqueen1967classification} and model-based clustering like MClust \citep{fraley1998mclust}. Alternatively, Bayesian nonparametric (BNP) models like the Dirichlet process (DP) \citep{ferguson1973bayesian} naturally induce clusters by allowing ties among observations.   These ``tied" values, which are random locations in the random probability measure of the BNP models, are also referred to as atoms in some literature, e.g., in \cite{denti2021common}. 
Hereafter, we use ``clusters" and ``atoms" interchangeably.

For complex problems and data structures   where multiple datasets are analyzed together , dependent clustering is often necessary. For example, in linguistic research, it is of interest to discover common themes across multiple documents \citep{teh2004sharing}, where the themes are modeled as shared clusters. In drug development, oftentimes different studies and corresponding data are pooled to increase the precision of statistical inference. However, drug effects might be heterogeneous and therefore subpopulations (clusters) of patients must be identified to better characterize the treatment effects. A common question for many dependent clustering problems is whether a clustering method can capture shared clusters across all or some groups while also identify unique ones that belong to a single group.

Various dependent clustering approaches have been proposed in   the BNP   literature.   Early pioneering work of  dependent Dirichlet process (DDP) is initiated in \cite{maceachern1999,maceachern2000}.   These BNP models generate different patterns of atoms \textit{a priori} on a spectrum that ranges from ``common-atoms model" to ``distinct-atoms model." Subsequently, common-atoms models, such as hierarchical DP (HDP) \citep{teh2004sharing}, Common Atoms Model (CAM) \citep{denti2021common}, and hidden-HDP \citep{lijoi2022flexible} assume all groups share the same set of atoms \textit{a priori}.   In contrast,   distinct-atoms models, such as the nested  DP (NDP) \citep{rodriguez2008nested}, assume that groups with different distributions all have  unique and distinctive atoms. Other methods in literature   like the hierarchical mixture of DP \citep{muller2004method}, the latent nested process (LNP) \citep{camerlenghi2019latent}, and the semi-HDP \citep{beraha2021semi} take the middle ground by mixing distinct- and common-atoms  processes. Consequently, a pattern of shared and unique clusters can be generated across groups under these models.

We consider a new approach to generate dependent clustering structure using a simple idea called atom skipping. Instead of mixing a distinct-atoms model with a common-atoms model, we construct random distributions by removing atoms from a common-atoms model in a group-specific fashion. This is realized by stochastically assigning the weight of an atom to be zero for each group. This effectively skips (removes) the atom from the group. For a single group or a single dataset, atom-skipping results in a new model called the Atom-Skipping Process (ASP). For multiple groups, it leads to the main proposal of the paper, the Plaid Atoms Model (PAM). When an atom is removed in all but one group, that atom becomes a unique cluster for that group. On the other hand, if an atom is not removed in any groups, it induces a common cluster shared for all groups. In-between is an atom that is removed in a fraction of groups, and the corresponding cluster is only shared by a subset of groups. The resulting dependent clustering pattern is slightly more flexible than some existing models. For example, when there are three or more groups, the set of overlapping clusters may vary between a pair of groups. 

Furthermore, due to group-specific atom skipping, PAM defines a generative model that explicitly defines overlapping (common) and non-overlapping (unique) clusters across groups, which  leads to more interpretable posterior inference. For example, PAM can perform inference on whether a cluster is absent in a group by reporting the posterior probability that the corresponding cluster has zero weight in the group. In contrast,  common-atoms models like HDP or CAM always produce a positive  cluster weight for any cluster in any group. 
An interesting by-product of atom-skipping is that the   marginal   mean of ASP and PAM follows a stochastic process that is called the Fractional Stick-Breaking Process (FSBP). This process is a simple modification of the stick-breaking representation of DP, and is linked to many random probability measures (RPM) and processes in the literature. 

The remainder of the article are organized as follows. In Section \ref{sec:review}, we review BNP models that are closely linked to PAM. In Section \ref{sec:pam}, we introduce   three related new models, ASP,  PAM, and FSBP. We discuss theoretical properties of the three new processes in Section \ref{sec:pam_theory}, highlighting their interconnections. In Section \ref{sec:sam_infer}, we discuss posterior inference and outline the slice sampler algorithm for PAM   and FSBP. Section \ref{sec:sim}   presents comparative simulation results of the proposed models and Section \ref{sec:exp} describes application of  PAM to publicly available datasets. Lastly,   Section \ref{sec:diss} concludes the paper with some discussion.

\section{Review of   Some   BNP Models for Clustering}
\label{sec:review}
\subsection{Methods for Clustering a Single Study or Dataset}

\begin{figure}
    \centering
    \includegraphics[scale=0.4]{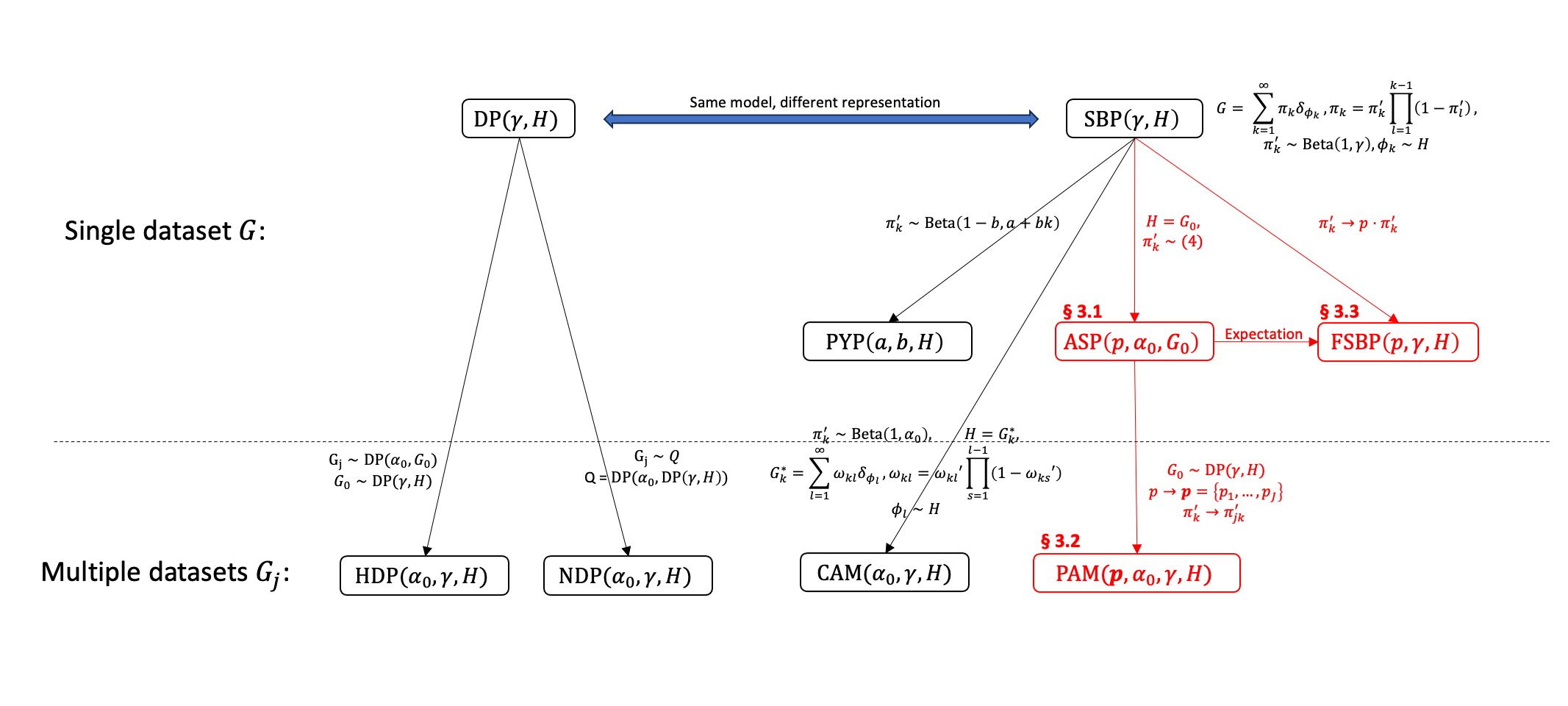}
    \caption{  A graphic   illustration of  relationship of selected BNP models. 
    A directed edge connecting two processes implies that the child process is an extension of the parent process.    The red nodes and edges represent the contribution of this work.   Section numbers of the manuscript are placed on the red nodes.   }
    \label{fig:illustration_relationships}
\end{figure}

We review   related   BNP   models to set the stage for the proposed new models. 
Figure \ref{fig:illustration_relationships} provides a graph illustrating the BNP models considered in our work. Specifically, nodes are BNP models, directed edges   describe 
the extension of the parent node to the child node,  black and red color represent existing and novel models respectively. 

Consider a dataset with $n$ observations   of $q$-dimensional vectors ($q  \geq 1$), with the $i$th observation denoted as $\bm{y}_i = (y_{i1}, \ldots, y_{iq})$, $i = 1, \ldots, n. $   Denote   the entire dataset $\bm{y} = \{\bm{y}_1, \ldots, \bm{y}_n\}$. Assume $\bm{y}_i$ takes a value from a suitable Polish space $X$ that is endowed with the respective Borel $\sigma$-field $\mathcal{X}$. The observations are assumed to arise from a nonparametric mixture model indexed by parameter $\bm{\theta}_i$ and a random distribution $G$ as follows:
\begin{equation}
\arraycolsep=1.4pt\def\arraystretch{1.25}
\begin{array}{l}
\bm{y}_i|\bm{\theta}_i \sim F(\bm{y}_i|\bm{\theta}_i), \quad \bm{\theta}_i|G \sim G, \quad i = 1, \ldots, n,\label{eq:y_i_ll}
\end{array}
\end{equation} 
where $F(.|\bm{\theta}_i)$ is a parametric distribution for $\bm{y}_i$ with parameter $\bm{\theta}_i$, and $G$ is assumed to have a nonparametric prior. 

\paragraph{Review of DP:}
The DP prior is denoted as   $G \sim DP(\gamma,H)$, where $\gamma > 0$ is the concentration parameter, and $H$ is the base measure. 
\cite{sethuraman1994constructive}   shows    that DP   generates random distributions with the stick-breaking representation:
\begin{equation}
\arraycolsep=1.4pt\def\arraystretch{1.25}
\begin{array}{l}
    G = \sum_{k=1}^{\infty} \pi_k\delta_{\bm{\phi}_k}, \,\, 
    {\pi_k} \sim \text{GEM}(\gamma), \,\, \text{and } \bm{\phi}_k \sim H,\label{eq:sbp}
    \end{array}
\end{equation} 
where GEM is the Griffiths-Engen-McCloskey distribution \citep{pitman2002poisson}. Specifically, $\pi_k \sim \text{GEM}(\gamma)$ means that $\pi_k = \pi_k' \prod_{l=1}^{k-1} (1-\pi_l')$, $\pi_k' \sim \mbox{Beta}(1, \gamma)$, where $\text{Beta}(a,b)$ denotes the beta distribution with mean $a/(a+b)$. This equivalent representation of DP is also referred to as the Stick-Breaking Process (SBP), denoted as $\text{SBP}(\gamma,H)$.

When applied to clustering, DP is often   known    for its ``rich-get-richer" characteristic   in that DP   tends to produce    few large  clusters with 
many tiny ones or singletons. To address this,   an alternative model is proposed known as the PYP.

\paragraph{Review of PYP:}
PYP$(a,b,H)$  extends and modifies DP by   assuming the  atom  weight follows $$\pi_k' \sim \text{Beta}(1-b,a+b\cdot k),$$
where $a > -b$  and $b \in [0,1)$. The construction of $G$ and the distribution of $\bm{\phi}_k$ remain the same as in equation \eqref{eq:sbp}. PYP reduces to DP if $b = 0$ and $a = \gamma$. 
Compare to DP, PYP has two desirable properties: 1) the expected number of clusters of PYP grows more rapidly (with a rate of $n^b$) than that of DP (which grows with a rate of $\text{log}(n)$), and 2) the rate of decay in terms of cluster-size follows a power law for PYP, but has an exponential tail in DP. However, due to non-i.i.d stick-breaking weights in PYP, many theoretical results of PYP are not available in closed form. For example, while the mean of DP is known to be the base-measure $H$ in equation \eqref{eq:sbp},   PYP does not have a  closed-form mean.

\subsection{Methods for Clustering Multiple   Studies or   Datasets}

Extend the previous setting   to $J>1$ studies or groups, each of which has a dataset of
$n_j$ observations.
The $i$th observation in group $j$ is denoted as   a $q$-dimensional ($q\ge 1$) vector   $\bm{y}_{ij}$. Let $\bm{y}_j = \{\bm{y}_{ij}; \; i = 1, \ldots, n_j\}$ represent the entire dataset for the $j$th group. Assume
\begin{equation}
\arraycolsep=1.4pt\def\arraystretch{1.25}
    \begin{array}{l} 
        \bm{y}_{ij}|\bm{\theta}_{ij} \sim F(\bm{y}_{ij}|\bm{\theta}_{ij}), \quad \bm{\theta}_{ij}|G_j \sim G_j, \,\,\,\,  i = 1, \ldots , n_j; \,\, j = 1, \ldots , J, \label{eq:y_ij_ll}
    \end{array}
\end{equation}
where $F(\cdot | \bm{\theta}_{ij})$ is a parametric distribution for $\bm{y}_{ij}$.
The models  reviewed  below assign priors to $G_j$ for $j = 1, \ldots, J$.   These models induce dependent partitions of $\bm{y}_j$'s, allowing for information borrowing between groups. Most BNP models differ in their construction of common or distinct atoms across groups. While one school chooses to build common-atoms models that share a common set of atoms for all groups, another school allows groups to have non-overlapping atoms known as distinct-atoms models. A third school mixes the two ideas so that more flexible patterns of atoms can be modeled. Our work belongs to the third school. 

\paragraph{Review of HDP:}
HDP \citep{teh2004sharing} is a common-atoms model. In this model, each $G_j$ is assigned a DP prior with a common base measure $G_0$, which itself is an instance of $DP$, i.e., 
$$G_j|\alpha_0,G_0 \sim \text{DP}(\alpha_0, G_0), \quad G_0|\gamma,H \sim \text{DP}(\gamma, H).$$
\noindent 
This    model    is denoted as $\text{HDP}(\alpha_0,\gamma,H)$. Using the stick-breaking representation \citep{sethuraman1994constructive} of DP, HDP can be rewritten as 
\begin{equation}\label{eq:HDP}
\arraycolsep=1.4pt\def\arraystretch{1.25}
\begin{array}{ll}
    G_j = \sum_{k=1}^{\infty} \pi_{jk}\delta_{\bm{\phi_k}}, &  \pi_{jk} = \pi_{jk}'\prod_{l=1}^{k-1}(1-\pi_{jl}') \\ 
    \multicolumn{2}{l}{ \pi_{jk}' \sim \text{Beta}\left( \alpha_0\beta_k, \alpha_0\left(1-\sum_{l=1}^k \beta_l\right) \right)}  \\
    \bm{\phi}_k  \sim H  \mbox{ and }  & \beta_k \sim \text{GEM}(\gamma)  
    \end{array}
\end{equation} 
where $\delta_{\{\cdot\}}$ is the indicator function. 
Note that although $G_0$ is not shown in the stick-breaking construction of HDP in equation \eqref{eq:HDP}, it can be reconstructed from $\bm{\beta} = \{\beta_k; k \geq 1\}$ and $\bm{\Phi} = \{\bm{\phi_k}; k \geq 1\}$, i.e., $G_0 = \sum_{k=1}^{\infty} \beta_k \delta_{\bm{\phi_k}}$.
Appropriate prior distributions, like the gamma distribution, can be specified for $\alpha_0$ and $\gamma$ to complete HDP. 

It is clear from equation \eqref{eq:HDP} that HDP is a common-atoms model, because 
all groups share   the same set of atoms in   $\bm{\Phi}$, and   the atom weights   $\pi_{jk} \neq 0.$  
  While the atoms are shared, their weights $\bm{\pi}_j = \{\pi_{jk}; k \geq 1\}$ are distinct for different groups.  
Consequently, for $G_j$ and $G_{j'}$ where $j \neq j'$, $G_j \neq G_{j'}$ with probability 1. 

\paragraph{Review of NDP:}
\cite{rodriguez2008nested} introduce the NDP, a model capable of clustering both subjects and groups.
In NDP, the group-level clusters are referred to as distributional clusters, and the group-specific distribution $G_j$ in NDP is defined as follows:
$$G_j|Q \sim Q, \quad Q = \text{DP}(\alpha_0,\text{DP}(\gamma,H)),$$
where the distribution of each group follows a DP with its base measure being another DP, rather than being a realization of DP as in HDP. We use $\text{NDP}(\alpha_0,\gamma,H)$ to denote this model. NDP is a distinct-atoms model, which can be seen in its stick-breaking representation:
\begin{equation}
\arraycolsep=1.4pt\def\arraystretch{1.25}
\begin{array}{lll}
    G_j = \sum_{k=1}^{\infty} \pi_{k}\delta_{G_k^*}, &  \multicolumn{2}{l}{\pi_{k} \sim \text{GEM}(\alpha_0)} \\
    G_k^* = \sum_{l=1}^{\infty} \omega_{kl}\delta_{\bm{\phi}_{kl}}, & \omega_{kl} \sim \text{GEM}(\gamma), \\   \bm{\phi}_{kl}   \sim H. 
    \label{eq:NDP}
    \end{array}
\end{equation} 
For $j \neq j'$, if $G_j$ and $G_{j'}$   are   not equal to the same $G_k^*$, none of   their    atoms will be the same.   In contrast, if they are equal to the same $G_k^*$, meaning  $G_j$ and $G_{j'}$ are two identical distributions, their atoms and atom weights will be identical. 
This phenomenon is known as ``degeneracy" \citep{camerlenghi2019latent}, where if two groups share   just  one atom they 
share all atoms and weights.   Otherwise, the atoms and weights must all be distinct for these two groups.  
This presents a challenge if we aim to find common clusters for two groups belonging to different distributional clusters. 

\paragraph{Review of CAM:}
\cite{denti2021common}   extend   NDP and   introduce   the Common Atoms Model, abbreviated as CAM. By definition, CAM is a common-atoms model. Building upon NDP, CAM provides distributional clustering similar to NDP. 
Specifically, CAM restricts the atoms in all $G_k^*$, $k \geq 1$ in NDP to a common set.   In other words, rather than assuming that for each group $k$ there is a distinct set of atoms $\{\bm{\phi}_{kl}\}$, CAM instead assumes all the groups share a common set of atoms $\{\bm{\phi}_l\}$. Mathematically,  the   $G_j$'s   
in CAM are defined as follows:
\begin{equation}
\arraycolsep=1.4pt\def\arraystretch{1.25}
\begin{array}{lll}
    G_j = \sum_{k=1}^{\infty} \pi_{k}\delta_{G_k^*}, &  \multicolumn{2}{l}{\pi_{k} \sim \text{GEM}(\alpha_0)} \\
    G_k^* = \sum_{l=1}^{\infty} \omega_{kl}\delta_{\bm{\phi}_{l}}, & \omega_{kl} \sim \text{GEM}(\gamma), &   \bm{\phi}_{l}   \sim H. 
    \label{eq:CAM}
    \end{array}
\end{equation} 
We use the notation $\text{CAM}(\alpha_0,\gamma,H)$ to denote this model. Another recent development building upon the modeling of common atoms can be found in hidden-HDP \citep{lijoi2022flexible}. Due to limited space, a review of this model is omitted. 

\paragraph{Other BNP models:}
The   aforementioned   common-atoms and distinct-atoms models represent the two extremes of a spectrum of BNP priors   for dependent random distributions.   
In the literature, many other models target the space in between, where the prior is allowed to contain both common and unique atoms \textit{a priori}.
Such models include the hierarchical mixture of DP \citep{muller2004method}, the latent nested process (LNP) \citep{camerlenghi2019latent}, and more recently,  the semi-HDP \citep{beraha2021semi}. 
All these models construct flexible priors by adding or mixing distinct-atoms and common-atoms models together, in a nonparametric or semi-parametric fashion. 
For a comprehensive review,  refer to \cite{quintana2022dependent}. In this work, we take a  different approach. We start from a common-atoms model, and using an idea of atom skipping 
in a probabilistic fashion for each group. The resulting model  provides common and unique atoms {\it a priori}   but with interesting theoretical properties and behavior in statistical inference.     

\section{Proposed BNP Models} 
\label{sec:pam}

\subsection{Atom-Skipping Process}

The proposed models utilize a simple idea of atom skipping by 
probabilistically setting the weight of an atom to be exactly zero.  
We first consider a model for a single random distribution (i.e., a single study or dataset). We denote such a model   the   ASP, standing for   the    atom-skipping process.  
Using the HDP in  \eqref{eq:HDP} as an example, atom skipping is implemented by assuming the prior for  $\pi_{jk}'$ to be
\begin{equation} \label{eq:zab}
    f(  \pi_{jk}'  ) = p \times  \underbrace{ f_{\text{Beta}}\left(\alpha_0\beta_k, \alpha_0\left(1-\sum_{l=1}^k \beta_l\right) \right) }_{\eqref{eq:HDP}} + (1 - p) \times   \underbrace{\delta_0}_{\mbox{atom skipping}}  ,
\end{equation}
 where $f_{\text{Beta}}(a,b)$ is the probability density function (p.d.f) of the beta distribution,   and $\delta_0$ is the indicator function at $0$. Then we define the atom-skipping process (ASP) for a single dataset as 
\begin{equation}
\arraycolsep=1.4pt\def\arraystretch{1.25}
\begin{array}{ll}
        G = \sum_{k=1}^{\infty} \pi_{k}\delta_{\bm{\phi_k}}, &   \pi_{k} = \pi_{k}'\prod_{l=1}^{k-1}(1-\pi_{l}'),
        \\
        \multicolumn{2}{l}{f(\pi_{k}'|\bm{\beta}, p, \alpha_0) = p \times f_{\text{Beta}}\left(\alpha_0\beta_k, \alpha_0\left(1-\sum_{l=1}^k \beta_l\right) \right) + (1 - p) \times \delta_0},  
        \label{eq:ASP}
    \end{array}
\end{equation}
where $\beta_k$ and $\bm{\phi}_k$ are assumed to be given. We let $G_0 =\sum_{k=1}^\infty \beta_k \delta_{\bm{\phi}_k}$ and denote  model \eqref{eq:ASP}  as  $G | p, \alpha_0, G_0 \sim \text{ASP}(p,\alpha_0,G_0).$ 
According to \eqref{eq:ASP}, since each $\pi'_k$ or $\pi_k$ has a probability to be zero, the corresponding atom $\bm{\phi}_k$ may be skipped when sampling from $G.$  

\subsection{Plaid Atoms Model}
Adding back the DP prior on $G_0$ and extending ASP to multiple datasets, we propose   the Plaid Atoms Model (PAM). Specifically, PAM is given in a hierarchical model as
\begin{equation}\label{eq:PAM}
\arraycolsep=1.4pt\def\arraystretch{1.25}
\begin{array}{ll}
    G_j|p_j, \alpha_0, G_0 \sim \text{ASP}(p_j,\alpha_0,G_0), \,\, 
    G_0|\gamma, H \sim \text{DP}(\gamma, H).
\end{array}
\end{equation}
Using  a stick-breaking representation, PAM can be shown to be equivalent to  
\begin{equation}
\arraycolsep=1.4pt\def\arraystretch{1.25}
\begin{array}{ll}
  \multicolumn{2}{l}{G_j = \sum_{k=1}^{\infty} \pi_{jk}\delta_{  \bm{\phi_{k}}  },} \\
  \multicolumn{2}{l}{\pi_{jk} = \pi_{jk}'\prod_{l=1}^{k-1}(1-\pi_{jl}'),}
        \\
  \multicolumn{2}{l}{f(\pi_{jk}'|\bm{\beta}, p, \alpha_0) = p_j \times f_{\text{Beta}}\left(\alpha_0\beta_k, \alpha_0\left(1-\sum_{l=1}^k \beta_l\right) \right) + (1 - p_j) \times   \delta_0  },\\
     \bm{\phi}_k \sim H 
        & \text{and } \beta_k   \sim \text{GEM}(\gamma). 
        \label{eq:PAM_g}
    \end{array}
\end{equation}
The proof of the equivalence between \eqref{eq:PAM} and \eqref{eq:PAM_g} is omitted and follows the same derivation in \cite{teh2004sharing}. Note that when $p_j=1, \; \forall j$, PAM is equivalent to HDP. This can be trivially shown by comparing models \eqref{eq:HDP} and \eqref{eq:PAM_g}.  

Let $\bm{p} = \{p_1, \ldots, p_J\}$. We denote this model as   $  G_1, \ldots, G_J   \sim \text{PAM}(\bm{p}, \alpha_0, \gamma, H)$. 
Additional pirors   can be placed on the parameters of $\bm{p}$, $\alpha_0$, and $\gamma$, for example, 
\begin{equation}\label{eq:PAM_hyper}
    p_j|a, b \sim \text{Beta}(a, b), \alpha_0 \sim \text{Gamma}(a_{\alpha}, b_{\alpha}), \gamma \sim \text{Gamma}(a_{\gamma}, b_{\gamma}).
\end{equation}

By construction, PAM is more versatile as a generative model. It allows different $G_j$'s to share some atoms but also possesses unique ones. A comparison of PAM and other dependent random distributions like HDP and CAM is given in Supplement A.1 as a reference. 

\paragraph{Continuous Data:}
  PAM in \eqref{eq:PAM} can be used as a prior for the random distribution in model  \eqref{eq:y_ij_ll}. 
If observations $y_{ij}$ are continuous and univariate ($q=1$), we 
use a Gaussian kernel by setting   $\bm{\phi}_k = (\mu_k,\sigma^2_k)$ and $F(\cdot|\bm{\phi}_k) = N(\cdot|\mu_k,\sigma^2_k)$.
To complete model specification for $\text{PAM}(\bm{p}, \alpha_0, \gamma, H),$   the base measure $H$ is modeled as the conjugate prior of normal-inverse-gamma (NIG), where $H = \text{NIG}(\mu_0, \kappa_0, \alpha_0, \beta_0)$, i.e., $\mu_k|\sigma_k^2 \sim N(\mu_0, \sigma_k^2/\kappa_0)$ and $\sigma_k^2 \sim \text{IG}(\alpha_0, \beta_0)$.   For multivariate observations ($q > 1$), the related model components are changed to multivariate normal and normal-inverse-Wishart distributions. The detail is ommitted for simplicity.  

\paragraph{Count Data:}
Following \cite{denti2021common}, we extend the proposed PAM to count data and refer to it as the Discrete Plaid Atoms Model (DPAM). We only consider univariate count data and hence $q=1$. 
Let $x_{ij} \in {\mathbb{N}}$ be the observed count data for observation $i = 1, \ldots , n_j$ in group $j =1,\ldots , J$, where ${\mathbb{N}}$ denotes the natural numbers. Thus the data vector $\bm{x}_j = (x_{1j},\ldots ,x_{n_jj})$ is the set of counts observed for the $j$th group. We apply the data augmentation framework in \cite{canale2011bayesian} and introduce latent continuous variables $y_{ij}$ so that 
 \begin{equation} \label{eq:8}
     \text{Pr}(x_{ij}= \omega) = \int_{a_\omega}^{a_{\omega+1}} g(y_{i,j}) dy_{ij}, \quad \omega=0, 1, 2, \cdots 
 \end{equation}
where $a_0 < a_1 < \cdots  < a_{\infty}$ is a fixed sequence of thresholds that take values $\{a_\omega; \omega \geq 0\} = \{-\infty, 0, 1, 2, \ldots , +\infty\}$, and $g(y_{ij})$ follows the PAM mixture model as in equations \eqref{eq:y_ij_ll} and \eqref{eq:PAM_g}. This construction allows posterior inference for $y_{ij}$ since it is trivial to see that 
$$
x_{ij} | y_{ij} = \sum_{\omega=0}^\infty \bm{1}_\omega(x_{ij}) \cdot \bm{1}_{[a_\omega, a_{\omega+1})}(y_{ij}),
$$
where $\bm{1}_a(b)$  equals 1 if $b=a$ or $b \in a$, and 0 otherwise. 

\subsection{Fractional Stick-Breaking Process} \label{subsec:fsbp}

Taking expected value of $\pi_k$ in \eqref{eq:ASP}, we derive a new process for a single group called the Fractional Stick-Breaking Process (FSBP). This new process gives an interesting and new solution for modeling a random distribution, and induces a clustering structure that is different from existing models like the DP or PYP. 

Let $p \in (0, 1)$, and $a,b > 0$ be fixed constants. The FSBP is an extension of the DP (or equivalently the SBP) and given by 
\begin{equation}
\arraycolsep=1.4pt\def\arraystretch{1.25}
\begin{array}{l}
    G = \sum_{k=1}^{\infty} \pi_k\delta_{\bm{\phi}_k}, \,\,  \pi_k = p\cdot{\pi_k}' \prod_{l = 1}^{k-1} (1 - p\cdot {\pi_l}'), \\  
    {\pi_k}'|\gamma \sim \text{Beta}(  a=1, b=\gamma  ), \,\, \bm{\phi}_k|H \sim H.\label{eq:fsbp_g}
    \end{array}
\end{equation}
We denote this model as $G \sim \text{FSBP}(p, \gamma, H).$ When $p=1$, $\text{FSBP}(p, \gamma, H)$ reduces to SBP$(\gamma, H)$ or equivalently DP$(\gamma, H).$ We show in Section \ref{sec:pam_theory} that the FSBP is the mean of the ASP and   has more expected number of clusters than that of DP with the same concentration parameter $\gamma$.  

\section{Properties   of   ASP, PAM, and FSBP}
\label{sec:pam_theory}

\subsection{Properties of ASP and PAM}
We start by showing   that   the   cluster   weights   in   ASP and PAM sum to 1.
\begin{prop} \label{prop:1}
  Assume $\bm{\beta} = \{\beta_k; k \geq 1\}$, $\beta_k \sim \text{GEM}(\gamma)$, $f(\pi_k'|\bm{\beta}, p, \alpha_0)$   is given in   \eqref{eq:ASP} and $f(\pi_{jk}'|\bm{\beta},p_j,\alpha_0)$   in   \eqref{eq:PAM_g}. 
  Furthermore, assume  $p,p_j \sim Beta(a, b)$. Then   
\begin{enumerate}
    \item $\sum_{k \geq 1}\pi_{k} = 1$, $\sum_{k \geq 1}\pi_{jk} = 1$, 
    \item $\text{E}[\pi_{k}|\bm{\beta},p] = p\cdot \beta_k' \prod_{l=1}^{k-1}(1-p\cdot \beta_l')$,  $\text{E}[\pi_{jk}|\bm{\beta},p_j] = p_j\cdot \beta_k' \prod_{l=1}^{k-1}(1-p_j\cdot \beta_l')$, and
    \item $\text{E}[\pi_{k}] = \text{E}[\pi_{jk}] = \frac{1}{1+\gamma'}\left(\frac{\gamma'}{1+\gamma'}\right)^{k-1}$ where $\gamma' = \frac{1 + \gamma - \bar{p}}{\bar{p}}$, $\bar{p} = \frac{a}{a+b}$.
\end{enumerate}
\end{prop}
The proof is in Supplement A.2.
This result shows that the random distributions $G$ from ASP and $G_j$ from PAM are proper discrete random distributions. Next, we show that the mean process of ASP is FSBP.

\begin{theo} \label{theo:ASP}
For an arbitrary set $A \subseteq X$, let $\alpha_0, \gamma > 0$, $H$ be a fixed probability measure,   $G_0 \sim DP(\gamma, H)$,  
  and   $G|G_0,p \sim ASP(p,\alpha_0,G_0)$   as in \eqref{eq:ASP}.   Then, conditional on $G_0$   and $p$,   the conditional mean of $G$   is  
$$\text{E}[G(A)|G_0,p] = G^*(A),$$
where $G^* \sim FSBP(p, \gamma, H)$.
\end{theo}
The proof of Theorem \ref{theo:ASP} is in Supplement A.3.
  Combining the results of Theorem \ref{theo:ASP}   and   Theorem \ref{theo:1}, the following corollary gives the maginal mean of the ASP. 
\begin{coro} \label{coro:1} If $G_0 \sim DP(\gamma, H)$,   $p \sim Beta(a,b)$,   and $G|G_0,p \sim ASP(p,\alpha_0,G_0),$   then   $\text{E}[G(A)] = \text{E}[\text{E}[G(A)|G_0,p]] = \text{E}[G^*(A)] = H(A).$
\end{coro}

Lastly, we look at properties related to PAM. Since PAM is   an extension of ASP to multiple groups, the results in Theorem \ref{theo:ASP} apply to the group-specific random distribution $G_j$ of PAM as well. Corollary \ref{coro:1} also   applies to a random distribution $G_j$ from   PAM. Moreover, in the next proposition, we show that \textit{a priori}, there is a positive probability for two observations from two different groups to be clustered together in PAM. 
\begin{prop} \label{prop:3}
Let $G_1, \ldots, G_J \sim \text{PAM}(\bm{p},\alpha_0,\gamma,H)$. Without loss of generality, for two groups $G_1$ and $G_2$, let $\bm{\theta}_{i1}|G_1 \sim G_1$ and $\bm{\theta}_{i'2}|G_{2} \sim G_{2}$, then
\begin{equation} \label{eq:posptt}
    \text{Pr}(\bm{\theta}_{i1} = \bm{\theta}_{i'2}) > 0.
\end{equation}
\end{prop}
\noindent The proof of Proposition \ref{prop:3} is given in Supplement A.4.

Unfortunately, closed-form results are unavailable for the variance, correlation structure, and partition probability functions of PAM. 
The expected number of clusters for PAM is  not available in closed form either.   
Consequently, we investigate the   clustering properties of PAM through a small simulation and compare   it   to  CAM and HDP. 

We assume there are 500 groups ($j=1, \ldots, 500$) and within each group $G_j$, we generate a random sample of $1,000$ observations from  CAM, HDP, or PAM. This leads to a total of $500,000$ observations for each process. When sampling an observation from these processes, computationally it is not feasible to sample  from an infinite mixture. Instead, we consider a finite mixture of $1,000$ atoms, which are sampled from  the base measure $H$: $\phi_k \sim H \text{ for } k = 1, \ldots,   1,000,  $ where $H = N(0,1)$.  We set the concentration parameters $\alpha_0 = \gamma = 1$ for CAM, HDP, and PAM.   Therefore, we use notation CAM$(1,1,H)$ and HDP$(1,1,H).$ We consider two versions of PAM, with $p_{j1} \sim \text{Beta}(80, 20)$ or $p_{j2} \sim \text{Beta}(20, 80), $ for $j=1, \ldots,500.$ This leads to PAM$(\bm{p}_1,1,1,H)$ and PAM$(\bm{p}_2,1,1,H),$ where $\bm{p}_1 = \{p_{j1}  ;   j=1, \ldots, 500\}$ and $\bm{p}_2 = \{p_{j2}   ;   j=1, \ldots, 500\}.$ We  sample the atom weights $\pi$'s in each group  based on their corresponding stick-breaking processes, model \eqref{eq:HDP} for HDP, \eqref{eq:CAM} for CAM, and \eqref{eq:PAM_g} for PAM. At the end, we obtain $1,000$ observations per group for 500 groups under each of the four processes, CAM$(1,1,H)$, HDP$(1,1,H)$, PAM$(\bm{p}_1,1,1,H)$ and PAM$(\bm{p}_2,1,1,H)$, with $H=N(0,1)$. 

\begin{figure}[hbtp]
    \centering
    \includegraphics[width=0.55\textwidth]{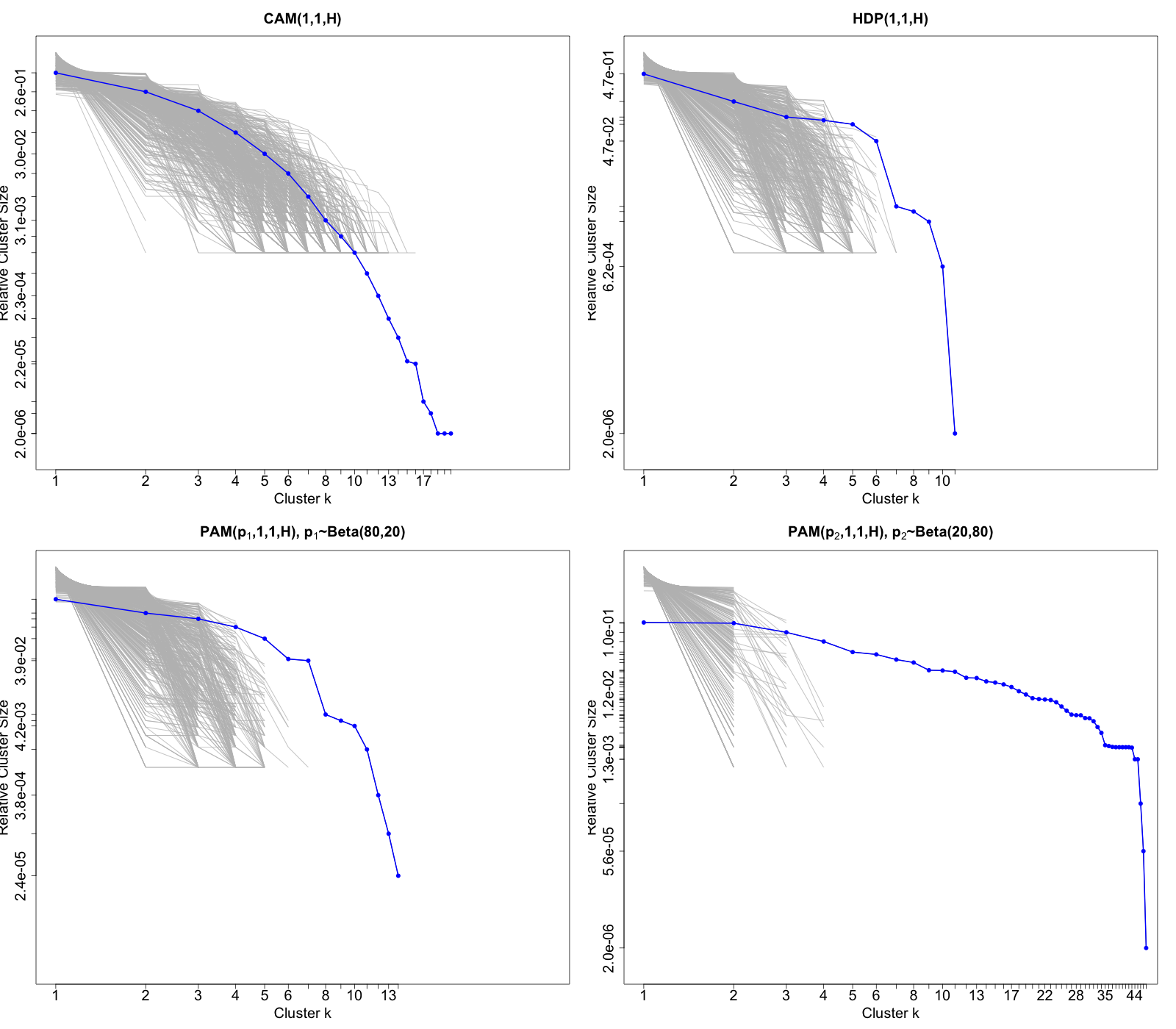}
    \caption{  Clustering pattern of   CAM, HDP, and PAM. 
    The four subplots present the relative cluster size against the number of clusters for the four processes, CAM$(1,1,H)$, HDP$(1,1,H)$, PAM$(\bm{p}_1,1,1,H)$ and PAM$(\bm{p}_2,1,1,H)$, with $H=N(0,1)$. The grey lines in each subplot correspond to the observations within each group and the blue   lines   correspond to   the relative cluster size of   all the observations aggregated across 500 groups.   } 
    \label{fig:cam_hdp_pam_prior}
\end{figure}

Figure \ref{fig:cam_hdp_pam_prior} summarizes the number of clusters and the relative cluster size, either for  a single group or for the entire 500,000 observations across all 500 groups, under each of the four processes.   The processes exhibit quite different behavior. First, the  average number of clusters in a group is 7.62 (SD 2.56), 3.00 (SD 0.86),   2.49 (SD 0.92), and 1.24 (SD 0.47)   for CAM$(1,1,H)$, HDP$(1,1,H)$, PAM$(\bm{p}_1,1,1,H)$ and PAM$(\bm{p}_2,1,1,H)$,   respectively.   This can be observed based on the average length of the grey lines in the subplots of Figure \ref{fig:cam_hdp_pam_prior}, each grey line representing a group.  However, aggregating all the observations, the total number of clusters is 18, 10, 13, and 43, for the four processes, respectively, corresponding to the length of the blue line in the figure. Therefore, HDP (top right) generates the smallest number of clusters while PAM$(\bm{p}_2,1,1,H)$  (bottom right) generates the largest number of clusters.  Interestingly, PAM$(\bm{p}_2,1,1,H)$ (bottom right plot) also   generates on average the smallest number of clusters per group (shortest grey lines). 
This means   that for PAM many clusters across groups are unique, a feature that is different from the other three processes. Lastly, PAM$(\bm{p}_1,1,1,H)$ (bottom left) behaves similar to HDP (top right),  which is expected, since when $p_j$ approaches 1, PAM is identical to HDP. Lastly, CAM (top left) generates on average the largest number of clusters per group (longest grey lines) without producing a large number of total clusters. This means CAM is inclined to generate more and overlapping clusters across groups. Additional results comparing the clustering behavior of the three models can be found in Supplement A.5.  


\subsection{Properties of FSBP}
We first show that the mean and variance of FSBP are available in closed forms and   is   related to DP.

\begin{theo} \label{theo:1}
For an arbitrary set $A \subseteq X$, let $  p   \in (0, 1),$  $\gamma > 0$ be fixed constants, and $H$   be   a fixed probability measure. For $G^* \sim \text{FSBP}(p, \gamma, H)$,  the mean and variance of $G^*$ on $A$ are 
$$\text{E}[G^*(A)] = H(A), \,\, \text{Var}\left(G^*(A)\right) = \frac{H(A)\{1-H(A)\}}{v}, \mbox{ where } v=\frac{1+\gamma}{p}+\frac{1-p}{p}.$$
\end{theo}
\noindent The proof of the theorem is in Supplement A.6.

\begin{remark}
    The mean and variance of $G^*$  match the mean and variance of a DP $G' \sim DP(v - 1, H)$,   respectively.  
\end{remark} 

\def\btheta{\bm{\theta}}
\def\bphi{\bm{\phi}}

We next derive the   exchangeable partition probability function (EPPF) of     FSBP $G^*$. Let $\bm{z} = \{z_1, \ldots, z_n\}$ represent   the vector of cluster memberships 
for $n$ observations   sampled from $G^*$. Without loss of generality, suppose $z_i \in \{1, \ldots, K\}$ which means there is a total of $K$ clusters indexed from $1$ to $K$. Then $\bm{z}$ defines a partition of the $n$ observations, denoted as $C(\bm{z}) = \{c_1, \ldots, c_K\}$ where $\cup_{k=1}^K c_k = \{1, \ldots, n\}$ and $c_k = \{i; z_i = k\}.$ 
For any partition $C$ of $\{1, \ldots, n\}$,   the EPPF of $G^*$ is defined as $\text{Pr}(C(\bm{z}) = C)$ \citep{pitman1995exchangeable}. Following the work of \cite{miller2019elementary}, we derive the expression for the EPPF of $G^*$.

\begin{theo} \label{theo:3}
    Let $p \in (0, 1)$ be a fixed constant, and let $H$ be a fixed probability measure.   Suppose   $G^* \sim FSBP(p, \gamma, H)$ and that $n$ observations are sampled from $G^*$.   Without loss of generality, denote $C=\{c_1, \ldots, c_K\}$ a partition of the $n$ observations, with $1\le K \le n.$ Furthermore, let $\bm{\lambda} = \{\lambda_1, \ldots, \lambda_K\}$ be a permutation of $\{1, \ldots, K\}$ and $S_K$ denote the set of all $K!$ possible permutations of $\{1, \ldots, K\}$.    The EPPF of $G^*$ for $n$  observations is given by
    \[
        \prod_{k=1}^K \Gamma(\gamma+1) p^{|c_k|} \frac{\Gamma(|c_k|+1)} {\Gamma(\gamma+|c_k|+1)}
        \times
     \left( \sum_{\bm{\lambda} \in S_K}\prod_{k=1}^K\left\{ \frac{_2F_1(-\alpha_{k+1}(\bm{\lambda}),|c_{\lambda_k}|+1;\gamma + |c_{\lambda_k}| + 1;p)}{1- \, _2F_1(-\alpha_{k+1}(\bm{\lambda}),|c_{\lambda_k}|+1;\gamma + |c_{\lambda_k}| + 1;p)} \right\} \right)
     \]
where $\Gamma(\cdot)$ is the gamma function, 
$|c|$ denotes the cardinality of the set $c$,  ${}_{2}F_1(a,b;c;d)$ is the hypergeometric function with parameters $a, b, c$ and $d$, $\alpha_k(\bm{\lambda}) = |c_{\lambda_k}| + |c_{\lambda_{k+1}}| + \cdots  + |c_{\lambda_K}|$, and $c_{\lambda_k}$ is the $\lambda_k$'s component of $C$. 
When $p \rightarrow 1$, the EPPF of $G^*$ converges to the EPPF of $G_0 \sim DP(\gamma, H)$, which is given by 
$$\frac{\gamma^{|C|}\Gamma(\gamma)}{\Gamma(n+\gamma)}  \prod_{k=1}^K \Gamma(|c_k|).$$
\end{theo}
\noindent The proof of theorem \ref{theo:3} is given in Supplement A.7.   Details of the hypergeometric function can be found in \cite{abramowitz1988handbook}.

We next explore the clustering property of the FSBP 
to   show that the expected number of clusters in $G^*$ is greater than the 
corresponding DP with $G_0 \sim DP(\gamma, H)$. 
The first lemma   derives    the probability of forming a new cluster   under FSBP.

\begin{lemma} \label{lemma:Gs1}
    Let $p \in (0, 1)$ and $\gamma > 0$ be fixed constants, and let $H$ be a fixed probability measure. Let $G^* \sim FSBP(p, \gamma, H)$, and let $\bm{\theta}_1, \cdots , \bm{\theta}_i|G^* \sim G^*$. Denote $w_i$ as a binary indicator for the $i$th sample $\bm{\theta}_i$, such that
    $$w_i = \left \{ \begin{matrix} 1 & \text{if }  \bm{\theta}_i \notin \{\bm{\theta}_1, \cdots , \bm{\theta}_{i-1}\} \\ 0 & \text{o.w.} \end{matrix} \right. .$$
    Then, for $i \geq 2$,
    $$Pr(w_i = 1|p, \gamma) = 1 - \sum_{k=2}^{i} (-1)^k {{i-1}\choose{k - 1}} \frac{(k-1)!}{\prod_{l = 1}^k (l +   \gamma  )} \frac{(\gamma+1)p^{k-1}}{ {}_{2}F_{1}(1, 1-k; \gamma+2; p)   }$$
    where ${}_{2}F_{1}(a,b;c;d)$ is the hypergeometric function. 
\end{lemma}
\noindent The proof of Lemma \ref{lemma:Gs1} is in Supplement A.8.

\begin{lemma} \label{lem:Gs2}
     It follows that  
    $$\lim_{p \rightarrow 1} \text{Pr}(w_i = 1|p, \gamma) = \frac{\gamma}{\gamma + i - 1}.$$
\end{lemma}
\noindent The proof of Lemma \ref{lem:Gs2} is in Supplement A.9.   Note that the right hand side of the equation in Lemma \ref{lem:Gs2} is the probability of forming a new cluster under DP \citep{muller2015bayesian}.   
Based on Lemma \ref{lemma:Gs1} and \ref{lem:Gs2}, we have the following theorem.

\begin{theo} \label{theo:Gs}
    Let $p \in (0, 1)$, $\gamma > 0$ be fixed constants, and $w_i$ be defined as in Lemma \ref{lemma:Gs1}. Then
    $$Pr(w_i = 1|p, \gamma) > \frac{\gamma}{\gamma + i - 1}.$$
\end{theo}
\noindent The proof of Theorem \ref{theo:Gs} is shown in Supplement A.10. The following corollary follows directly from Theorem \ref{theo:Gs}.

\begin{coro} \label{coro:Gs}
    Let $n^*$ be the   expected   number of clusters of $G^* \sim FSBP(p, \gamma, H)$ on $n$ samples.   Then  
    $$E[n^*|p,\gamma] = 1 + \sum_{i=2}^n Pr(w_i = 1|p, \gamma).$$
    Let $n_0$ be the   expected   number of clusters of $G_0 \sim DP(\gamma, H)$ on $n$ samples.   Then  
    $$E[n_0|\gamma] = \sum_{i=1}^n \frac{\gamma}{\gamma + i - 1}.$$
    Additionally, we have
    $$E[n^*|p,\gamma] > E[n_0|\gamma] \approx \gamma\log\left(\frac{\gamma+n}{\gamma}\right).$$
\end{coro}

\begin{remark} \label{remark:Gs}
    The FSBP  has a   larger    expected number of clusters than  DP   with the same concentration parameter . 
\end{remark}

In summary, FSBP in \eqref{eq:fsbp_g} can be considered as a ``truncated" DP
with a factor of $p$. When $p=1$, FSBP is the same as DP. 

\section{Posterior Inference}
\label{sec:sam_infer}
\subsection{Overview}
We develop computational algorithms for sampling  PAM and FSBP. We do not consider sampler for ASP since ASP can be viewed as PAM for a single group and thus can be sampled similarly as PAM.   
For PAM, we modify an efficient slice sampler  in \cite{denti2021common} and illustrate the new algorithm using univariate data. The modified sampler  
can be easily extended to accommodate multivariate observations (i.e., $q > 1$)   and discrete data. Alternative approaches like the Gibbs sampler based on the Chinese restaurant franchise process \cite{teh2004sharing} or blocked-Gibbs sampler \cite{rodriguez2008nested} by truncating the infinity mixture in PAM are not considered, as they are either not feasible or prone to inferential errors.   

\subsection{Slice Sampler for   PAM and FSBP  }

To facilitate the development of the   slice   sampler for PAM, we adopt the parametrization in \cite{denti2021common} and \cite{teh2004sharing}, adding the sampling model for observation $y_{ij}$.   
  Specifically,   the proposed PAM can be represented using a set of latent indicator variables   $\bm{Z} = \{z_{ij}; \, i \geq 1, \, j = 1, \ldots, J\}$ 
as cluster memberships for the observations. In other words,  $z_{ij} = k$ if observation $i$ in group $j$ is assigned to cluster $k$. Denoting $\bm{\pi}_j = \{\pi_{jk}; k \geq 1\}$   and adding the sampling model for $y_{ij}$, we consider a PAM mixture model  as   
\begin{equation} \label{eq:PAM-mix}
\arraycolsep=1.4pt\def\arraystretch{1.25}
    \begin{array}{ll}
        y_{ij}|z_{ij},\bm{\Phi} \sim F(\bm{y}_{ij}| \bm{\phi}_{z_{ij}}), \,\,\,\,  \\ 
        z_{ij}|\bm{\pi}_j \sim \sum_{k=1}^{\infty} \pi_{jk}\delta_{k}(z_{ij}), \,\,\,\, 
        \pi_{jk} = \pi_{jk}'\prod_{l=1}^{k-1}(1-\pi_{jl}'), 
        \\
        f(\pi_{jk}'|\bm{\beta}, p_j, \alpha_0) = p_j \times f_{\text{Beta}}\left( \alpha_0\beta_k, \alpha_0\left(1-\sum_{l=1}^k \beta_l\right) \right) + (1 - p_j) \times   \delta_0.   \\
    \end{array}
\end{equation}
The other components of PAM are the same as \eqref{eq:PAM_g} and \eqref{eq:PAM_hyper}.   This reparameterization is routinely used to facilitate posterior inference \citep{denti2021common,teh2004sharing}.
 
By integrating out $z_{ij}$ in   model    \eqref{eq:PAM-mix}, we can rewrite the density  function for $y_{ij}$ as an infinite mixture as 
\begin{equation}\label{eq:like}
    f(y_{ij}|\bm{\Phi},\bm{\pi}_j) = \sum_{k \geq 1}\pi_{jk} \cdot p(y_{ij}|\bm{\phi}_{k}), 
\end{equation}
where $\bm{\Phi} = \{\bm{\phi}_k   ; k \geq 1 \}  $.
Following \cite{kalli2011slice}, we use a set of uniformly distributed random variables $\bm{u} = \{u_{i,j}  ; i = 1, \ldots, n_j, j = 1, \ldots, J  \}$ to separate the ``active" mixture components from the other ``inactive" components, which will become clear next. By definition, each $u_{ij} \sim \text{Unif}(0,1)$. Additionally, we consider $J$ deterministic probabilities $\bm{\xi}_j = \{\xi_{jk}  ; k \geq 1\}  $ for a fixed $j$,  where $\xi_{jk} \equiv \xi_k = (1-\zeta)\zeta^{k-1}$, $\zeta \in (0,1)$ is a fixed parameter with a default value of $0.5$, and $\bm{\xi}_j \equiv \bm{\xi} = \{\xi_k  ; k \ge 1\} $. A more complicated construction may allow different $\zeta_j$ for different groups $j$, which we do not consider here.
As a result, the augmented likelihood for observation $y_{ij}$ can be expressed as:
\begin{equation} \label{eq:9}
    f_{\bm{\xi}}(y_{ij}, u_{ij}|\bm{\Phi},\bm{\pi}_j) = \sum_{k\geq 1} 1_{\{u_{ij} < \xi_k\}} \frac{\pi_{jk}}{\xi_k} p(y_{ij}|\bm{\phi}_k),
\end{equation}
  where $1_{\{A\}}$ equals 1 if condition A is satisfied, and 0 otherwise.  
Integrating   out 
$u_{ij}$ in \eqref{eq:9}   returns $f(y_{ij}|\bm{\Phi},\bm{\pi}_j)$ in \eqref{eq:like}. 
Now adding the cluster indicator $z_{ij}$ in \eqref{eq:PAM-mix}, we express \eqref{eq:9} as 
\begin{equation} \label{eq:10}
    f_{\bm{\xi}}(y_{ij}, u_{ij}|z_{ij},\bm{\Phi},\bm{\pi}_j) = \sum_{k\geq 1} 1_{\{z_{ij}=k\}} 1_{\{u_{ij} < \xi_{z_{ij}}\}} \frac{\pi_{jz_{ij}}}{\xi_{z_{ij}}} p(y_{ij}|\bm{\phi}_{z_{ij}})
\end{equation} 

The proposed slice sampler follows a Gibbs-sampler style, in which it iteratively samples the following parameters, 
\begin{enumerate}
    \itemsep0em 
    \item $u_{ij}|\cdots  \propto   \text{Unif}(0, \xi_{z_{ij}})  $,
    \item the stick-breaking weights $\beta_k'$, $\pi_{jk}'$, and $p_j$,
    \item the indicator $z_{ij}$ with $\text{Pr}(z_{ij} = k|\cdots ) \propto 1_{\{u_{ij}<\xi_k\}}\frac{\pi_{jk}}{\xi_k}p(y_{ij}|\bm{\phi}_k)$, and
    \item  the atom location parameter $\bm{\phi}_k|\cdots  \propto \prod_{z_{ij} = k}N(y_{ij}|\bm{\phi}_k)p_H(\bm{\phi}_k)$.
\end{enumerate}
In the last step, since $\bm{\phi}_k \sim H$, $p_H(\bm{\phi}_k)$ denotes the prior density of $H$. The entire sampler is  presented in  Algorithm \ref{alg:cap}. Below we describe the details of sampling $\pi_{jk}'$ in step 2 above. The   other   details of the entire slice sampler are in Supplement A.11. 

In each iteration of the slice sampler, due to the introduction of   the   latent uniform variate $u_{ij}$ and the truncation on $\xi_k$, the infinite summation in equation \eqref{eq:9} can be reduced to a finite sum through ``stochastic truncation". To see this, first notice that  $\{\xi_k  ; k \geq 1  \}$ is a descending sequence, and therefore only finitely many $\xi_k$'s can meet the condition   $u_{ij} < \xi_k$.  
In other words, given $\bm{u}$, there exists a $K' \ge 1$ such that when $k \ge K'$,  $\min(  u_{ij}) \geq \xi_{k}$, where the min is taken over all $i$ and $j$. 
This means that up to $K'$ of the $\xi_k$'s will be larger than  $u_{ij}$. Let $K^* = K' - 1$. Then, noticing that   $\xi_{K^*} = (1-\zeta)\zeta^{K^*}$,   we can easily show that  

\begin{equation} \label{eq:Kstar}
K^* = \floor*{\frac{\log(\min(\bm{u}))-\log(1-\zeta)}{\log(\zeta)}}.
\end{equation}

\noindent Here, $K^*$ is called the ``stochastic truncation" in the slice sampler. Given $K^*$, sampling $\beta_k'$ is straightforward but requires a Metropolis-Hastings (MH) step (See Supplement A.11 for details). To sample $\pi_{jk}'$, again conditional on $K^*$, let

$\bm{Z}_j = \{z_{ij}  ; i = 1, \ldots, n_j  \}$, $m_{jk} = \sum_{i=1}^{n_j} 1(z_{ij} = k)$, 
and refer to the stick-breaking representation. The full conditional distribution of $\pi_{jk}'$ is given by
$$p(\pi_{jk}'|\cdots ) = p(\pi_{jk}'|\bm{Z}_j,\bm{\beta},p_j,\alpha_0) \propto \left[{\pi_{jk}'}^{m_{jk}}(1-\pi_{jk}')^{\sum_{s=k+1}^{K^*}m_{js}}\right] f(\pi_{jk}')$$
where $f(\pi_{jk}')$ is defined in equation \eqref{eq:zab}. When $m_{jk} > 0$, 
it means cluster $k$ in group $j$ is not empty, and therefore $\pi'_{jk} \ne 0$ (otherwise, it would not be possible to have a non-empty cluster $k$ in group $j$). Hence, the full conditional of $\pi_{jk}'$ is  
\begin{equation} \label{eq:11}
    p(\pi_{jk}'|\cdots ) = f_{\text{Beta}} \left(\alpha_0\beta_k + m_{jk}, \alpha_0\left(1-\sum_{l=1}^k \beta_l \right) + \sum_{s = k+1}^{K^*} m_{js}\right).
\end{equation}
Recall $f_{\text{Beta}}(,)$ denotes a beta distribution density. When $m_{jk} = 0$, which could mean $\pi_{jk}' = 0$ or $\pi_{jk}' \neq 0$ but the atom is not sampled, we have 
$$p(\pi_{jk}'|\cdots ) \propto (1-\pi_{jk}')^{\sum_{s=k+1}^{K^*} m_{js}} f(\pi_{jk}'). $$
This can be expressed as 
\begin{equation} \label{eq:12}
     p(\pi_{jk}'|\cdots)  = p_j^* \times f_{\text{Beta}}\left(\alpha_0\beta_k, \alpha_0\left(1-\sum_{l=1}^{k}\beta_l\right) + \sum_{s=k+1}^{K^*}m_{js}\right) + (1-p_j^*) \times \delta_0 
\end{equation}
where 
$$p_j^* = \frac{p_j}{p_j + (1-p_j)\times \frac{B\left(\alpha_0\beta_k,\alpha_0\left(1-\sum_{l=1}^{k}\beta_l\right)\right)}{B\left(\alpha_0\beta_k,\alpha_0\left(1-\sum_{l=1}^{k}\beta_l\right)+\sum_{s=k+1}^{K^*}m_{js}\right)}}$$
and $B(a,b)$ is the beta function.

Lastly, sampling $p_j$ and the concentration parameters follow standard MCMC simulation \citep{escobar1995bayesian},  details of which is provided in Supplement A.11.

\paragraph{Additional step for count data} Finally, for DPAM an additional step is added to update the latent continuous variable. Denote $\text{TN}(\mu, \sigma^2;a,b)$ the truncated normal distribution with mean $\mu$, variance $\sigma^2$, and boundaries $a$ and $b$, the full conditional distribution of $y_{ij}$ is
\begin{equation} \label{eq:14}
    y_{ij}|\cdots  \sim \text{TN}(\mu_{z_{ij}},\sigma_{z_{ij}}^2; a_{x_{ij}}, a_{x_{ij}+1}).
\end{equation}

\paragraph {Computation Algorithm} 
Algorithm \ref{alg:cap} introduces the proposed  slice sampler. For multivariate observations, step 9 of Algorithm \ref{alg:cap} can be replaced with a conjugate NIW prior, and multivariate normal can be used for $p(y_{ij}|\bm{\phi}_k)$ in step 8. On the other hand, the extension to DPAM can be achieved by adding steps to sample the latent $y_{ij}$ according to equation \eqref{eq:14} after step 7, and modifying the likelihood $p(y_{ij}|\bm{\phi}_k)$ in step 8 with 
$$p(x_{ij}|\bm{\phi}_k) = \Delta\Phi(a_{x_{ij}}|\bm{\phi}_k) =  \Phi(a_{x_{ij}+1}|\bm{\phi}_k) - \Phi(a_{x_{ij}}|\bm{\phi}_k),$$
where $\Phi(\cdot)$ denotes the cumulative distribution function (c.d.f) of the Gaussian distribution. 

\begin{algorithm}
\caption{Slice-Efficient Sampler for PAM}\label{alg:cap}
\begin{algorithmic}[1]
\For{  $m = 1, \ldots, M$  }
\State Sample each $u_{ij}$ from $u_{ij} \sim \text{Unif}(0, \xi_{z_{ij}})$ and find $K^*$ in \eqref{eq:Kstar}.
\State Sample all $\beta_k'$ for $k = 1, \cdots , K^*$ with MH step.
\For{each $\pi_{jk}'$ for $j = 1, \cdots , J$ and $k = 1, \cdots , K^*$}
    \State \textbf{if} $m_{jk} > 0$, sample $\pi_{jk}'$ from \eqref{eq:11}. \textbf{otherwise}, sample $\pi_{jk}'$ from \eqref{eq:12}.
\EndFor
\State Sample $\bm{p} = \{p_j  ; j = 1, \ldots, J  \}$: denote $m_{j0} = \sum_{k=1}^{K^*} 1(\pi_{jk}' = 0)$,
$$p_j|\cdots  \sim \text{Beta}(a + K^* - m_{j0}, b + m_{j0})$$ 
\State Sample $\bm{Z} = \{z_{ij}  ; i = 1, \ldots, n_j, j = 1, \ldots, J  \}$ from the following full condition:
$$p(z_{ij} = k|\cdots ) \propto 1_{\{u_{ij} < \xi_{k}\}}\frac{\pi_{jk}}{\xi_{k}}p(y_{ij}|\bm{\phi}_k)$$
\State Sample $\bm{\phi}_k$ from a conjugate NIG.
\EndFor
\end{algorithmic}
\end{algorithm}

\paragraph {Label Switching}

As PAM involves an infinite mixture model, the issue of label switching can arise in MCMC samples \citep{papastamoulis2015label}.
To address the problem of label switching, we use the Equivalence Classes Representatives (ECR) algorithm described in \cite{papastamoulis2010artificial}. Details of label-switching with the ECR method are in Supplement A.11.

\paragraph{Slice sampler for FSBP}

The slice sampler for FSBP follows the same flow as the one for PAM above. We simply need to add the sampling model   \eqref{eq:y_i_ll} and rewrite the FSBP in  \eqref{eq:fsbp_g} using latent indicator variables $\bm{Z} = \{z_i; i \geq 1\}$ in a mixture model given by   
\begin{equation}
\arraycolsep=1.4pt\def\arraystretch{1.25}
\begin{array}{l}
\bm{y}_i|z_i,\bm{\Phi} \sim F(\bm{\phi}_{z_i}), \,\, z_i|\bm{\pi} \sim  \sum_{k \geq 1} \pi_k \delta_k, \\
\pi_k = p\cdot{\pi_k}' \prod_{l = 1}^{k-1} (1 - p\cdot {\pi_l}'), \\
\pi_k' \sim \text{Beta}(1,\gamma), \,\, \bm{\phi}_k \sim H,
\end{array}
\nonumber
\end{equation} 
where $\bm{\pi} = \{\pi_k; k \geq 1\}$.  The detail of the sampler is almost identical to PAM and left for  Supplement A.12. 

\subsection{Inference on Clusters}
\label{subsec:inf_c_u}

Like all BNP models,   both   PAM   and FSBP   produce    random clusters and their associated posterior distributions.   The slice sampler in the previous section produces Markov chain Monte Carlo (MCMC) samples that eventually converge to the true joint posterior distribution of all the parameters. These samples are used for posterior inference, including estimating a single clustering outcome of the observations, even though the posterior distribution of the clusters is available. 
We discuss the corresponding inference under PAM next. We consider two approaches but only present one of them below, leaving the other approach to the Supplement A.11.

First, for the $m$th MCMC sample, denote the  
matrix of cluster memberships of all the observations as   $\bm{Z}^{(m)} = \{z_{ij}^{(m)}; \; i = 1, \ldots, n_j, j = 1, \ldots, J \}$, and the vector of observations in the $j$th group as $  \bm{Z}^{(m)}_j   = \{z^{(m)}_{1j}, \ldots, z^{(m)}_{n_j j}\}.$ These $z$ values can be sampled in Step 8 of Algorithm \ref{alg:cap}. 
Let $\bm{t}^{(m)}_j = \left\{t^{(m)}_1, \ldots,   t^{(m)}_{K_j^{(m)}}  \right\}$ denote the labels of these clusters, which are the unique values of  the cluster memberships in $\bm{Z}^{(m)}_j$.   Here $K_j^{(m)}$ represents the number of clusters in group $j$ for the $m$th sample.   Then the set and number of common clusters between groups $j$ and $j'$ are given by $\bm{t}^{(m)}_j \cap \bm{t}^{(m)}_{j'}$ and its cardinality, respectively, and the set and number of unique clusters for group $j$ are given by $\bm{t}^{(m)}_j \; mod  \;\bm{Z}^{(m)}\backslash \bm{Z}^{(m)}_j$ and its cardinality, respectively. Here, operation $A \; mod \; B$ for two sets $A$ and $B$ is redefined as the unique elements in $A$ but not $B$, and $\bm{Z} \backslash \bm{Z}_j$ means the set after removing $\bm{Z}_j$ from $\bm{Z}.$   Through these operations, for every MCMC sample $m$ we obtain clustering results for the observations. Together, all the MCMC samples constitute an approximation of the posterior distributions of the clusters. 

To produce a point estimate of the clustering result, we follow the approach in \cite{wade2018bayesian} to estimate an optimal partition through a decision-theoretic approach that minimizes the variation of information \citep{meilua2007comparing}. This optimal partition is then used as a ``point estimate" of the random clusters obtained from PAM   or FSBP   posterior inference. 

\section{Simulation Study}
\label{sec:sim}

\subsection{Simulation Setup}

We assess the performance of PAM and FSBP via simulation. The ASP model is not evaluated since it is simply a PAM for a single group. In the simulation, we generate data from a Gaussian finite mixture model with specific clustering patterns, and apply BNP models as a prior for data analysis. Posterior inference from the BNP models is then compared to the simulation truth. We compare PAM with CAM and HDP in Scenarios 1 and 2, and FSBP with DP in   Scenario   3. In all simulations, the variance is $\sigma^2 = 0.6$ in the Gaussian mixture.

\paragraph{Scenario 1 - PAM Univariate data} We consider three cases under   Scenario   1 to assess the performance of PAM under various clustering patterns. 

\noindent \textbf{Case 1: Unique Clusters} We generate data from groups that have non-overlapping clusters. This extreme case provides an evaluation of models' performance to capture unique clusters.   
We assume $J = 2$ groups, each with $n_j = n = 200$ samples. Within each group, the observations are generated from a mixture of four Gaussian distributions with distinct means. 
In mathematical terms, we have
$$f(y_{ij}) \propto   \sum_{k =1}^4   \frac{1}{4}N(m_{jk},\sigma^2), \, i = 1, \ldots, n, \, j = 1, 2,$$
where $m_{jk}$ represents the cluster mean for Group $j$ and cluster $k$. 
There is a total of 8 clusters across two groups. For Group 1, the cluster means are $m_{1k} \in \{0, 4, 8, 12\}$, and for Group 2, the cluster means are $m_{2k} \in \{-16, -12, -8, -4\}$. 

\noindent \textbf{Case 2: A Single Common Cluster}   In this case, we assume the presence of one common cluster between  groups. Specifically, we consider $J = 3$ groups, each comprising $n_j = n = 100$ samples. The observations in each group   again   follow a mixture of Gaussian distributions. 
A common cluster with mean 0 is shared across all three groups, while each group possesses its own unique clusters.   Details regarding the cluster means and weights in each group can be found in Table A.2 in Supplement A.13. 

\noindent \textbf{Case 3: Nested Clusters}   In this case, taken from \cite{denti2021common}, nested clusters are generated across groups.  
Specifically,  
let $J = 6$ groups.   Ascending number and overlapping clusters are generated via the mixture of Gaussian distributions given by   
$$f({y}_{ij}) \propto \sum_{k=1}^j \frac{1}{  j  }N(m_k, \sigma^2), \quad i=1, \ldots, n_j, \; j = 1, \cdots , 6, $$
where the cluster means $m_k \in \{0, 5, 10, 13, 16, 20\}$ for $j=1, \ldots, 6$.   Therefore, there are $j$ true clusters in group $j$ and clusters in group $j$   is nested in group $(j+1),$   
with only the first cluster $N(m_1, \sigma^2)$ shared across all six groups. 
We test two sub-cases of Case 3 by setting the number of observations in group $n_j = n_A$, where $n_A \in \{50, 100, 150\}$, or by setting $n_j = n_B\times j$, where $n_B \in \{10, 20, 40\}$.

\paragraph{Scenario 2 - PAM Multivariate data} In this scenario,   each observation $y_{ij}$ is assumed to be a $3$-dimensional vector.    Additionally, we consider   $J = 3$ groups, each with $n_j = n$ subjects, where $n \in \{50, 100, 200\}$. The multivariate   observations are generated from a mixture of multivariate Gaussian distributions, with the cluster means and weights shown in Table A.3 in Supplement A.13. 
The true covariance matrix is assumed to be the identity matrix. 
There are a total of five clusters across all groups:
Group 1 possesses all five  clusters, Group 2 has three clusters (clusters 1, 3 and 4), and Group 3 has two clusters (clusters 2 and 3).   Note that cluster 3 is the only common cluster across all three groups.  

In both scenarios,   
we compare the performance of PAM with HDP and CAM.   We obtain a point-estimate of clustering results based on the procedure in \cite{wade2018bayesian} and  assess the models' performance based on  the following criteria.   
\begin{enumerate}
    \item The total number of clusters,  number of common clusters, and number of unique clusters based on   the  estimated clustering results.   
    \item The adjusted Rand index (ARI) \citep{hubert1985comparing} between   the estimated clustering results    
    and the ground truth, with a value closer to 1 indicating better performance. 
    \item The normalized Frobenius distance (NFD) \citep{horn1990norms} between the estimated posterior pairwise co-clustering matrices and the true co-clustering structure, with a value closer to 0 indicating better performance. 
\end{enumerate}
These metrics have been routinely adopted in the literature, e.g., in \cite{denti2021common}.
 
\paragraph{Scenario 3 - FSBP Univariate data} 
In this scenario, we evaluate the performance of FSBP. We consider $n = 300$ observations, each following a mixture of five Gaussian distributions with distinct means, given by 
$$f(y_i) \propto \sum_{k=1}^5 \frac{1}{5}N(m_k,\sigma^2), \; i = 1, \ldots, n,$$
where $m_k \in \{0, 3, 6, 9, 12\}$. FSBP is then compared to DP, and the performance is assessed based on the estimated posterior density function, as well as the number of clusters inferred with each method.  

\subsection{Simulation Results for PAM and FSBP}
\label{sec:sim_1}

\paragraph{  Scenario 1 }
 
We generate 30 datasets for each available sample size in each case. For each simulated dataset, we adopt standard prior settings for the hyperparameters in   model    \eqref{eq:PAM_g}. Specifically, we use the NIG distribution as the base measure $H$, with hyperparameters $\mu_0 = 0$, $\kappa_0 = 0.1$, $\alpha_0 = 3$ and $\beta_0 = 1$. We use Jeffrey's prior for $p_j$'s, i.e., $a = b = 0.5$. Lastly, we set $a_{\alpha_0} = b_{\alpha_0} = a_{\gamma} = b_{\gamma} = 3$ for the gamma priors of the concentration parameters $\alpha_0$ and $\gamma$. We collect an MCMC sample of 10,000 iterations after 10,000 iterations of burn-in. The Markov chains mix well. 
  
We present the simulation results for all three cases in Table \ref{tab:sc1case12}. The winning performance is highlighted in bold font. We use notation $G_j$ to denote the group $j$. Full results in terms of cluster numbers are presented for cases 1 and 2. For Case 3, results from one sample size $n=150$ is presented, and we selected clustering results for G5 and G6 as they are representative of the models' distinct behavior. Full results are reported in Supplement A.13.   Overall, PAM exhibits competitive performance in terms of identifying the correct number of clusters and ARI/NFD scores. PAM also is the most stable method consistently producing the smallest standard deviations. In Case 1, PAM is superior in capturing the special clustering structure where no clusters are shared across groups. In contrast, HDP seems to struggle in identifying the unique clusters in this case. These can be found in ``Number of clusters" for ``All groups" in the table. To further examine the model fitting in   Case   1, Figure A.6 in Supplement A.13 shows that HDP (middle panel) sometimes merge two different clusters in the posterior inference, leading to under-estimated cluster numbers. CAM and PAM appear to be able to avoid this and report mostly the correct clustering structure.   In   Case   2, CAM and HDP are the better methods, both able to capture the sole common cluster more often than PAM. These two cases seem to show distinct behavior of PAM vs CAM and HDP. We confirm this in case 3. In particular, PAM is  more likely to  identify the correct number of clusters across all groups as the average number of clusters under PAM is 5.97 compared to 4.97 for CAM and 4.27 for HDP.  However, CAM is better at identifying common clusters, say between G5 and G6, while PAM is more capable of finding the unique cluster in G6. 

\begin{table}[!htbp]
\footnotesize
    \centering
    \resizebox{\linewidth}{!}{
    \begin{tabular}{c|lc|ccc|c}
       \hline
       Case & \multicolumn{2}{|c|}{Metrics}  & CAM & HDP & PAM & Truth \\
       \hline
       \multirow{8}{*}{Case 1} &\multirow{3}{*}{\# of clusters} & All groups & 7.87 (0.35) & 5.80 (0.66) & \textbf{7.97 (0.18)} & 8 \\
       & & G1 & 4.07 (0.25) & 3.37 (0.56) & \textbf{3.97 (0.18)} & 4 \\
       & & G2 & 3.87 (0.35) & 2.43 (0.50) &\textbf{4.00 (0.00)} & 4 \\
       \cline{2-7}
       & \# of common clusters & & 0.07 (0.25) & \textbf{0.00 (0.00)} & \textbf{0.00 (0.00)} & 0 \\
       \cline{2-7}
       & \multirow{2}{*}{\# of unique clusters} & G1 & \textbf{4.00 (0.00)} & 3.37 (0.56) & 3.97 (0.18) & 4 \\
       & & G2 & 3.80 (0.48) & 2.43 (0.50) & \textbf{4.00 (0.00)} & 4 \\
       \cline{2-7}
       &  & ARI  & 0.96 (0.04) & 0.67 (0.09) & \textbf{0.97 (0.02)} & \\
       & & NFD  & \textbf{0.01 (0.01)} & 0.09 (0.03) & \textbf{0.01 (0.01)} & \\
       \hline
       \multirow{13}{*}{Case 2} & \multirow{4}{*}{\# of clusters} & All groups & \textbf{5.00 (0.00)} & 5.03 (0.18) & 5.07 (0.25) & 5 \\
       & & G1 & \textbf{2.27 (0.91)} & 1.70 (0.47) & 1.67 (0.48) & 2 \\
       & & G2 & 3.80 (0.71) & \textbf{2.83 (0.53)} & 2.60 (0.50) & 3 \\
       & & G3 & 2.97 (0.76) & 2.13 (0.35) & \textbf{2.00 (0.00)} & 2 \\
       \cline{2-7}
       & \multirow{4}{*}{\# of common clusters} & All groups & \textbf{1.20 (0.71)} & 0.47 (0.51) & 0.30 (0.47) & 1 \\
       & & G1 and G2 & 1.80 (0.96) & \textbf{0.60 (0.56)} & 0.33 (0.48) & 1 \\
       & & G1 and G3 & 1.40 (0.77) & \textbf{0.80 (0.61)} & 0.63 (0.49) & 1 \\
       & & G2 and G3 & 2.03 (0.93) & \textbf{0.70 (0.47)} & 0.53 (0.51) & 1 \\
       \cline{2-7}
       & \multirow{4}{*}{\# of unique clusters} & G1 & 0.27 (0.45) & 0.77 (0.50) & \textbf{1.00 (0.26)} & 1 \\
       & & G2 & 1.17 (0.65) & \textbf{2.00 (0.00)} & 2.03 (0.18) & 2 \\
       & & G3 & 0.73 (0.52) & \textbf{1.10 (0.31)} & 1.13 (0.35) & 1 \\
       \cline{2-7}
       &  & ARI  & 0.85 (0.04) & \textbf{0.87 (0.05)} & \textbf{0.87 (0.04)} & \\
       &  & NFD  & 0.04 (0.01) & \textbf{0.03 (0.01)} & \textbf{0.03 (0.01)} & \\
       \hline
       \multirow{6}{*}{\makecell{Case 3 \\ ($  n_j   = 150$)}} & \multirow{4}{*}{\# of clusters} & All groups &  4.97 (0.49) &  4.27 (0.58) & \textbf{5.97 (0.62)} & 6 \\
       & & G5 & \textbf{4.43 (0.50)} & 3.33 (0.48) & 4.13 (0.57) & 5 \\
       & & G6 & \textbf{4.60 (0.62)} & 3.17 (0.46) & 4.53 (0.68) & 6 \\
       \cline{2-7}
       & \# of common clusters & G5 and G6 & \textbf{4.43 (0.50)} & 3.13 (0.35) & 3.47 (0.68) & 5 \\
       \cline{2-7}
       & \multirow{2}{*}{\# of unique clusters} & G5 & \textbf{0.00 (0.00)} & \textbf{0.00 (0.00)} & 0.53 (0.51) & 0 \\
       & & G6 & 0.13 (0.35) & 0.03 (0.18) & \textbf{0.90 (0.40)} & 1 \\
       \cline{2-7}
       & & ARI  & \textbf{0.95 (0.02)} &  0.90 (0.04) & 0.95 (0.03) & \\
       & & NFD & 0.07 (0.02) &  0.03 (0.01) & \textbf{0.02 (0.01)} & \\
       \hline
    \end{tabular}}
    \caption{Simulated univariate data in Scenario   1. Clustering performance of CAM, HDP, and PAM is evaluated based on the following metrics: number of clusters across all and individual groups, number of common clusters across all groups and pairwise groups, number of unique clusters within each group, Adjusted Rand Index (ARI), and normalized Forbenius distance (NFD). Entries represent the Mean (SD) over 30 datasets. Bold entries mean the corresponding model performs the best with the corresponding metric. Note that the notation G1 to G6 refers to Group 1 to Group 6, respectively.} 
    \label{tab:sc1case12}
\end{table}


 Since by definition PAM allows the weight $\pi_{jk}$ of cluster $k$ in group $j$ to be zero, it can output $\text{Pr}({\pi}_{jk} = 0|\text{Data})$ which can be interpreted as cluster $k$ is absent from  group $j$. 
 In addition, $\text{Pr}({\pi}_{jk} > 0|\text{Data})$ describes the posterior probability that cluster $k$ is present in group $j$, and $\text{Pr}({\pi}_{jk} > 0, \pi_{j'k}>0|\text{Data})$  the posterior probability that 
 cluster $k$ is shared between groups $j$ and $j'$. More generally, a posterior probability of different configurations of $\pi$'s can be used to estimate more complex clustering patterns. In contrast,  common-atoms models like HDP and CAM assign $\text{Pr}({\pi}_{jk} = 0|\text{Data}) \equiv 0$ and $\text{Pr}({\pi}_{jk} > 0|\text{Data}) \equiv 1$ by definition. A work-around might be to report the frequency of a cluster sampled in the MCMC iterations in a group, which can be used as an approximation to the probability a cluster belongs to the group.

To illustrate our point, in Table \ref{tab:prob_c}, we present posterior summaries of PAM and CAM using a simulated dataset under Case 1, in which  clusters 1-4 belong to group 2 (G2) and 5-8 to group 1 (G1). Both PAM and CAM report small, $<0.01$, but non-zero  estimated cluster weights (posterior mean). However, PAM  reports large posterior probabilities of ``Unique in G1" for clusters 5-8 and of ``Unique in G2" for clusters 1-4, while those posterior probabilities are 0's for CAM. Therefore, PAM gives a more interpretable summary based on the posterior probability of atom weights equal to or greater than 0.

\begin{table}[!htbp]
    \centering
    \resizebox{\linewidth}{!}{
    \begin{tabular}{lc|cccc|cccc}
    \hline
         & & Cluster 1 & Cluster 2 & Cluster 3 & Cluster 4 & Cluster 5 & Cluster 6 & Cluster 7 & Cluster 8 \\
         \hline
         \multicolumn{2}{l|}{True mean} & -16 & -12 & -8 & -4 & 0 & 4& 8 & 12 \\ 
         \hline
         \multirow{2}{*}{True weight} & G1 & \multicolumn{4}{c|}{0.00} 
         & 0.25 & 0.25 & 0.25 & 0.25 \\
         & G2 & 0.25 & 0.25 & 0.25 & 0.25 & \multicolumn{4}{c}{0.00} 
         \\
         \hline
          & & \multicolumn{8}{c}{PAM Estimates (CAM Estimates)} \\
         \hline
         \multicolumn{2}{l|}{Mean} & -15.99 (-15.98) & -11.81 (-11.81) & -7.74 (-4.82) & -4.07 (-3.74) & 0.11 (0.13) & 3.91 (3.92) & 7.85 (7.84) & 11.96 (11.93) \\
         \hline
         \multirow{2}{*}{Weight} & G1 & \multicolumn{4}{c|}{$< 0.01 (<0.01)$ } 
         & 0.19 (0.24) & 0.24 (0.24) & 0.23 (0.21) & 0.25 (0.25) \\
         & G2 & 0.22 (0.22) & 0.27 (0.26) & 0.22 (0.12) & 0.29 (0.22) & \multicolumn{4}{c}{$< 0.01 (<0.01)$ } 
         \\
         \hline
         \multicolumn{2}{l|}{\makecell{Unique in G1 \\ $\widehat{\text{Pr}}({\pi}_{1k} > 0,$ $\pi_{2k}=0 | \text{Data})$}}
         & 0.00 (0.00) & 0.00 (0.00) & 0.00 (0.00) & 0.00 (0.00) & 0.75 (0.00) & 0.84 (0.00) & 0.81 (0.00) & 0.86 (0.00) \\ \cline{1-2}
         \multicolumn{2}{l|}{\makecell{ Unique in G2 \\ $\widehat{\text{Pr}}({\pi}_{1k} = 0,$  $\pi_{2k}>0 | \text{Data})$}}
         & 0.78 (0.00) & 0.78 (0.00) & 0.68 (0.00) & 0.79 (0.00) & 0.02 (0.00) & 0.00 (0.00) & 0.00 (0.00) & 0.00 (0.00) \\
         \hline
        
    \end{tabular}}
    \caption{  Posterior summaries of CAM and PAM for a randomly selected dataset in Case 1 of Scenario 1. Reported estimates for Mean and Weight are posterior means. The last two rows correspond to MCMC-estimated posterior probabilities of a cluster has zero weight in one group and positive in the other. }
    \label{tab:prob_c}
\end{table}

\paragraph{Scenario Two} For the multivariate data scenario, we use the following prior settings for the  hyperparameters in \eqref{eq:PAM_g}. The NIW distribution   ($\text{NIW}(\bm{\mu}_0,\kappa_0,\nu_0,\bm{\Psi})$)   is used as the base measure $H$, with hyperparameters $\bm{\mu}_0 = \bm{0} = \{0, 0, 0\}$, $\kappa_0 = 0.1$, $\nu_0 = 4$ and $\bm{\Psi} = I_3$, where $I_3$ is the $3 \times 3$ identity matrix. Similar to Scenario   1  , we use Jeffrey's prior for $p_j$. We also set $a_{\alpha_0} = b_{\alpha_0} = a_{\gamma} = b_{\gamma} = 3$ in the gamma priors for the concentration parameters $\alpha_0$ and $\gamma$.
For simplicity, we only report the model accuracy on the number of clusters, ARI, and NFD for this simulation. We generate 30 datasets for each sample size, and summarize the results in Table A.4 in Supplement A.13.
The results indicate that all three methods have high accuracy in the multivariate data simulation. PAM performs competitively with the other two methods in terms of the ARI and NFD metrics when the sample size is large ($n \geq 100$). 

\paragraph{Scenario 3} We generate 30 datasets for the simulation of FSBP, each with 300 observations. Standard priors are adopted for the hyperparameters. Specifically, we use NIG distribution as the base measure $H$, set $\mu_0 = 0$, $\kappa_0 = 0.1$, $\alpha_0 = 3$, $\beta_0 = 1$, use Jeffrey's prior for $p$ ($p \sim \text{Beta}(0.5,0.5)$), and use $\text{Gamma}(3,3)$ for the hyperparameter $\gamma$. We collect an MCMC sample of 5,000 iterations after 5,000 iterations of burn-in. The Markov chains mix well.
The results are presented in Figure A.7 in Supplement A.13. On average, both methods successfully capture the five   true  clusters. However, FSBP provides more accurate posterior inference on the cluster distributions (top panel) and cluster numbers (bottom panel).


\section{Case Studies}
\label{sec:exp}

We apply PAM to two real-life datasets, one from a microbiome project and the other related to treatment of warts. The microbiome data demonstrate PAM's performance for count data and the warts data consists of multivariate observations.

\subsection{Microbiome Dataset}
\label{sec:exp1}

The   microbiome   dataset, reported   in    \cite{o2015fat},   measures 
microbiota abundance    for 38 healthy middle-aged African Americans (AA) and rural Africans (AF). The study   aims    to investigate the effect of diet swap between individuals of AF and AA, as traditional foods for these populations differ. The 38 study participants   are   instructed to follow their characteristic diet, such as a low-fat and high-fiber diet for AF and a high-fat and low-fiber diet for AA, for two weeks, and then swap diets for another two weeks. We consider cluster the subjects based on the measured microbiota abundance, in terms of counts of 
operational taxonomic units (OTUs), which reflect the recurrences of the corresponding OTUs in a particular ecosystem \citep{jovel2016characterization,kaul2017analysis}. For more background, refer to \cite{o2015fat} and Section 4 of \cite{denti2021common}. Hereafter, we use the term ``expression" and ``counts" interchangeably in this application.

To apply PAM, or more specifically DPAM (due to the discrete data of OTU counts), we treat each subject as a group, and counts of different OTUs as observations within a group. Following the same data-preprocessing steps as in \cite{denti2021common}, we obtain 38 subjects (17 AF and 21 AA) with 119 OTUs. Note that all the OTUs are the same,   and so a cluster here refers to a group of OTU counts, just like in \cite{denti2021common}.   
When applying to demonstrate the CAM model, \cite{denti2021common} use the entire dataset with a goal to generate nested clusters of subjects and OTU counts within subjects. 
The proposed PAM cannot cluster subjects and therefore  we randomly select four subjects from the dataset for analysis.  In a future work, we will consider extend PAM to allow nested clustering and will apply the new model to the full dataset.

We randomly select four subjects   as four groups,  two AAs (individuals 5 and 22) and two AFs (individuals 13 and 14), from the dataset.    We remove the OTUs that had zero expression in all four individuals from the selected data. In the end, we obtain a dataset with  $J = 4$ individuals (groups) and $n_j = 109$ OTUs (observations). The histograms of the microbiome populations of the four selected individuals are shown in Supplement A.14. 

Let $x_{ij}$ denote the observed OTU count for OTU $i$ from individual $j$. For inference, similar to  \cite{denti2021common}, we  incorporate the average OTU frequencies for subject $j$, denoted as $\eta_j = \frac{1}{n}\sum_{i=1}^n x_{ij}$, as a scaling factor in the latent variable $y_{ij}$ of the DPAM model in \eqref{eq:8}.  This leads to the following distribution for the change of variables:  
\begin{equation}
    y_{ij}|\bm{Z}, \bm{\mu}, \bm{\sigma}^2 \sim N(\eta_j\mu_{z_{ij}}, \eta_j^2\sigma_{z_{ij}}^2) \leftrightarrow  \frac{y_{ij}}{\eta_j}|\bm{Z}, \bm{\mu}, \bm{\sigma}^2 \sim N(\mu_{z_{ij}}, \sigma_{z_{ij}}^2)
\end{equation}
The prior hyperparameters follow the same settings as in Scenario   1  of the   simulation study, and we present the   analysis results  in Table \ref{tab:microbio}.   

PAM reports a total of eight estimated clusters across the four individuals: clusters 1 and 2 are shared by all four individuals (with posterior probabilities of 1.00 and 0.95, respectively), cluster 7 is shared among individuals 5 (AF), 13 (AA), and 14 (AA) (with posterior probability of 0.96), and cluster 8 is shared among individuals 5 and 22 (both from AF, with posterior probability of 0.93). The other clusters are unique to a specific individual (with posterior probabilities of 0.60, 0.60, 0.42, and 0.51, respectively, for clusters 3 to 6). Based on the optimal partition of OTUs, we plot the taxa counts (TC) of OTUs grouped by all eight estimated clusters as well as by both clusters and individuals in Figure A.9 in Supplement A.14. Note that for easy demonstration of clusters across individuals, we have manually reordered the clusters in ascending order based on the cluster mean. The boxplots illustrate the clusters and their distributions across individuals.

\begin{table}
\centering
\resizebox{\linewidth}{!}{
\begin{tabular}{ll|cccccccc}
    \hline 
    & & Cluster 1 & Cluster 2 & Cluster 3 & Cluster 4 & Cluster 5 & Cluster 6 & Cluster 7 & Cluster 8\\
    \hline
    \multicolumn{2}{l|}{Mean} & 0.07(0.01) & 0.53(0.04) & 1.75(0.20) & 1.50(0.26) & 2.21(0.27) & 3.73(0.36) & 9.89(1.21) & 74.21(8.99) \\
    \hline
    \multirow{4}{4em}{Weight} & ID 5 & 0.56 & 0.26 & 0.11 & $< 0.01$ & $< 0.01$ & $< 0.01$ & 0.05 & 0.02  \\
    & ID 22 & 0.84 & 0.12 & $< 0.01$ & $< 0.01$ & $< 0.01$ & 0.02 & $< 0.01$ & 0.02  \\
    & ID 13 & 0.77 & 0.11 & $< 0.01$ & $< 0.01$ & 0.10 & $< 0.01$ & 0.02 & $< 0.01$  \\
    & ID 14 & 0.74 & 0.10 & $< 0.01$ & 0.11 & $< 0.01$ & $< 0.01$ & 0.05 & $< 0.01$ \\
    \hline
    \multirow{4}{*}{Unique in} & ID 5 & 0.00 & 0.00 & 0.60 & 0.00 & 0.00 & 0.00 & 0.00 & 0.06 \\
    & ID 22 & 0.00 & 0.00 & 0.00 & 0.00 & 0.00 & 0.51 & 0.00 & 0.00 \\
    & ID 13 & 0.00 & 0.00 & 0.00 & 0.00 & 0.42 & 0.00 & 0.00 & 0.00 \\
    & ID 14 & 0.00 & 0.00 & 0.00 & 0.60 & 0.00 & 0.00 & 0.00 & 0.00 \\
    \hline
\end{tabular}}
\caption{Estimated clusters based on \cite{wade2018bayesian} using posterior samples from PAM. A total of eight OTU count clusters is estimated. ``Mean" and ``Weight" are the posterior mean estimates of the cluster mean
and   weight. Parantheses are the standard deviations. An entry in a row corresponding to ``Unique in" is the posterior probability that a cluster (column) is only present in the individual (row) but not in other individuals (rows). } 
\label{tab:microbio}
\end{table}

We report an interesting finding related to the PAM clustering of OTU counts. 
Specifically,   the counts of the   OTU \textit{Prevotella melaninogenica} is in cluster 8, which has the highest expression and is shared (both the cluster and the OTU   counts) only by AF individuals 5 and 22. This finding is consistent with previous studies that have shown that the individuals  
with a predominance of \textit{Prevotella spp.} are more likely to consume fiber, which is a typical component of an African diet \citep{graf2015contribution,preda2019effects}. 


\subsection{Warts Dataset}
\label{subsec:warts}

In this example, we consider a publicly available dataset   reporting treatment of patients with  warts. Two groups of patients are considered, treated with  immunotherapy or cryotherapy.   Each treatment group contains medical records for 90 patients, and for each patient, six baseline characteristics (covariates) are reported, including the patient's gender, age (Age), time elapsed before treatment (Time), the number of warts (NW), the type of warts, 
and the surface area of warts in $\text{mm}^2$ (Area). Additionally, patients' responses to the corresponding treatments are also recorded.

To better understand potential differences between responders to the two treatments, we use PAM to cluster the covariate values of the responders.   The sets of responders contain 71 patients for the immunotherapy group and 48 for  the cryotherapy group.  We construct an  observation $y_{ij}$ as    $q=4$ -dimensional vector including four continuous baseline covariates, Age, Time (time of sickness before treatment)
, NW (number of warts)
, and Area (surface area of warts)
. We set the hyperparameters of the priors to be the same  as in Scenario 2 of the simulation and apply PAM to the dataset of two groups of warts patients.   

Table \ref{tab:warts} reports inference results.  PAM identifies a total of seven clusters, three of which  are shared between the immunotherapy and cryotherapy groups, and the remaining four unique to a group.
\begin{table}
\centering
\resizebox{\linewidth}{!}{
\begin{tabular}{ lc|ccc|cc|cc }
    \hline
    & & Cluster 1 & Cluster 2 & Cluster 3 & Cluster 4 & Cluster 5 & Cluster 6 & Cluster 7 \\
    \hline
    \multirow{4}{4em}{Mean} & Age & 18.53 & 31.66 & 23.68 & 27.36 & 19.64 & 24.51 & 16.55 \\
    & Time & 6.19 & 6.71 & 8.63 & 6.96 & 7.38 & 4.41 & 3.80 \\
    & NW & 2.44 & 7.13 & 8.44 & 2.75 & 7.98 & 7.54 & 4.28 \\
    & Area & 68.41 & 40.82 & 195.16 & 389.20 & 312.65 & 87.78 & 6.41 \\
    \hline
    \multirow{2}{4em}{Weight} & Immunotherapy (G1) & 0.15 & 0.68 & 0.02 & 0.12 & 0.03 & $< 0.01$ & $< 0.01$  \\
    & Cryotherapy (G2) & 0.31 & 0.17 & 0.06 & $< 0.01$ & $< 0.01$ & 0.36 & 0.10 \\
    \hline
    \multicolumn{2}{l|}{$\widehat{\text{Pr}}(\pi_{1k} > 0, \pi_{2k} > 0|\text{Data})$} & 1.00 & 0.81 & 0.66 & 0.31 & 0.32 & 0.28 & 0.40 \\
    \multicolumn{2}{l|}{$\widehat{\text{Pr}}(\pi_{1k} > 0, \pi_{2k} = 0|\text{Data})$} & 0.00 &  0.19 & 0.00 & 0.69 & 0.68 & 0.00 & 0.00 \\
    \multicolumn{2}{l|}{$\widehat{\text{Pr}}(\pi_{1k} = 0, \pi_{2k} > 0|\text{Data})$} & 0.00 &  0.00 & 0.34 & 0.00 & 0.00 & 0.72 & 0.60 \\
    \hline
\end{tabular}}
\caption{Estimated clusters based on \cite{wade2018bayesian} using posterior samples from PAM. 
Reported are the cluster means, weights, probabilities of common, $\widehat{\text{Pr}}(\pi_{1k} > 0, \pi_{2k} > 0|\text{Data})$, and unique clusters in either the immunotherapy (G1) or the cryotherapy (G2), $\widehat{\text{Pr}}(\pi_{jk} > 0, \pi_{j'k} = 0|\text{Data})$, for the inferred seven clusters. Observed data consist of  4-dimensional covariate vectors for all the patients. The covarites are, ``Age", ``Time" referring to the time elapsed before treatment, ``NW" referring to number of warts, and ``Area" referring to the surface area of warts of the patient.   }
\label{tab:warts}  
\end{table}
The table reveals that, among all responders, individuals with younger age, a short time elapsed from treatment (less than five months), and small surface area of warts form unique clusters in the cryotherapy group. On the other hand, those who were not treated for a longer time and had a large surface area of warts (over 300 $\text{mm}^2$) form distinct clusters in the immunotherapy group.
Furthermore, it seems that the number of warts does not provide much information in determining a better treatment option for warts patients. 

These findings are consistent with results from previously published studies. For instance, \cite{khozeimeh2017intralesional} found that patients younger than 24 years old showed a better response to cryotherapy, and patients who received cryotherapy within six months had a very high probability of being cured. This is consistent with the information implied by clusters 6 and 7, which are unique to the cryotherapy group. Moreover, another study by \cite{khozeimeh2017expert} developed an expert system with fuzzy rules, and one such rule for immunotherapy is ``If (types of wart is Plantar) and (time elapsed before treatment is VeryLate) then (response to treatment is Yes)." In \cite{khozeimeh2017expert}'s expert system, time elapsed before treatment longer than six months is considered ``VeryLate". This rule echoes the common and unique clusters for the immunotherapy group found by PAM. In the unique clusters 4 and 5, and the common clusters 1 to 3, the time before treatment was 6.96, 7.38, 6.19, 6.71 and 8.63 months, respectively, all  larger than six months. 
Additional  results are illustrated in Supplement A.15, which shows the cluster membership of each patient. The figure indicates that patients with a large area of warts are unique to the immunotherapy group, while those with a younger age are mostly from the cryotherapy group.

\section{Discussion}
\label{sec:diss}

We have introduced   a   novel BNP 
model   constructed with a novel technique called Atom Skipping. A stochastic process that uses atom-skipping on single datasest is ASP, which has a mean process of FSBP, an extension of DP   that has higher expected number of clusters than DP with the same concentration parameter.   
Extending ASP to multiple groups forms the proposed 
PAM, where the weights of clusters in PAM are allowed to be exactly zero in some groups, effectively removing these clusters from those groups. Thus, PAM generates an interpretable
clustering structure. Additionally, PAM accommodates count data and multivariate observations. Efficient slice samplers are developed for PAM, with substantial modifications due to atom-skipping. In simulation studies, PAM demonstrated its robustness across different simulation scenarios. In particular, 
it performed the best when there are many unique clusters with little or no common ones among the groups.   In the case studies, PAM also produces sensible results. 

There are some limitations to our current work. Firstly, the model is unable to cluster groups (i.e., distributional clusters), unlike NDP and CAM. However, we are currently working on a separate model that extends PAM to cluster nested data at both group and observational levels. Secondly, the model has not been applied to real datasets consisting of different types of covariates, such as binary and multinomial covariates. Finally, 
longitudinal data is another interesting direction for extending the model.

\bigskip

\noindent \textbf{Acknowledgement}

\noindent Dehua Bi's effort is partly supported by  NIH RO1 NS114552 and NSF DMS1953340. Yuan Ji's effort is partly supported by NSF DMS1953340.

\bibliographystyle{plainnat}
\bibliography{Bibliography-MM-MC}


\clearpage
\newpage
\appendix

\section{}


\renewcommand\thefigure{\thesection.\arabic{figure}} 
\setcounter{figure}{0}

\renewcommand\thetable{\thesection.\arabic{table}} 
\setcounter{table}{0}

\renewcommand\theequation{\thesection.\arabic{equation}} 
\setcounter{equation}{0}

\subsection{Features of BNP models}
\label{sec:BNP_comp}

Table \ref{tab:feat} summarizes the features of some BNP models, along with the proposed PAM. A feature is checked based on the definition of the model, not the posterior inference. 

\begin{table}[h!]
\begin{center}
\resizebox{\linewidth}{!}{
\begin{tabular}{ c|cccc }
    \hline
     BNP & Common Atoms / & Common Atoms / &  Distinct Atoms / & Plaid$^*$ Atoms / \\
     Models & Common Weights & Distinct Weights & Distinct Weights & Distinct Weights \\
    \hline
    CAM & \checkmark & \checkmark & &   \\ \hline
    HDP & & \checkmark & & \\ \hline
    LNP & \checkmark & & & \checkmark \\ \hline
    NDP & \checkmark & & \checkmark & \\ \hline
    PAM & & \checkmark & \checkmark & \checkmark \\
    \hline
    \multicolumn{5}{l}{ \footnotesize $*$ ``Plaid" atoms means groups can share common atoms but can also possess unique atoms. }
\end{tabular}}
\end{center}
\caption{Features supported by various BNP models. A check-mark means the model supports such feature. }
\label{tab:feat}
\end{table}

\subsection{Proof of Proposition 1} \label{sec:proofp1}

We show the results for $\pi_{jk}'$ as the result for $\pi_k'$ is the same, with index $j$ removed. Conditional on $\bm{\beta} = \{\beta_k; k \geq 1\}$ (which is equivalent as conditional on $\bm{\beta}' = \{\beta_k'; k \geq 1\}$ because $\beta_k$ is constructed from $\beta_k'$ deterministically),  we have

\begin{equation}
    E[\pi_{jk}'|\bm{\beta},p_j] = \frac{p_j\beta_k}{1 - \sum_{l=1}^{k-1}\beta_l}.
\end{equation}

Then

$$E[E[\pi_{jk}|\bm{\beta},p_j]] = E\left[E\left [\pi_{jk}'\prod_{l=1}^{k-1}(1-\pi_{jl}')|\bm{\beta}, p_j \right]\right] $$
$$= E\left[ \frac{p_j\beta_k}{1 - \sum_{l=1}^{k-1}\beta_l} (1 - p_j\beta_1) \prod_{l=2}^{k-1} \left ( \frac{1 - \sum_{w=1}^{l-1}\beta_w - p_j\beta_l}{1-\sum_{w=1}^{l-1}\beta_w} \right ) \right]$$ 
$$= E\left[ p_j\beta_k \prod_{l=1}^{k-1} \left (  \frac{1 - \sum_{w=1}^{l-1}\beta_w - p_j\beta_l}{1-\sum_{w=1}^{l}\beta_w} \right ) \right]$$
$$= E\left[ p_j\beta_k \prod_{l=1}^{k-1} \left (  \frac{1 - \sum_{w=1}^{l}\beta_w + \beta_l - p_j\beta_l}{1-\sum_{w=1}^{l}\beta_w} \right ) \right]$$ 
$$= E\left[ p_j\beta_k \prod_{l=1}^{k-1} \left \{ \frac{\sum_{w=l+1}^{\infty}\beta_w + (1-p_j)\beta_l}{\sum_{w=l+1}^{\infty}\beta_w} \right \} \right]$$
$$=E\left[ p_j\beta_k \prod_{l=1}^{k-1} \left \{ 1 + \frac{(1-p_j)\beta_l}{\sum_{w=l+1}^{\infty}\beta_w} \right \} \right]$$

Expanding the 
term in the expectation, 
we have

$$p_j\beta_k \prod_{l=1}^{k-1} \left\{ 1 + \frac{(1-p_j)\beta_l}{\sum_{w=l+1}^{\infty}\beta_w} \right\} = p_j \beta_k'\prod_{l=1}^{k-1} (1-\beta_l') \prod_{l=1}^{k-1} \left \{ 1 + \frac{(1-p_j)\beta_l'\prod_{s=1}^{l-1}(1-\beta_s')}{\sum_{w=l+1}^{\infty}\beta_w'\prod_{s=1}^{w-1}(1-\beta_{s}')} \right \}$$
$$= p_j \beta_k'\prod_{l=1}^{k-1} (1-\beta_l') \prod_{l=1}^{k-1} \left \{ 1 + \frac{(1-p_j)\beta_l'\prod_{s=1}^{l-1}(1-\beta_s')}{\beta_{l+1}'\prod_{s=1}^{l}(1-\beta_s') + \beta_{l+2}'\prod_{s=1}^{l+1}(1-\beta_s') + \beta_{l+3}'\prod_{s=1}^{l+2}(1-\beta_s') + \cdots } \right \}$$
$$= p_j \beta_k'\prod_{l=1}^{k-1} (1-\beta_l') \prod_{l=1}^{k-1} \left \{ 1 + \frac{(1-p_j)\beta_l'}{\beta_{l+1}'(1-\beta_l') + \beta_{l+2}'\prod_{s=l}^{l+1}(1-\beta_s') + \beta_{l+3}'\prod_{s=l}^{l+2}(1-\beta_s') + \cdots }\right \}$$
$$= p_j \beta_k'\prod_{l=1}^{k-1} (1-\beta_l') \prod_{l=1}^{k-1} \left \{ 1 + \frac{(1-p_j)\beta_l'}{(1-\beta_l')}\frac{1}{\beta_{l+1}' + \beta_{l+2}'(1-\beta_{l+1}') + \beta_{l+3}'\prod_{s=l+1}^{l+2}(1-\beta_s') + \cdots } \right \}$$
\begin{equation}\label{eq:epi}
= p_j \beta_k'\prod_{l=1}^{k-1} (1-\beta_l') \prod_{l=1}^{k-1}  \left \{  1 + \frac{(1-p_j)\beta_l'}{(1-\beta_l')}\frac{1}{\sum_{w=l+1}^{\infty} \beta_w'\prod_{s=l+1}^{w-1}(1-\beta_s')} \right \}
\end{equation}

Denote $\Gamma = \sum_{w=l+1}^{\infty} \beta_w'\prod_{s=l+1}^{w-1}(1-\beta_s')$ in \eqref{eq:epi}. Then it follows 

\begin{equation*} 
1 - \Gamma = (1 - \beta_{l+1}')(1 - \beta_{l+2}')\cdots  = \prod_{w=l+1}^\infty (1 - \beta_w') = 0. 
\end{equation*}
Therefore, $\Gamma = 1$ and the expectation of \eqref{eq:epi} becomes 

$$E[E[\pi_{jk}|\bm{\beta}',p_j]] = E\left[ p_j \beta_k'\prod_{l=1}^{k-1} (1-\beta_l') \prod_{l=1}^{k-1} \left\{\frac{1-\beta_l'+(1-p_j)\beta_l'}{(1-\beta_l')} \right\}\right]$$
\begin{equation} \label{eq: usedintheo}
= E\left[p_j\beta_k'\prod_{l=1}^{k-1}(1-p_j\beta_l') \right] = E[p_j]E[\beta_k']\prod_{l=1}^{k-1}(1-E[p_j]E[\beta_l'])
\end{equation}

Since $\beta_k' \sim \text{Beta}(1, \gamma)$ and $p_j \sim \text{Beta}(a, b)$, we have 

\begin{equation*} 
    E[\pi_{jk}] = \frac{\bar{p}}{1+\gamma}\left( \frac{1 + \gamma - \bar{p}}{1+\gamma} \right)^{k-1} = \frac{1}{1+\gamma'}\left(\frac{\gamma'}{1+\gamma'}\right)^{k-1}
\end{equation*}

\noindent where $\gamma' = \frac{1+\gamma-\bar{p}}{\bar{p}}$, $\bar{p} = \frac{a}{a+b}$. This proves the second and third claims in Proposition 1.

To show the first claim, 
we first show $E\left[\sum_{k \geq 1}\pi_{jk} \right] = 1$. Notice that

$$E\left[\sum_{k \geq 1}\pi_{jk} \right] = \sum_{k \geq 1} E[\pi_{jk}] = \sum_{k \geq 1} \frac{\bar{p}}{1+\gamma}\left( \frac{1 + \gamma - \bar{p}}{1+\gamma} \right)^{k-1} $$
$$= \sum_{k^* \geq 0} \frac{\bar{p}}{1+\gamma}\left( 1-\frac{\bar{p}}{1+\gamma} \right)^{k^*} = \frac{\bar{p}}{1+\gamma}\times \frac{1+\gamma}{\bar{p}} = 1.$$


Next, we show $0 < \sum_{k\geq 1}\pi_{jk} \leq 1$. It is trivial to see that $\sum_{k\geq 1}\pi_{jk} > 0$. We now show $\sum_{k\geq 1}\pi_{j,k} \leq 1$. Notice
$$1 - \sum_{k\geq 1}\pi_{jk} = 1-\pi_{j1}' - \pi_{j2}'(1-\pi_{j1}') - \pi_{j3}'(1-\pi_{j1}')(1-\pi_{j2}') - \cdots  = \prod_{k=1}^\infty (1-\pi_{jk}') \geq 0$$
since $0 \leq \pi_{jk}' < 1$. Therefore, $\sum_{k\geq 1}\pi_{jk} \leq 1$. Thus, we have shown $0 < \sum_{k\geq 1}\pi_{jk} \leq 1$ and $E\left[\sum_{k\geq 1}\pi_{jk}\right] = 1$, and we conclude $\sum_{k \geq 1}\pi_{jk} = 1$ almost surely. This proves the first claim of Proposition 1. \qed

\bigskip

\subsection{Proof of Theorem 1}
\label{sec:prooftheopam}

For $G|G_0,p \sim ASP(p,\alpha_0,G_0)$,
we derive the mean of $G$. Recall $G_0 = \sum_{k=1}^\infty \beta_k\delta_{\bm{\phi}_k}$. Conditional on $G_0$ is equivalent as conditional on $\bm{\beta}' = \{\beta_k'; k \geq 1\}$ and $\bm{\Phi} = \{\bm{\phi}_k; k \geq 1\}$. From equation \eqref{eq: usedintheo} in subsection \ref{sec:proofp1}, we have
$$\text{E}[G(A)|G_0,p] = \text{E}[G(A)|\bm{\beta}', \bm{\Phi},p] = \sum_{k=1}^\infty \text{E}[\pi_k|\bm{\beta}',p]\delta_{\bm{\phi}_k}(A)$$ 
$$= \sum_{k=1}^\infty p\beta_k' \prod_{l=1}^{k-1}(1 - p\beta_l')\delta_{\bm{\phi}_k}(A) = G^*(A),$$
where $G^* = \sum_{k=1}^{\infty} \omega_k\delta_{\phi_k}$, $\omega_k = \omega_k'\prod_{l = 1}^{k-1}(1 - \omega_l')$, $\omega_k' = p\cdot\beta_k'$.
As $G_0 \sim DP(\gamma,H)$, we have $\beta_k' \sim \text{Beta}(1, \gamma)$, and $\bm{\phi}_k \sim H$. 
Plugging in the priors for $\beta_k'$ and $\bm{\phi}_k$, we see that the stick-breaking construction of $G^*$ is equal to that of FSBP in Section 3.3. \qed 


\bigskip

\subsection{Proof of Proposition 2} \label{sec:proofeqp}

Let $\bm{\theta}_{i1}|G_1 \sim G_1$ and $\bm{\theta}_{i'2}|G_{2} \sim G_{2}$, without loss of generality, 
$$\text{Pr}(\bm{\theta}_{i1} = \bm{\theta}_{i'2}) = \int \text{Pr}(\bm{\theta}_{i1} = \bm{\theta}_{i'2}|G_1, G_2)p(G_1)p(G_2)dG_1 dG_2 > \int 0 p(G_1)p(G_2)dG_1 dG_2 = 0$$
if and only if $\text{Pr}(\bm{\theta}_{i1} = \bm{\theta}_{i'2}|G_1, G_2) > 0$. We next show $\text{Pr}(\bm{\theta}_{i,1} = \bm{\theta}_{i'2}|G_1, G_2) > 0$.
Denote the set $A^s = \{\bm{\phi}_k;\pi_{jk} \neq 0 \text{ and } \pi_{j'k} \neq 0\}$ and $A^j = \{\bm{\phi}_k; \pi_{jk} \neq 0\}$ for $j \neq j'$, $j = 1, 2$. Then 
\begin{equation} \label{eq:thetaeqtheta}
    \text{Pr}(\bm{\theta}_{i1} = \bm{\theta}_{i'2}|G_1, G_2) = \text{Pr}(\bm{\theta}_{i1} = \bm{\theta}_{i'2}|\bm{\theta}_{i1} \in A^s, \bm{\theta}_{i'2} \in A^s)\text{Pr}(\bm{\theta}_{i1} \in A^s, \bm{\theta}_{i'2} \in A^s|G_1,G_2)
\end{equation}
The second term in \eqref{eq:thetaeqtheta} is
$$\text{Pr}(\bm{\theta}_{i1} \in A^s, \bm{\theta}_{i'2} \in A^s|G_1,G_2) $$
$$= \text{Pr}(\bm{\theta}_{i1} \in A^s, \bm{\theta}_{i'2} \in A^s|A^s \neq \emptyset,G_1,G_2)\text{Pr}(A^s \neq \emptyset) + $$
$$\text{Pr}(\bm{\theta}_{i1} \in A^s, \bm{\theta}_{i'2} \in A^s|A^s = \emptyset,G_1,G_2)\text{Pr}(A^s = \emptyset)$$
$$= \text{Pr}(\bm{\theta}_{i1} \in A^s, \bm{\theta}_{i'2} \in A^s|A^s \neq \emptyset,G_1,G_2)\text{Pr}(A^s \neq \emptyset)$$
Then $\text{Pr}(A^s \neq \emptyset) = 1 - \text{Pr}(A^s = \emptyset) = 1 - \prod_{k=1}^{\infty} \{p_1(1-p_2)+p_2(1-p_1)\} = 1$. This is because  at each atom $k$, $G_1$ selects the atom with probability $p_1$ and $G_{2}$ does not select the atom, with probability $(1 - p_{2})$, or vice versa.
Denote $K^s = \{k; \bm{\phi}_k \in A^s\}$ and $K^j = \{k; \bm{\phi}_k \in A^j\}$. The term $\text{Pr}(\bm{\theta}_{i1} \in A^s, \bm{\theta}_{i'2} \in A^s|A^s \neq \emptyset,G_1,G_2)$ is evaluated as
$$\text{Pr}(\bm{\theta}_{i1} \in A^s, \bm{\theta}_{i'2} \in A^s|A^s \neq \emptyset,G_1,G_2) = \left[\sum_{k \in K^s}\pi_{1k}\right]\left[\sum_{k \in K^s}\pi_{2k}\right].$$
Since $\text{Pr}(A^s \neq \emptyset) = 1$, $|K^s| \geq 1$, and since $\pi_{1k} > 0$ and $\pi_{2k} > 0$ for $k \in K^s$, for some arbitrary $k^* \in K^s$, we have
$$\text{Pr}(\bm{\theta}_{i1} \in A^s, \bm{\theta}_{i'2} \in A^s|A^s \neq \emptyset,G_1,G_2) \geq \pi_{1k^*}\pi_{2k^*} > 0$$
Therefore,
$$\text{Pr}(\bm{\theta}_{i1} \in A^s, \bm{\theta}_{i'2} \in A^s|G_1,G_2) = \text{Pr}(\bm{\theta}_{i1} \in A^s, \bm{\theta}_{i'2} \in A^s|A^s \neq \emptyset,G_1,G_2)\times 1 > 0.$$
And the first term in \eqref{eq:thetaeqtheta} is
$$ \text{Pr}(\bm{\theta}_{i1} = \bm{\theta}_{i'2}|\bm{\theta}_{i1} \in A^s, \bm{\theta}_{i'2} \in A^s) = E[\text{Pr}(\bm{\theta}_{i1} = \bm{\theta}_{i'2}|G_1, G_2, \bm{\theta}_{i1} \in A^s, \bm{\theta}_{i'2} \in A^s)]$$
$$= E\left[\left( \sum_{\bm{\phi}_k \in A^s} I(\bm{\theta}_{i1} = \bm{\theta}_{i'2}=\bm{\phi}_k)p(\bm{\phi}_k)\right)|G_1, G_2, \bm{\theta}_{i1} \in A^s, \bm{\theta}_{i'2} \in A^s\right]$$
$$ = E\left[ \left( \sum_{\bm{\phi}_k \in A^s} \pi_{1k}\pi_{2k}p(\bm{\phi}_k) \right) |  G_1, G_2, \bm{\theta}_{i1} \in A^s, \bm{\theta}_{i'2} \in A^s\right]$$
$$\stackrel{(a)}{=} \sum_{k \in K^s} E[\pi_{1k}]\text{E}[\pi_{2k}] = \sum_{k \in K^s}E\left\{\pi_{1k}'\prod_{l \in K^1, l < k}(1 - \pi_{1l}')\right\}E\left\{\pi_{2k}'\prod_{l \in K^2, l < k}(1 - \pi_{2,l}')\right\}$$
$$ \geq \sum_{k \in K^s}E\left\{\pi_{1k}'\prod_{l \in K^1, l < k}(1 - \pi_{1l}')\prod_{l \in {K^1}^c, l < k}(1 - {\pi_{1l}'}^*)\right\}E\left\{\pi_{2k}'\prod_{l \in K^2, l < k}(1 - \pi_{2l}')\prod_{l \in {K^2}^c, l < k}(1 - {\pi_{2l}'}^*)\right\}$$
$$ \stackrel{(b)}{=} \sum_{k \in K^s}[E[\beta_k]^2] \stackrel{(c)}{=} \sum_{k \in K^s} \left[ \frac{1}{1+\gamma}\left( \frac{\gamma}{1+\gamma} \right)^{k-1} \right]^2$$
where I(A) is the indicator function that equals to 1 if condition A is satisfied, ${\pi_{jl}'}^* \sim \text{Beta}\left(\alpha_0\beta_k, \alpha_0\left(1-\sum_{l=1}^k \beta_l\right)\right)$, and ${K^j}^c$s are the complement sets of $K^j$, for $j = 1, 2$. In addition, (a) is true because 
$$p(\bm{\phi}_k|G_j) = \left\{ \begin{matrix} 1 & \text{if }\bm{\phi}_k \in G_j \\ 0 & \text{o.w.} \end{matrix} \right. ,$$ and (b) is true because the term $\pi_{jk}'\prod_{l \in K^j, l < k}(1 - \pi_{jl}')\prod_{l \in {K^j}^c, l < k}(1 - {\pi_{jl}'}^*) = {\pi_{jk}'}^*\prod_{l < k}(1 - {\pi_{jl}'}^*)$ for $k \in K^s$ (i.e., equation (4)), with conditional expectation (conditional on $\bm{\beta}$) equals to $\beta_k$, and (c) is true because $\beta_k = \beta_k'\prod_{l < k}(1 - \beta_k')$, $\beta_k' \sim \text{Beta}(1, \gamma)$.

Again since $|K^s| \geq 1$, for some arbitrary $k^* \in K^s$, we have
$$\sum_{k \in K^s} \left[ \frac{1}{1+\gamma}\left( \frac{\gamma}{1+\gamma} \right)^{k-1} \right]^2 \geq \left[ \frac{1}{1+\gamma}\left( \frac{\gamma}{1+\gamma} \right)^{k^*-1} \right]^2 > 0.$$
Thus, we have
$$\text{Pr}(\bm{\theta}_{i,1} = \bm{\theta}_{i'2}|\bm{\theta}_{i1} \in A^s, \bm{\theta}_{i'2} \in A^s) > 0.$$
Combine with $\text{Pr}(\bm{\theta}_{i1} \in A^s, \bm{\theta}_{i'2} \in A^s|A^s \neq \emptyset,G_1,G_2) > 0$, we have now shown that
$$\text{Pr}(\bm{\theta}_{i1} = \bm{\theta}_{i'2}|G_1, G_2) > 0,$$
which completes the proof. \qed

\clearpage
\newpage

\subsection{Additional Simulation Plots of Expected Number of Clusters for CAM, HDP, and PAM}\label{sec:prior_chp_sim}

\begin{figure}[H]
    \centering
    \includegraphics[width = \textwidth]{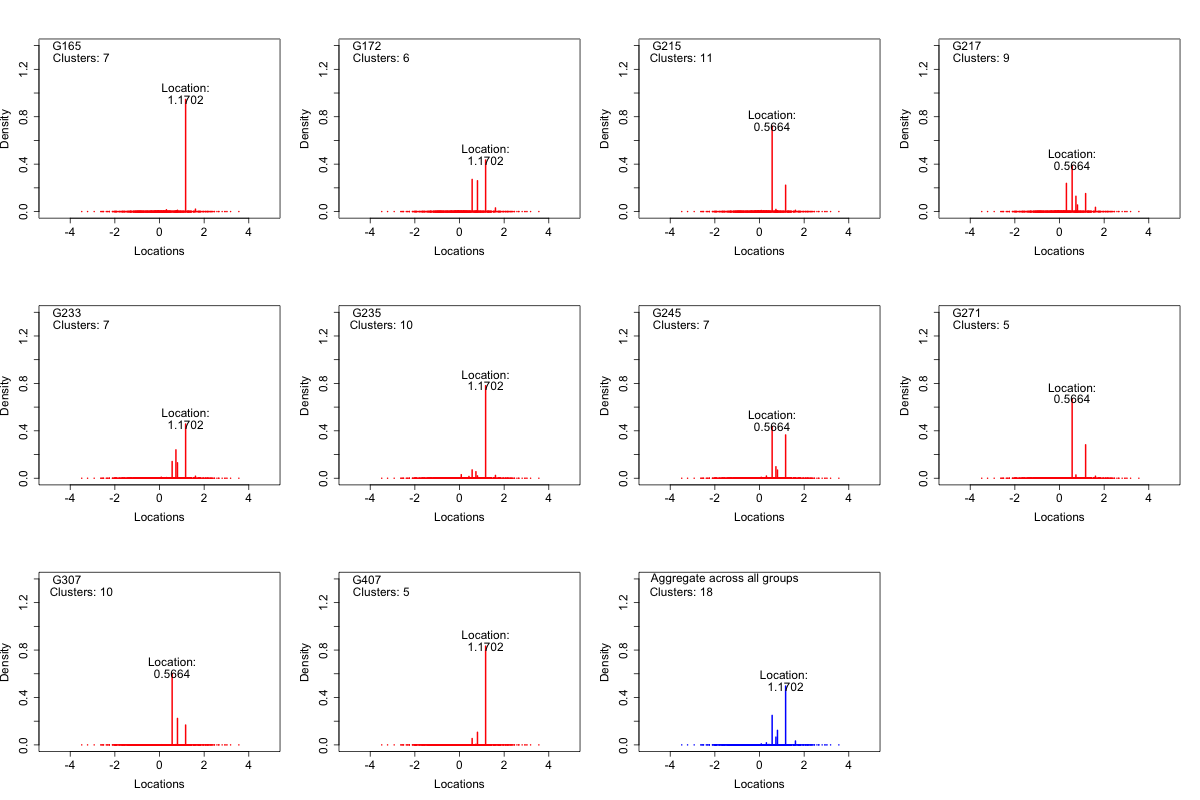}
    \caption{Plots of simulated $G_j$ for 10 randomly selected samples (subplots with red sticks) and the random distribution aggregating all 500 groups (bottom right subplot with blue sticks) for CAM$(1,1,H)$, $H = N(0,1)$. In each plot, the text ``G$j$" represents group $j$ for $j \in \{1, \ldots, 500\}$, ``Cluster:$K_j$" represents the number of clusters $K$ in group $j$, and ``Location:$\phi_k$" represents the location that has the highest probability in the random discrete distribution $G_j$. }
    \label{fig:CAM_prop}
\end{figure}

\begin{figure}[H]
    \centering
    \includegraphics[width = \textwidth]{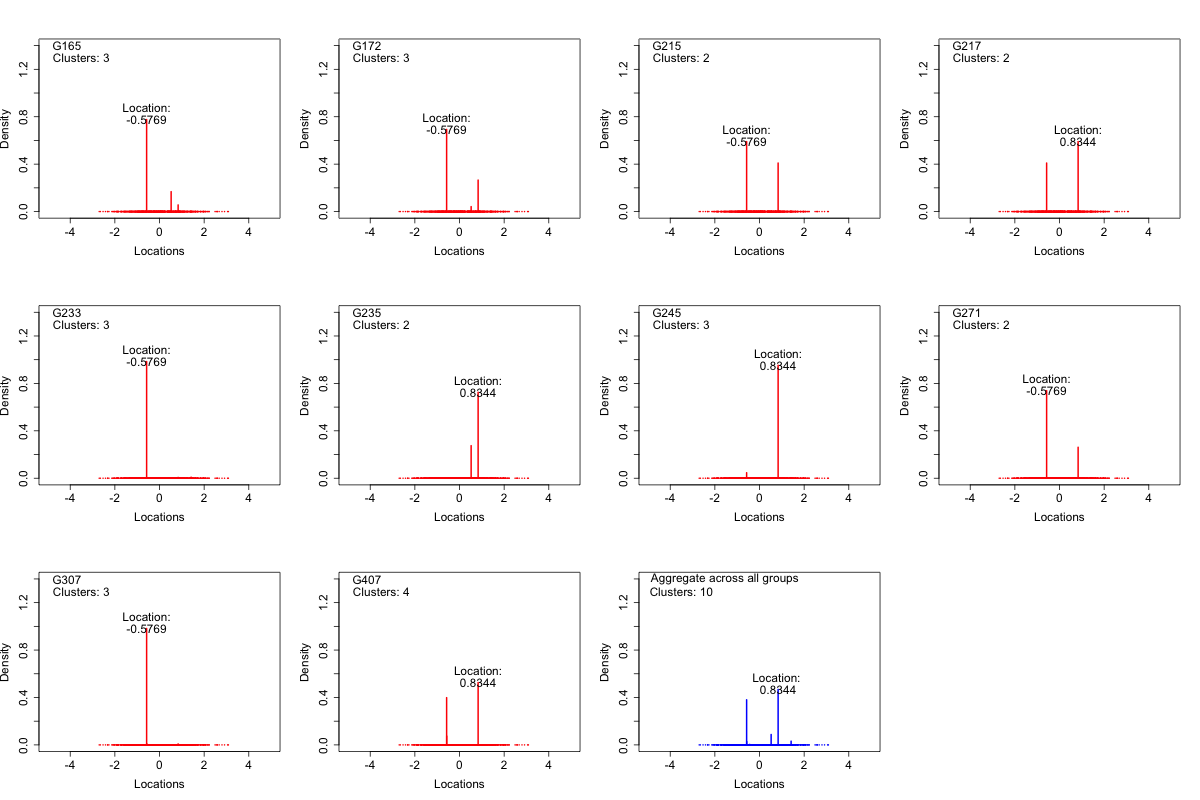}
    \caption{Plots of simulated $G_j$ for 10 randomly selected samples (subplots with red sticks) and the random distribution aggregating all 500 groups (bottom right subplot with blue sticks) for HDP$(1,1,H)$, $H = N(0,1)$. In each plot, the text ``G$j$" represents group $j$ for $j \in \{1, \ldots, 500\}$, ``Cluster:$K_j$" represents the number of clusters $K$ in group $j$, and ``Location:$\phi_k$" represents the location that has the highest probability in the random discrete distribution $G_j$. }
    \label{fig:HDP_prop}
\end{figure}

\begin{figure}[H]
    \centering
    \includegraphics[width = \textwidth]{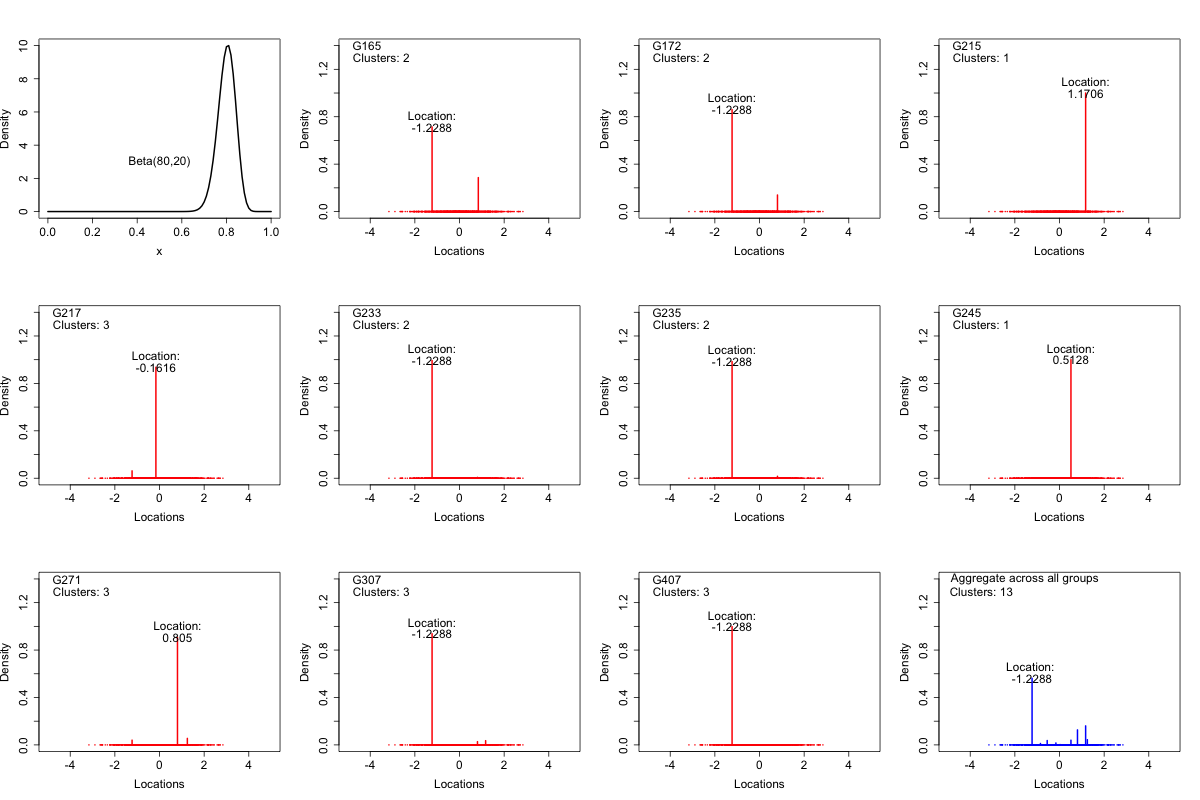}
    \caption{Plots of prior for $\bm{p}_1$ (top left subplot), simulated $G_j$ for 10 randomly selected samples (subplots with red sticks), and the random distribution aggregating all 500 groups (bottom right subplot with blue sticks) for PAM$(\bm{p}_1, 1,1,H)$, $H = N(0,1)$, $p_{j1} \sim \text{Beta}(80,20)$. In each plot, the text ``G$j$" represents group $j$ for $j \in \{1, \ldots, 500\}$, ``Cluster:$K_j$" represents the number of clusters $K$ in group $j$, and ``Location:$\phi_k$" represents the location that has the highest probability in the random discrete distribution $G_j$.}
    \label{fig:PAM_p1_prop}
\end{figure}

\begin{figure}[H]
    \centering
    \includegraphics[width = \textwidth]{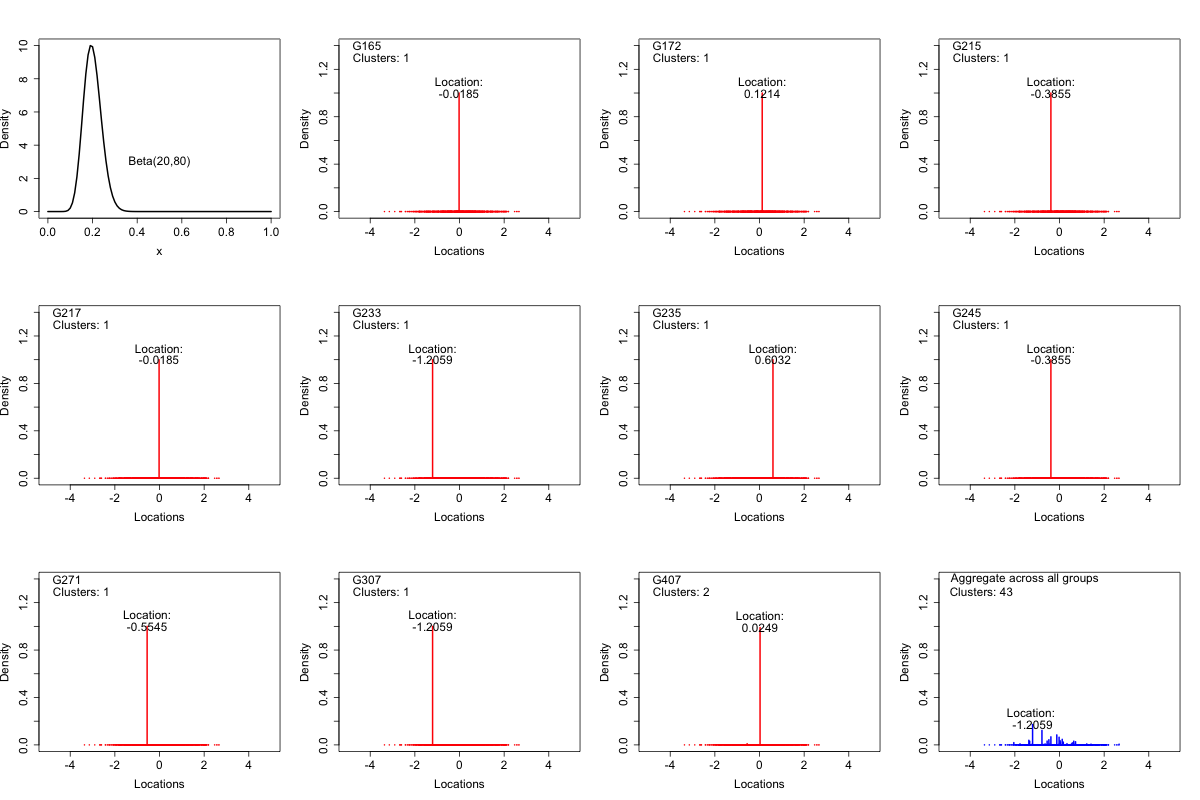}
    \caption{Plots of prior for $\bm{p}_2$ (top left subplot), simulated $G_j$ for 10 randomly selected samples (subplots with red sticks), and the random distribution aggregating all 500 groups (bottom right subplot with blue sticks) for PAM$(\bm{p}_2, 1,1,H)$, $H = N(0,1)$, $p_{j2} \sim \text{Beta}(20,80)$. In each plot, the text ``G$j$" represents group $j$ for $j \in \{1, \ldots, 500\}$, ``Cluster:$K_j$" represents the number of clusters $K$ in group $j$, and ``Location:$\phi_k$" represents the location that has the highest probability in the random discrete distribution $G_j$.}
    \label{fig:PAM_p2_prop}
\end{figure}


\clearpage
\newpage

\bigskip 

\subsection{Proof of Theorem 2}
\label{sec:prooftheo1}

The FSBP is a special case of the kernel stick-breaking process of \cite{dunson2008kernel}. Using their notation, the kernel function $K(\bm{x}, \Gamma_k) = p$, i.e., constant over $k$ and independent of covariates. Thus, their theoretical results are applicable in our case. From equation (4) of \cite{dunson2008kernel}, the mean of $G^*$ is immediate and given by 
$$E[G^*(A)] = E\left[E[G^*(A)|\bm{\beta}',p]\right] = E[H(A)] = H(A),$$
where $\bm{\beta}' = \{\beta_k'; k \geq 1\}, \beta_k' \sim \text{Beta}(1,\gamma)$. 
To find the variance of $G^*$, apply equation (7) of Theorem 1 of \cite{dunson2008kernel}
\begin{equation} \label{eq:varGjp}
    \text{Var}(G^*(A)) = \frac{\mu^{(2)}Var_{Q(A)}}{2\mu-\mu^{(2)}}
\end{equation}
where 
$$Var_{Q(A)} = Var_H\{\delta_{\bm{\phi}_k}(A)\} = H(A)(1 - H(A)),$$
$$\mu = 
p \cdot E[\beta_k'] = \frac{p}{1+\gamma},$$
and
$$\mu^{(2)} = 
p^2 \cdot E[{\beta_k'}^2] = \frac{2p^2}{(1+\gamma)(2+\gamma)}.$$
Substituting the expression for $Var_{Q(A)}$, $\mu(x)$, and $\mu^{(2)}(x)$ into equation \eqref{eq:varGjp}, we obtain
$$Var(G^*(A)) = \frac{H(A)(1-H(A))}{\frac{1+\gamma}{p} + \frac{1-p}{p}}.$$
\qed
\bigskip

\subsection{Proof of Theorem 3} \label{sec:eppf}

Denote $\bm{\Phi} = \{\bm{\phi}_1, \bm{\phi}_2, \ldots\}$ the atoms in $G^*$. 
Consider $n$ samples generated from $G^*$, $\bm{\Theta} = \{\bm{\theta}_i; i = 1, \ldots, n\}$, $\bm{\theta}_i|G^* \sim G^*$, and $\bm{\theta}_i$ takes a value in $\bm{\Phi}$ with a probability. 
Assume there are $K$  clusters, denote the atoms associated with the $K$ clusters
by $\bm{\Phi}_K = \{\bm{\phi}_{r_1}, \ldots, \bm{\phi}_{r_K}\}$ where each $r_k$ indexes the $k$th cluster and $r_k \in \mathbb{N}$, where $\mathbb{N}$ denotes the set of all natural numbers, i.e., $\mathbb{N} = \{1, 2, \ldots\}$. Denote $\mathcal{r} = \{r_1, \ldots, r_K\}$ the index set in ascending order of the $K$ clusters, i.e., $r_1 < r_2 < \ldots < r_K$. Let $\bm{z}=\{z_1, \ldots, z_n\}$ be the cluster label where $\{z_i = k\}$ means observation $\bm{\theta}_i$ belongs to cluster $k$, i.e., $\{\bm{\theta}_i = \bm{\phi}_{r_k}\}$. Further, denote $c_k = \{i;z_i = k\}$ the indices of $\bm{\theta}_i$'s belonging to cluster $k$. 
It is important to note that the cluster label $k$'s do not need to be consecutive integers. 
For example, $K=3$ and $\mathcal{r}=\{1,3,5\}$ or $K=5$ and $\mathcal{r}=\{2,5,6,20, 100\}.$ Lastly, assume the unique value of the $k$th cluster is the atom $\bm{\phi}_k$, i.e., $\{\bm{\theta}_i=\bm{\phi}_k\}$ if  $\{z_i=k\}$, for $k\in\mathcal{r}$. 

\def\bz{\bm{z}}
\def\bzz{\bm{z}^*}

Let $m = \max(z_1, \cdots, z_n)$. It follows that $K \leq m$ due to the fact that the cluster labels do not need to be consecutive integers. A partition $\bm{z}$ of the $n$ samples $\bm{\Theta}$ is then denoted as $C(\bz) = \{c_k; k \in \mathcal{r}\}$, the collection of $c_k$'s, where $c_k \cap c_{k'} = \emptyset$ for $k \neq k'$, $|C(\bm{z})| = K$, and $\cup_{k \in \mathcal{r}} c_k = \{1, \ldots , n\}$. Here,  $|.|$ refers to the cardinality of a set. The EPPF of $G^*$ evaluated at a specific partition $C$ is given by
\begin{equation} \label{eq:PCzeqC}
    \text{Pr}(C(\bz)=C) = \sum_{\bzz \in \mathbb{N}^n} \text{Pr}(C(\bzz)=C|\bm{z}=\bzz)\text{Pr}(\bm{z}=\bzz) = \sum_{\bzz \in \mathbb{N}^n} I(C({\bzz}) = C)\text{Pr}(\bm{z}=\bzz)
\end{equation}
where $\mathbb{N}^n$ is the $n-$dimensional space of positive integers. The second equality is true since given $\bz=\bzz$, $C(\bzz)$ is fixed and is either equal to $C$ or not. 

We first find $\text{Pr}(\bm{z} = \bzz)$. For a specific $\bzz = \{z_1^*, \ldots, z_n^*\}$, denote $e_k(\bzz) = |\{i; z_i^* = k\}|$, $f_k(\bzz) = |\{i; z_i^* > k\}|$, and $g_k(\bzz) = |\{i; z_i^* 
 \geq k\}|$. Also let $m(\bzz) = \max(z_1^* ,\ldots, z_n^*)$. Recall the definition of FSBP in Section 3.3, with $a = 1$ and $b = \gamma$, $\pi_k' \sim \text{Beta}(1,\gamma)$, and we have
$$\text{Pr}(\bm{z} = \bzz) = \int \text{Pr}(\bm{z}^*|\pi_1', \cdots, \pi_{m(\bzz)}')p(\pi_1')\cdots p(\pi_{m(\bzz)}')d\pi_1'\cdots d\pi_{m(\bzz)}'$$
$$= \int \left[ \prod_{k=1}^{m(\bzz)} \left\{p\pi_k'\prod_{l < k}(1-p\pi_l')\right\}^{e_k(\bzz)} \right]p(\pi_1')\cdots p(\pi_{m(\bzz)}')d\pi_1'\cdots d\pi_{m(\bzz)}'$$
$$= \int \left[ \prod_{k=1}^{m(\bm{z}^*)} (p\pi_k')^{e_k(\bzz)}(1-p\pi_k')^{f_k(\bzz)}\right]p(\pi_1')\cdots p(\pi_{m(\bzz)}')d\pi_1'\cdots d\pi_{m(\bzz)}'$$
$$= \prod_{k=1}^{m(\bm{z}^*)} \left\{ \frac{p^{e_k(\bm{z}^*)}}{B(1,\gamma)} \int {\pi_k'}^{e_k(\bm{z}^*)}(1-p\pi_k')^{f_k(\bm{z}^*)}(1-\pi_k')^{\gamma-1} d\pi_k' \right\}$$
where $B(a,b)$ is the Beta function with parameters $a$ and $b$. If we re-write the integral of the last step as the follows:
$$\int {\pi_k'}^{(e_k(\bm{z}^*) + 1) - 1}(1-p\pi_k')^{-(-f_k(\bm{z}^*))}(1-\pi_k')^{(\gamma + e_k(\bm{z}^*) + 1) - (e_k(\bm{z}^*) + 1) - 1} d\pi_k',$$
it is easy to see that this integration can be written as the Euler type hypergeometric function. Thus, we have
$$\int {\pi_k'}^{e_k(\bm{z}^*)}(1-p\pi_k')^{f_k(\bm{z}^*)}(1-\pi_k')^{\gamma-1} d\pi_k' = B(e_k(\bm{z}^*)+1, \gamma) {_2}F_1(-f_k(\bm{z}^*),e_k(\bm{z}^*)+1;\gamma + e_k(\bm{z}^*) + 1;p)$$
where ${_2}F_1(a,b;c;d)$ is the hypergeometric function with parameters $a, b, c$ and $d$. Consequently, we have
$$\text{Pr}(\bm{z} = \bm{z}^*) = \prod_{k=1}^{m(\bm{z}^*)}\left\{ \frac{p^{e_k(\bm{z}^*)}}{B(1,\gamma)}B(e_k(\bm{z}^*)+1, \gamma) {_2}F_1(-f_k(\bm{z}^*),e_k(\bm{z}^*)+1;\gamma + e_k(\bm{z}^*) + 1;p) \right\}$$
$$= \prod_{k=1}^{m(\bm{z}^*)}\left\{ \frac{p^{e_k(\bm{z}^*)}\Gamma(\gamma+1)}{\Gamma(\gamma)}\frac{\Gamma(e_k(\bm{z}^*)+1)\Gamma(\gamma)}{\Gamma(\gamma+e_k(\bm{z}^*)+1)} \, {_2}F_1(-f_k(\bm{z}^*),e_k(\bm{z}^*)+1;\gamma + e_k(\bm{z}^*) + 1;p) \right\}$$
$$= \prod_{k=1}^{m(\bm{z}^*)}\left\{{p^{e_k(\bm{z}^*)}}\frac{\Gamma(\gamma+1)\Gamma(e_k(\bm{z}^*)+1)}{\Gamma(\gamma+e_k(\bm{z}^*)+1)} \, {_2}F_1(-f_k(\bm{z}^*),e_k(\bm{z}^*)+1;\gamma + e_k(\bm{z}^*) + 1;p) \right\}$$
$$=\left\{ \prod_{k=1}^{m({\bm{z}^*})} \Gamma(\gamma+1) p^{e_k(\bm{z}^*)} \frac{\Gamma(e_k(\bm{z}^*)+1)}{\Gamma(\gamma+e_k(\bm{z}^*)+1)}\right\} \left\{\prod_{k=1}^{m(\bm{z}^*)} {_2}F_1(-f_k(\bm{z}^*),e_k(\bm{z}^*)+1;\gamma + e_k(\bm{z}^*) + 1;p) \right\}$$
\begin{equation} \label{eq:pzeqz}
    = \left\{ \prod_{c \in C(\bm{z}^*)} \Gamma(\gamma+1) p^{|c|} \frac{\Gamma(|c|+1)}{\Gamma(\gamma+|c|+1)}\right\} \left\{\prod_{k=1}^{m(\bm{z}^*)} {_2}F_1(-g_{k+1}(\bm{z}^*),e_k(\bm{z}^*)+1;\gamma + e_k(\bm{z}^*) + 1;p) \right\}
\end{equation}
where $f_k(\bm{z}^*) = g_{k+1}(\bm{z}^*)$.

Back to equation \eqref{eq:PCzeqC} and substituting in equation \eqref{eq:pzeqz}, we have 

$$\text{Pr}(C(\bz) = C) = \left\{ \prod_{c \in C} \Gamma(\gamma+1) p^{|c|} \frac{\Gamma(|c|+1)}{\Gamma(\gamma+|c|+1)}\right\} \times $$
\begin{equation} \label{eq:PCzCsubPzeqz}
    \underbrace{ \sum_{\bm{z}^* \in \mathbb{N}^n}I(C({\bm{z}^*}) = C)\left\{ \underbrace{ \prod_{k=1}^{m(\bm{z}^*)} {_2}F_1(-g_{k+1}(\bm{z}^*),e_k(\bm{z}^*)+1;\gamma + e_k(\bm{z}^*) + 1;p) }_{(A)} \right\}}_{(B)}.
\end{equation}

Now, recall $K = |C|$ is the number of unique clusters in the $n$ samples, and $C = \{c_1, \ldots, c_K\}$. 
Denote $S_K$ the set of all $K!$ permutations of $\{1, \ldots, K\}$, and denote $\bm{\lambda} = \{\lambda_1, \ldots, \lambda_K\} \in S_K$ a permutation of $\{1, \ldots, K\}$. 
For any $\bm{\lambda} \in S_K$, define $\alpha_k(\bm{\lambda}) = |c_{\lambda_k}| + \cdots + |c_{\lambda_K}|$. By definition, $\alpha_{K+1}(\bm{\lambda}) = 0$. Consider a given $\bm{z}^*$ 
such that $C({\bm{z}}^*) = C$, recall that $r_1, \ldots, r_K$ are the distinct values of $\bm{z}^*$ in ascending order, i.e., $r_1 < r_2 < \cdots < r_k < \cdots < r_K$, $r_k \in \mathbb{N}$, we can rewrite the (A) term in \eqref{eq:PCzCsubPzeqz} as

$$\prod_{k=1}^{m(\bm{z}^*)} {_2}F_1(-g_{k+1}(\bm{z}^*),e_k(\bm{z}^*)+1;\gamma + e_k(\bm{z}^*) + 1;p)$$
$$= \left[ {_2}F_1(-g_{r_2}(\bm{z}^*),e_{r_1}(\bm{z}^*)+1;\gamma + e_{r_1}(\bm{z}^*) + 1;p) \right]^{r_1} \times $$
$$\left[ {_2}F_1(-g_{r_3}(\bm{z}^*),e_{r_2}(\bm{z}^*)+1;\gamma + e_{r_2}(\bm{z}^*) + 1;p) \right]^{r_2 - r_1} \times \cdots \times $$
$$\left[ {_2}F_1(-g_{r_K+1}(\bm{z}^*),e_{r_K}(\bm{z}^*)+1;\gamma + e_{r_K}(\bm{z}^*) + 1;p) \right]^{r_K - r_{K-1}}$$
$$= \left[ {_2}F_1(-\alpha_2(\bm{\lambda}),|c_{\lambda_1}|+1;\gamma + |c_{\lambda_1}| + 1;p) \right]^{d_1} \times \left[ {_2}F_1(-\alpha_3(\bm{\lambda}),|c_{\lambda_2}|+1;\gamma + |c_{\lambda_2}| + 1;p) \right]^{d_2} \times \cdots \times $$
$$\left[ {_2}F_1(-\alpha_{K+1}(\bm{\lambda}),|c_{\lambda_K}|+1;\gamma + |c_{\lambda_K}| + 1;p) \right]^{d_K} = \prod_{k=1}^K \left[ {_2}F_1(-\alpha_{k+1}(\bm{\lambda}),|c_{\lambda_k}|+1;\gamma + |c_{\lambda_k}| + 1;p) \right]^{d_k}$$
where $\bm{d} = (d_1, \ldots , d_K)$, $d_1 = r_k$, and $d_k = r_k - r_{k-1}$ for $k = 2, \ldots, K$. 
For any $\bzz \in \mathbb{N}^n$, 
note that the definition of $\bm{d}$ and $\bm{\lambda}$ sets up a one-to-one correspondence, which is a bijection, between $\{\bm{z}^* \in \mathbb{N}^n;C({\bm{z}^*}) = C\}$ and $\{(\bm{\lambda},\bm{d}); \bm{\lambda} \in S_K, \bm{d} \in \mathbb{N}^K\}$, and the expression in (B) in \eqref{eq:PCzCsubPzeqz} can then be rewritten as
$$\sum_{\bm{z}^* \in \mathbb{N}^n}I(C({\bm{z}^*}) = C)\left\{ \prod_{k=1}^{m(\bm{z}^*)} {_2}F_1(-g_{k+1}(\bm{z}^*),e_k(\bm{z}^*)+1;\gamma + e_k(\bm{z}^*) + 1;p) \right\}$$ 
$$= \sum_{\bm{\lambda} \in S_K}\sum_{\bm{d} \in \mathbb{N}^K}\prod_{k=1}^K \left[{_2}F_1(-\alpha_{k+1}(\bm{\lambda}),|c_{\lambda_k}|+1;\gamma + |c_{\lambda_k}| + 1;p) \right]^{d_k}$$
$$\stackrel{(a)}{=} \sum_{\bm{\lambda} \in S_K}\prod_{k=1}^K\sum_{d_k \in \mathbb{N}}\left[{_2}F_1(-\alpha_{k+1}(\bm{\lambda}),|c_{\lambda_k}|+1;\gamma + |c_{\lambda_k}| + 1;p) \right]^{d_k} $$
\begin{equation} \label{eq:final_re}
\stackrel{(b)}{=}  \sum_{\bm{\lambda} \in S_K}\prod_{k=1}^K\left\{ \frac{{_2}F_1(-\alpha_{k+1}(\bm{\lambda}),|c_{\lambda_k}|+1;\gamma + |c_{\lambda_k}| + 1;p)}{1- \, {_2}F_1(-\alpha_{k+1}(\bm{\lambda}),|c_{\lambda_k}|+1;\gamma + |c_{\lambda_k}| + 1;p)} \right\}
\end{equation}
where the second equality (a) can be shown as the follows: let $f(\alpha_k(\bm{\lambda})) = \, {_2}F_1(-\alpha_{k+1}(\bm{\lambda}),|c_{\lambda_k}|+1;\gamma + |c_{\lambda_k}| + 1;p)$, then
$$\sum_{\bm{d} \in \mathbb{N}^K}\prod_{k=1}^K f(\alpha_k(\bm{\lambda}))^{d_k} = \sum_{d_1 \in \mathbb{N}} \cdots \sum_{d_K \in \mathbb{N}} \left[ f(\alpha_1(\bm{\lambda}))^{d_1} \cdots f(\alpha_K(\bm{\lambda}))^{d_K}\right]$$
$$= \sum_{d_1 \in \mathbb{N}} \cdots \sum_{d_{K-2} \in \mathbb{N}} \left\{\sum_{d_{K-1} \in \mathbb{N}} f(\alpha_1(\bm{\lambda}))^{d_1} \cdots f(\alpha_{K-1}(\bm{\lambda}))^{d_{K-1}} \left(\sum_{d_K \in \mathbb{N}} f(\alpha_K(\bm{\lambda}))^{d_K}\right)\right\}$$
$$= \left(\sum_{d_K \in \mathbb{N}} f(\alpha_K(\bm{\lambda}))^{d_K}\right) \left\{ \sum_{d_1 \in \mathbb{N}} \cdots \sum_{d_{K-2} \in \mathbb{N}} \left\{\sum_{d_{K-1} \in \mathbb{N}} f(\alpha_1(\bm{\lambda}))^{d_1} \cdots f(\alpha_{K-1}(\bm{\lambda}))^{d_{K-1}}\right\} \right\}$$
$$= \left(\sum_{d_1 \in \mathbb{N}} f(\alpha_1(\bm{\lambda}))^{d_1}\right)\times \cdots \times \left(\sum_{d_K \in \mathbb{N}} f(\alpha_K(\bm{\lambda}))^{d_K}\right) = \prod_{k=1}^K \sum_{d_k \in \mathbb{N}}f(\alpha_k(\bm{\lambda}))^{d_k}.$$
And the last equality (b) of equation \eqref{eq:final_re} is due to geometric series: $\sum_{d=1}^{\infty}(r^d) = 1/(1-r)-1 = r/(1-r)$. Moreover, ${_2}F_1(-\alpha_{k+1}(\bm{\lambda}),|c_{\lambda_k}|+1;\gamma + |c_{\lambda_k}| + 1;p)$ is between 0 and 1. This can be seen from the derivative of the hypergeometric function:
$$\frac{d}{dp} {_2}F_1(-\alpha_{k+1}(\bm{\lambda}),|c_{\lambda_k}|+1;\gamma + |c_{\lambda_k}| + 1;p)$$
$$= -\frac{\alpha_{k+1}(\bm{\lambda})(|c_{\lambda_k}|+1)}{(\gamma + |c_{\lambda_k}| + 1)} \, {_2}F_1(-\alpha_{k+1}(\bm{\lambda})+1,|c_{\lambda_k}|+2;\gamma + |c_{\lambda_k}| + 2;p)$$
$$= -\frac{\alpha_{k+1}(\bm{\lambda})(|c_{\lambda_k}|+1)}{(\gamma + |c_{\lambda_k}| + 1)}(1-p)^{\gamma + \alpha_{k+1}(\bm{\lambda})} \, {_2}F_1(\gamma + |c_{\lambda_k}| + \alpha_{k+1}(\bm{\lambda}) + 1,\gamma; \gamma + |c_{\lambda_k}| + 2;p) < 0.$$
Since the derivative is less than zero, the function monotonically decrease with $p$. For $p \in (0,1]$, the hypergeometric function ${_2}F_1(-\alpha_{k+1}(\bm{\lambda}),|c_{\lambda_k}|+1;\gamma + |c_{\lambda_k}| + 1;p)$ equals 1 when $p = 0$ and equals 
$$0 < \frac{(\gamma)_{\alpha_{k+1}(\bm{\lambda})}}{(\gamma + |c_{\lambda_k}| + 2)_{\alpha_{k+1}(\bm{\lambda})}} < 1$$
when $p = 1$, where $(a)_b$ is the rising Pochhammer symbol defined as $(a)_b = 1$ if $b = 0$ and $(a)_b = a(a+1)\cdots(a+b-1)$ if $b > 0$. Substituting \eqref{eq:final_re} into (B) of \eqref{eq:PCzCsubPzeqz}, we have proved the EPPF of Theorem 3.

Lastly, for the claim of the EPPF of $G^*$ converging to the EPPF of $G_0 \sim DP(1,H)$ when $p \rightarrow 1$, 
the hypergeometric function 
$${_2}F_1(-f_k(\bm{z}^*),e_k(\bm{z}^*)+1;\gamma+e_k(\bm{z}^*)+1;1) = \frac{\Gamma(\gamma+e_k(\bm{z}^*)+1)\Gamma(\gamma+f_k(\bm{z}^*))}{\Gamma(\gamma)\Gamma(\gamma+e_k(\bm{z}^*)+f_k(\bm{z}^*) + 1)} $$
$$= \frac{\Gamma(\gamma+e_k(\bm{z}^*)+1)\Gamma(\gamma+f_k(\bm{z}^*))}{\Gamma(\gamma)\Gamma(\gamma+g_k(\bm{z}^*)+1))}$$
where $g_k(\bm{z}^*) = f_k(\bm{z}^*) + e_k(\bm{z}^*)$. And
equation \eqref{eq:pzeqz} becomes
$$\prod_{k=1}^{m(\bm{z}^*)} \frac{\Gamma(\gamma+1)\Gamma(e_k(\bm{z}^*)+1)\Gamma(\gamma+e_k(\bm{z}^*)+1)\Gamma(\gamma+f_k(\bm{z}^*))}{\Gamma(\gamma+e_k(\bm{z}^*)+1)\Gamma(\gamma)\Gamma(\gamma+g_k(\bm{z}^*)+1)}$$
$$= \prod_{k=1}^{m(\bm{z}^*)} \frac{\gamma\Gamma(e_k(\bm{z}^*)+1)\Gamma(f_k(\bm{z}^*) + \gamma)}{\Gamma(g_k(\bm{z}^*)+\gamma+1)},$$
which then equals the right-hand side of the sixth equal sign of equation $\text{Pr}(\bm{z} = z)$ in the proof of Lemma 2.2 in
\cite{miller2019elementary}. Then the author shows that (Proof of Theorem 2.1 therein) the EPPF of $G_0 \sim DP(\gamma, H)$ can be written as
$$\text{Pr}(C(\bm{z}) = C) = \frac{\gamma^{|C|}\Gamma(\gamma)}{\Gamma(n+\gamma)}\prod_{c\in C}\Gamma(|c|).$$

\bigskip

\subsection{Proof Lemma 1} \label{appd:lemma1}

Since $G^* \sim FSBP(p, \gamma, H)$, consider the following prediction rule for samples $\bm{\theta}_i|\bm{\theta}_1, \cdots , \bm{\theta}_{i-1}$, where $\bm{\theta}_1, \cdots , \bm{\theta}_i|G^* \sim G^*$:
$$\text{Pr}(\bm{\theta}_i|\bm{\theta}_1, \cdots , \bm{\theta}_{i-1})  = W_{\text{base}_i} H + \sum_{l=1}^{i-1} W_{i_l} \delta_{\bm{\theta}_l}$$
where $W_{\text{base}_i}$ corresponds to the probability $\bm{\theta}_i$ sampled from the base probability measure $H$ (and not equal to any $\bm{\theta}_l \in \{\bm{\theta}_1, \ldots , \bm{\theta}_{i-1}\}$) when there are $i$ samples, and $W_{i_l}$ corresponds to the probability of $\bm{\theta}_i$ sampled from a previously seen $\bm{\theta}_l$ for $l = 1, \ldots , i-1$. Then, we have
$$\text{Pr}(w_i = 1|p,\gamma) = \text{Pr}(\bm{\theta}_i \notin \{\bm{\theta}_1, \ldots , \bm{\theta}_{i-1}\}|G^*) = W_{\text{base}_i}.$$  

\noindent $W_{\text{base}_i}$ can be evaluated by (using the prediction rule in Theorem 2 of \cite{dunson2008kernel})
$$W_{\text{base}_i} = \left\{1 - \sum_{k = 2}^i (-1)^k \sum_{I \in N_i^{(k,i)}} \omega_I\right\},$$
where $N_i^{(k,i)}$ is a set contains all possible $k$-dimensional subsets of $\{1, \cdots , i\}$ that includes index $i$, with $I$ an element (a set) in the set, $\omega_I = \mu_I\cdot \left(\sum_{l=1}^{|I|}(-1)^{l-1}\sum_{m\in I_l}\mu_m\right)^{-1}$, $\mu_I = E[\prod_{k \in I} p{\pi_{k}}']$, and $I_l$ the set of length-$l$ subsets of the set $I$. The cardinality of the sets $N_i^{(k,i)}$, $I$, and $I_l$ are $|N_i^{(k,i)}| = {{i-1}\choose{k-1}}$, $|I| = k$, and $|I_l| = {{k}\choose{l}}$, respectively. For example, let $i = 3$, $k = 2$, and $l = 1$. $N_{i=3}^{(k=2,i=3)} = \{I_1, I_2\} = \{\{1,3\}, \{2,3\}\}$, with $|N_{i=3}^{(k=2,i=3)}| = 2$. Also, $|I_1| = |I_2| = 2$. And when $I = I_1$, $I_{l=1} = \{\{1\},\{3\}\}$, and when $I = I_2$, $I_{l=2} = \{\{2\},\{3\}\}$. Both have cardinality $|I_l| = 2$, $l = 1, 2$. 

For $G^*$, recall ${\pi_k'} \sim \text{Beta}(1,\gamma)$. For a set $I$, $\mu_I = E\left[ \prod_{k \in I} p\pi_k' \right]$, which can be shown to be
$$\mu_I = p^{|I|}\prod_{l=1}^{|I|}\frac{l}{l+\gamma}.$$
Thus, $\mu_I$ depends on the cardinality of the set $I$ only. Furthermore, for $\sum_{m \in I_l}\mu_m$ in the denominator of $\omega_I$, $\mu_m$ can be similarly computed, and the values are the same for all $m \in I_l$ (since $\mu_m$ depends only on $|m|$, and all $m \in I_l$ are of the same cardinality that is equal to $l$). Plugging in $\mu_I$ and $\sum_{m\in I_l}\mu_m$ to the theorem, we have 
$$\omega_I = \frac{p^{|I|}\prod_{l=1}^{|I|}\frac{l}{l+\gamma}}{\sum_{l=1}^{|I|} (-1)^{l-1} {{|I|}\choose{l}} p^l \prod_{m=1}^l \frac{m}{m+\gamma}},$$
which again only depends on the cardinality of the set $I$, i.e., $|I|$. Let $|N_i^{(k,i)}| = {{i-1}\choose{k-1}} = B$. Further notice that the sets in $N_i^{(k,i)}$, denoted as $I_1, \ldots, I_{b'}, \ldots, I_B$, have the same cardinality for a given $k$, i.e., $|I_{b'}| = k$ for all $b' \in \{1, ..., B\}$. Thus, we have
$$W_{\text{base}_i} = 1 - \sum_{k = 2}^i (-1)^k \sum_{I 
 \in N_i^{(k,i)}} \omega_I = 1 - \sum_{k = 2}^i (-1)^k {{i-1}\choose{k-1}} \frac{p^{k}\prod_{l=1}^{k}\frac{l}{l+\gamma}}{\sum_{l=1}^{k} (-1)^{l-1} {{k}\choose{l}} p^l \prod_{m=1}^l \frac{m}{m+\gamma}}$$
 $$= 1 - \sum_{k = 2}^i (-1)^k {{i-1}\choose{k-1}} \frac{k!}{\prod_{l=1}^{k}(l+\gamma)} \frac{p^{k-1}}{\sum_{l=1}^{k} (-1)^{l-1} {{k}\choose{l}} p^{l-1} \frac{l!}{\prod_{m=1}^l (m+\gamma)}}$$
 $$= 1 - \sum_{k = 2}^i (-1)^k {{i-1}\choose{k-1}} \frac{k!}{\prod_{l=1}^{k}(l+\gamma)} \frac{(\gamma + 1)p^{k-1}}{k\times {}_{2}F_{1}(1, 1-k; \gamma+2; p)}$$
 \begin{equation} \label{eq:propGs}
     = 1 - \sum_{k = 2}^i (-1)^k {{i-1}\choose{k-1}} \frac{(k-1)!}{\prod_{l=1}^{k}(l+\gamma)} \frac{(\gamma + 1)p^{k-1}}{{}_{2}F_{1}(1, 1-k; \gamma+2; p)}.
 \end{equation}
 where ${}_{2}F_{1}(a, b; c; z)$ is the hypergeometric function.
 
Since FSBP is a special case of KSBP, and in KSBP, $W_{\text{base}_i} \in (0,1)$, we have
\begin{equation*} 
0 < \sum_{k = 2}^i (-1)^k {{i-1}\choose{k-1}} \frac{(k-1)!}{\prod_{l=1}^{k}(l+\gamma)} \frac{(\gamma + 1)p^{k-1}}{{}_{2}F_{1}(1, 1-k; \gamma+2; p)} < 1.
\end{equation*}
 
 \qed

\subsection{Proof Lemma 2} \label{appd:lemma2}

Setting let $p \rightarrow 1$ in equation \eqref{eq:propGs}, we have
$$\lim_{p \rightarrow 1}\text{Pr}(w_i = 1|p, \gamma) = 1 - \sum_{k = 2}^i (-1)^k {{i-1}\choose{k-1}} \frac{(k-1)!}{\prod_{l=1}^{k}(l+\gamma)} \frac{(\gamma + 1)}{{}_{2}F_{1}(1, 1-k; \gamma+2; 1)}$$
$$\stackrel{(a)}{=} 1 - \sum_{k = 2}^i (-1)^k {{i-1}\choose{k-1}} \frac{(k-1)!}{\prod_{l=1}^{k}(l+\gamma)} \frac{(\gamma + 1)}{\frac{\gamma+1}{\gamma+k}}$$
$$= 1 - \sum_{k = 2}^i (-1)^k \frac{\Gamma(i)}{\Gamma(i-k+1)}\frac{\Gamma(\gamma+1)}{\Gamma(\gamma+k)} = 1 - \frac{i-1}{\gamma+i-1} = \frac{\gamma}{\gamma + i - 1},$$
where the second equality (a) is because $${}_{2}F_{1}(1, 1-k; \gamma+2; 1) = \frac{\Gamma(\gamma+2)\Gamma(\gamma+k)}{\Gamma(\gamma+1)\Gamma(\gamma+1+k)} = \frac{(\gamma+1)\Gamma(\gamma+1)\Gamma(\gamma+k)}{\Gamma(\gamma+1)(\gamma+k)\Gamma(\gamma+k)} = \frac{\gamma+1}{\gamma+k}.$$ 
Notice that $\frac{\gamma}{\gamma + i - 1}$ is the probability of generating a new sample $\bm{\theta}_i \notin \{\bm{\theta}_1, \cdots , \bm{\theta}_{i-1}\}$, i.e., from the base measure, in DP. \qed

\subsection{Proof Theorem 4} \label{appd:theoGs}

To show $\text{Pr}(w_i = 1|p, \gamma) > \frac{\gamma}{\gamma+i-1}$, it is sufficient to show that $1 - \frac{\gamma}{\gamma+i-1} > 1 - \text{Pr}(w_i = 1|p, \gamma)$, or
$$\frac{i-1}{\gamma+i-1} > \sum_{k=2}^{i} (-1)^k {{i-1}\choose{k - 1}} \frac{(k-1)!}{\prod_{l = 1}^k (l + \gamma)} \frac{(\gamma+1)p^{k-1}}{ {}_{2}F_{1}(1, 1-k; \gamma+2; p)   }.$$

\noindent First, notice that the hypergeometric function ${}_{2}F_{1}(1,1-k;\gamma+2;p)$ is monotonically decreasing with respect to $p$ since
$$\frac{d}{dp} {}_{2}F_{1}(1,1-k;\gamma+2;p) = -\frac{(k-1){}_{2}F_{1}(2,2-k;\gamma+3;p)}{\gamma+2}$$
$$= -\frac{(k-1)(1-p)^{\gamma+k-1}{}_{2}F_{1}(\gamma+1,\gamma+k+1;\gamma+3;p)}{\gamma+2} < 0,$$
with ${}_{2}F_{1}(1,1-k;\gamma+2;0) = 1$ and ${}_{2}F_{1}(1,1-k;\gamma+2;1) = \frac{\gamma + 1}{\gamma + k}$.
As a result, $\frac{1}{{}_{2}F_{1}(1,1-k;\gamma+2;p)}$ is monotonically increasing with $p$, with maximum at $p \rightarrow 1$, and
$$\lim_{p \rightarrow 1} \frac{1}{{}_{2}F_{1}(1,1-k;\gamma+2;p)} = \frac{\gamma + k}{\gamma + 1}.$$

\noindent Next, when substituting this maximum for ${}_{2}F_{1}(1,1-k;\gamma+2;p)$, it can be shown that

$$\frac{i-1}{\gamma+i-1} - \sum_{k = 2}^i (-1)^k {{i-1}\choose{k-1}} \frac{(k-1)!}{\prod_{l=1}^{k}(l+\gamma)}\frac{(\gamma + 1)p^{k-1}}{{}_{2}F_{1}(1,1-k;\gamma+2;p)} $$
$$ > \frac{i-1}{\gamma+i-1} - \sum_{k = 2}^i (-1)^k {{i-1}\choose{k-1}} \frac{(k-1)!}{\prod_{l=1}^{k}(l+\gamma)}\frac{(\gamma + 1)p^{k-1}(\gamma + k)}{\gamma + 1}$$
$$= \frac{i-1}{\gamma+i-1} - \sum_{k = 2}^i (-1)^k \frac{\Gamma(i)}{\Gamma(i-k+1)}\frac{\Gamma(\gamma+1)}{\Gamma(\gamma+k)} p^{k-1}$$
$$= \frac{i-1}{\gamma+i-1} - \frac{(i-1){}_{2}F_{1}(1,2-i;\gamma+2;p)p}{\gamma+1}.$$

Now, for $ p\cdot {}_{2}F_{1}(1,2-i;\gamma+2;p)$, from the property of hypergeometric function, we have $0\cdot {}_{2}F_{1}(1,2-i;\gamma+2;0) = 0 \cdot 1 = 0$, and $1 \cdot {}_{2}F_{1}(1,2-i;\gamma+2;1) = \frac{\Gamma(\gamma+2)\Gamma(\gamma+i-1)}{\Gamma(\gamma+1)\Gamma(\gamma+i)} = \frac{\gamma + 1}{\gamma + i - 1}$. In addition, we have 
$$\frac{d}{dp}{}_{2}F_{1}(1,2-i;\gamma+2;p)p = {}_{2}F_{1}(2,2-i;\gamma+2;p)$$
$$= (1-p)^{\gamma+i-2}{}_{2}F_{1}(\gamma,\gamma+i;\gamma+2;p) > 0,$$
and therefore, $p\cdot {}_{2}F_{1}(1,2-i;\gamma+2;p)$ monotonically increases with $p$, is equal to 0 if $p \rightarrow 0$, and is equal to $\frac{\gamma+1}{\gamma+i-1}$ if $p \rightarrow 1$. Consequently, for $p \in (0, 1)$, we have
$$\frac{i-1}{\gamma+i-1} - \frac{(i-1){}_{2}F_{1}(1,2-i;\gamma+2;p)p}{\gamma+1} > \frac{i-1}{\gamma+i-1} - \frac{i-1}{\gamma+i-1} = 0.$$ \qed

\bigskip



\subsection{Additional Details on Posterior Inference}
\label{sec:additional_detail} 

\paragraph{More details on the slice-efficient sampler} To sample $\beta_k'$ conditional on the other parameters and data,  we use an Metropolis-Hastings (MH) step to sample  from
\begin{equation} \label{eq:13}
    \begin{array}{c}
         p(\beta_k'|\cdots ) \propto \prod_{\{(j,l); j = 1, \ldots, J, \, l \geq k, \, \pi_{jl}' \neq 0\}} 
         \left[ \frac{{\pi_{jl}'}^{\alpha_0\beta_l - 1}(1 - \pi_{jl}')^{\alpha_0\left(1-\sum_{s=1}^{l}\beta_s\right)-1}}{B(\alpha_0\beta_l,\alpha_0\left(1-\sum_{s=1}^{l}\beta_s\right)} \right]
         \times (1-{\beta_k'})^{\gamma-1} 
    \end{array}
\end{equation}

\noindent where 
$\beta_k = \beta_k'\prod_{l=1}^{k-1}(1-\beta_l')$. 
In addition, we use a uniform distribution as the proposal density function: $\beta_{k_{\text{prop}}}' \sim \text{Unif}(\beta_{k_{\text{curr}}}' - \epsilon, \beta_{k_{\text{curr}}}' + \epsilon)$, where $\beta_{k_{\text{prop}}}'$ is the proposal, $\beta_{k_{\text{curr}}}'$ is the $\beta_k'$ in current iteration, and $\epsilon \in (0, 1)$ is the step size. If $\beta_{k_{\text{prop}}}' < 0$, we set $\beta_{k_{\text{prop}}}' = |\beta_{k_{\text{prop}}}'|$, and if $\beta_{k_{\text{prop}}}' > 1$, we set $\beta_{k_{\text{prop}}}' = 2 - \beta_{k_{\text{prop}}}'$. It can be shown the proposal density is symmetric. 

To sample $p_j$ with a prior of $p_j \sim \text{Beta}(a, b)$, we have
$$ p(p_j|\cdots)  \propto p_j^{\sum_k 1(\pi_{jk}' \neq 0)+a-1}(1-p_j)^{\sum_k I(\pi_{jk}' = 0)+b-1}.$$
Denoting $m_{j0} = \sum_{k=1}^{K^*} I(\pi_{jk}' = 0)$ the number of zero weights, we can sample $p_j$ as
\begin{equation*} 
    p_j|\cdots  \sim \text{Beta}(a+K^*-m_{j0}, b + m_{j0})
\end{equation*}

\noindent If we assume that the concentration parameters $\alpha_0$ and $\gamma$ are random with gamma priors, we can sample them 
using the procedure described in \cite{escobar1995bayesian} and \cite{teh2004sharing}.
In \cite{teh2004sharing}, the authors show that the full conditional of $\alpha_0$ and $\gamma$ is based on a matrix $\bm{W} = \{w_{jk}; j = 1, \ldots, J, k \geq 1 \}$ that records the number of tables in restaurant $j$ serving dish $k$ according to the Chinese restaurant franchise process, and the posterior of this matrix depends only on $\bm{Z}$ and $\bm{\beta}$. We use equation (40) of \cite{teh2004sharing} to construct a latent matrix $\bm{W}$ and then follow the same method as the HDP to sample both concentration parameters.

\paragraph{Label switching} As shown in the manuscript, 
we use the ECR algorithm of \cite{papastamoulis2010artificial} to resolve the issue of label switching.
This algorithm post-processes the MCMC samples using label permutations. The idea behind ECR is based on the invariance of likelihood with respect to the permutation of component labels. 

For each MCMC iteration with label matrix $\bm{Z}^{(m)} = \{z_{ij}^{(m)}; i = 1, \ldots, n_j, j = 1, \ldots, J\},\, z_{ij}^{(m)} \in \{1, \cdots , K^{(m)} \}$, where the superscript $(m)$ denotes the $m$th MCMC iteration, we can form a partition of the $N = \sum_{j=1}^J n_j$ observations based on $\bm{Z}^{(m)}$. With slightly abuse of notation, we denote the corresponding unique labels of $\bm{Z}^{(m)}$ as $\bm{t}^{(m)} = \{t_1^{(m)}, \cdots , t_{K^{(m)}}^{(m)}\}, \, t_k^{(m)} \in \{1, \cdots , K^{(m)}\}$. 
For example, suppose we have a sample of $N = 7$ observations across $J = 2$ groups, $\bm{y} = \begin{bmatrix} y_{11} & y_{21} & y_{31} & \\  y_{12} & y_{22} & y_{32} & y_{42} \end{bmatrix}$, and two iterations of MCMC samples, i.e., $m = 1$ and $m = 2$. Assume in the MCMC samples, both partition the observations into the same 3 clusters, i.e., $K^{(1)} = K^{(2)} = 3$, $\text{Cluster A} = \{y_{11}, y_{12}, y_{22}\}, \text{Cluster B} = \{y_{21}, y_{31}\}, \text{and } \text{Cluster C} = \{y_{32}, y_{42}\}$, according to their corresponding $\bm{Z}^{(1)}$ and $\bm{Z}^{(2)}$.
However, in each of the two MCMC iterations, different labels of $\bm{t}^{(1)} = \{1,2,3\}$, with $\bm{Z}^{(1)} = \begin{bmatrix} 1 & 2 & 2 & \\ 1 & 1 & 3 & 3 \end{bmatrix}$,
and $\bm{t}^{(2)} = \{2,1,3\}$, with $\bm{Z}^{(2)} = \begin{bmatrix} 2 & 1 & 1 & \\ 2 & 2 & 3 & 3 \end{bmatrix}$ are assigned to the observations.
Thus, there is a switched label of Cluster A and Cluster B through $m = 1$ and $m = 2$. To resolve the label-switching issue, the method finds a permutation of labels at each MCMC iteration, denote as $\bm{\tau}^{(m)}(\bm{t}^{(m)})$, such that, compare to a reference label, say $\bm{t}^{(1)}$, $\bm{\tau}^{(2)}(\bm{t}^{(2)}) = \bm{t}^{(1)} = \{1,2,3\}$.

Specifically, the ECR method first picks an MCMC sample from one iteration (e.g., one close to MAP) as the reference label. Then, the  method iterates over each MCMC sample of parameters of interest to find a random permutation of labels corresponding to the equivalent allocation of the reference label. We then switch the labels accordingly for all model parameters related to the cluster labels, i.e., label matrix $\bm{Z}$, MCMC samples of cluster weights $\{\pi_{jk}\}$, and cluster means $\{\bm{\phi}_k\}$. 
The  ECR method is implemented in \textbf{R} package \textbf{label.switching} \citep{papastamoulis2015label}.
We use ECR to relabel the MCMC samples of the weights. After permuting the weights according to the result of ECR, 
we then explore the MCMC samples of the permuted weights for all $j$ groups to learn the common and unique clusters in the groups.

\paragraph{Additional method to summarize common and unique clusters} 

The second approach to summarize common and unique clusters is to use the posterior sample of the group-specific weights $\bm{\pi}^{(m)}_j$, $j=1,\ldots, J$. Specifically, in the $m$th MCMC iteration, denote the number of common clusters between groups $j$ and $j'$ as $n_{\text{comm}}(\{\bm{\pi}_j^{(m)},\bm{\pi}_{j'}^{(m)}\})$, and denote the number of unique clusters in group $j$ as $n_{\text{uniq}}(\bm{\pi}_j^{(m)})$, we have
\begin{equation*}
    \begin{aligned}
        n_{\text{comm}}(\{\bm{\pi}_j^{(m)},\bm{\pi}_{j'}^{(m)}\}) = \sum_{k = 1}^{|\bm{\pi}_j^{(m)}|} 1(\pi_{jk}^{(m)} \neq 0 \text{ and } \pi_{j'k}^{(m)} \neq 0), \\
        n_{\text{uniq}}(\bm{\pi}_j^{(m)}) = \sum_{k = 1}^{|\bm{\pi}_j^{(m)}|} 1\left(\pi_{jk}^{(m)} \neq 0 \text{ and } \sum_{j' \in \{1,\cdots ,j-1,j+1,\cdots ,J\}} \pi_{j'k}^{(m)} = 0 \right)
    \end{aligned}
\end{equation*}
where $|\cdot|$ denotes the cardinality of the corresponding vector. Thus, the weight approach is able to learn the same information as the $\bm{Z}$ matrix method. 

\subsection{Slice Sampler for FSBP}\label{sec:slice_fsbp}


Follow \cite{kalli2011slice}, we derive the slice sampler for PAM. From the model in manuscript, the density function for observation $y_i$ can be rewritten as an infinite mixture 
\begin{equation*} 
f_{\bm{\xi}}(y_i,u_i|z_i,\{\phi_k; k \geq 1\}, \{\pi_k; k \geq 1\}) = \sum_{k \geq 1}1_{\{z_i = k\}}1_{\{u_i < \xi_k \}}\frac{\pi_{z_i}}{\xi_{z_i}}p(y_i|\phi_{z_i}),
\end{equation*}
where $u_i$ is the latent variable for observation $i$, and $\xi_k$ is the same quantity as defined in the slice sampler of PAM. Thus, stochastic truncation $K^*$ can be similarly computed following that of PAM. 

To sample from FSBP, we iteratively sample the following parameters:
\begin{enumerate}
    \item $u_i \sim \text{Unif}(0,\xi_{z_i})$,
    \item stick-breaking weight $\pi_k'$ for $k = 1, \ldots, K^*$,
    \item the indicator $z_i$ with $\text{Pr}(z_i=k|\ldots) \propto 1_{\{u_i < \xi_k\}}\frac{\pi_k}{\xi_k}p(y_i|\bm{\phi}_k)$, and
    \item the atom locations $\bm{\phi}_k|\cdots \propto \prod_{\{i;z_i = k, i = 1, \ldots, n\}}N(y_i|\bm{\phi}_k)p_H(\bm{\phi}_k)$.
\end{enumerate}

To sample $\pi_k'$, we can use a MH step, where the full condition is
$$p(\pi_k'|\cdots) \propto \left[(p\pi_k')^{m_k}(1-p\pi_k')^{\sum_{s=k+1}^{K^*}m_s}\right]f(\pi_k') $$
$$\propto (\pi_k')^{m_k + a - 1}(1-p\pi_k')^{\sum_{s=k+1}^{K^*}m_s}(1 - \pi_k')^{b-1}.$$
Here, $m_k = \sum_{i=1}^n 1(z_i = k)$, and $f(\pi_k')$ is the density function of the prior for $\pi_k'$ as defined in Section 3.3. The same proposal density for $\beta_k'$ (discussed in subsection \ref{sec:additional_detail}) can be used. 

Lastly, if we place a Beta prior on $p$, then conditional on $\{\pi_k'\}$, $p$ can be similarly sampled with another MH step and the same proposal. The other hyperparameter, $\gamma$, can also be straight-forwardly sampled as in PAM (discussed in subsection \ref{sec:additional_detail}). The entire sampler is presented in Algorithm \ref{alg:slice_fsbp}.

\begin{algorithm}[H]
\caption{Slice Sampler for FSBP}\label{alg:slice_fsbp}
\begin{algorithmic}[1]
\For{$m = 1, \cdots , M$}
\State Sample each $u_{i}$ from $u_{i} \sim \text{Unif}(0, \xi_{z_{i}})$ and find $K^*$.
\State Sample all $\pi_k'$ for $k = 1, \cdots , K^*$ with MH step.
\State Sample $p$ with MH step.
\State Sample $z_i$ from the following full condition:
$$p(z_{i} = k|\cdots ) \propto 1_{\{u_{i} < \xi_{k}\}}\frac{\pi_{k}}{\xi_{k}}p(y_{i}|\bm{\phi}_k)$$
\State Sample $\bm{\phi}_k$ from a conjugate NIG.
\EndFor
\end{algorithmic}
\end{algorithm}

\subsection{Additional Distributions of Simulated Data in Section 6} \label{sec:additional_fig_sim}

\paragraph{Additional simulation data and results}

Table \ref{tab:Sc1Case2_truth} shows the cluster mean and weight in the simulation setup for Case 2 of Scenario 1. 

\begin{table}[H]
    \centering
    \begin{tabular}{lc|ccccc}
    \hline
       &  & Cluster 1 & Cluster 2 & Cluster 3 & Cluster 4 & Cluster 5 \\
    \hline
       Mean &  & -4 & -2 & 0 & 2 & 4 \\
    \hline
       \multirow{3}{*}{Weight} & Group 1 & 0 & 0.8 & 0.2 & 0 & 0 \\
       & Group 2 & 0.3 & 0 & 0.1 & 0.6 & 0 \\
       & Group 3 & 0 & 0 & 0.2 & 0 & 0.8 \\
    \hline
    \end{tabular}
    \caption{Simulation truth of cluster means and weights for Case 2 of Scenario 1. Note that cluster 3 is the common cluster among all groups, while the other clusters are unique to their corresponding groups.}
    \label{tab:Sc1Case2_truth}
\end{table}
    
\begin{figure}[H]
\begin{center}
\includegraphics[width=\textwidth]{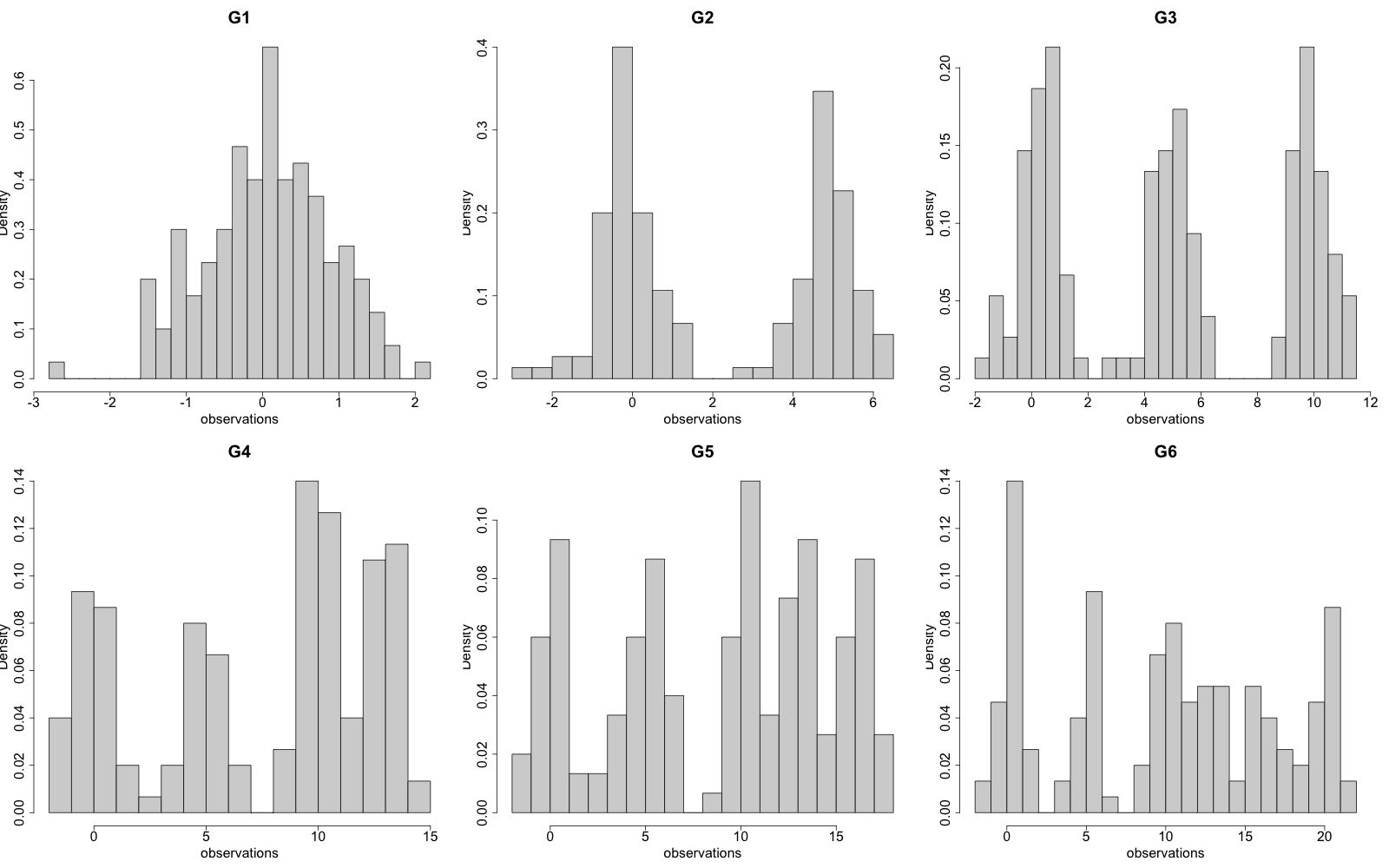}
\end{center}
\caption{Histogram of data distribution for randomly selected sample with sample size of 150 in Case 3 of Scenario 1. G1 to G6 means Groups 1 to 6, respectively.}
\label{fig:a0}
\end{figure}

Figure \ref{fig:a0} shows the data distribution of one randomly selected sample, with a sample size of 150, in Case 3 of Scenario 1. Note that the titles G1 to G6 refer to Groups 1 to 6, respectively.

Table \ref{tab:multv} shows the cluster means and weights in the simulation setup for the multivariate data in Scenario 2. 

\begin{table}[H]
\begin{center}
\begin{tabular}{ lc|ccccc }
    \hline
    & & Cluster 1 & Cluster 2 & Cluster 3 & Cluster 4 & Cluster 5 \\
    \hline
    Mean & & $\left(\begin{matrix} -6 \\ 4 \\ -6 \end{matrix}\right)$ & $\left(\begin{matrix} -3 \\ 2 \\ -3 \end{matrix}\right)$ & $\left(\begin{matrix} 0 \\ 0 \\ 0 \end{matrix}\right)$ & $\left(\begin{matrix} 3 \\ -2 \\ -3 \end{matrix}\right)$ & $\left(\begin{matrix} 6 \\ -4 \\ -6 \end{matrix}\right)$ \\
    \hline
    \multirow{3}{4em}{Weight} & Group 1 & 0.2 & 0.2 & 0.2 & 0.2 & 0.2 \\
    & Group 2 & 0.3 & 0 & 0.5 & 0.2 & 0 \\
    & Group 3 & 0 & 0.6 & 0.4 & 0 & 0 \\
    \hline
\end{tabular}
\end{center}
\caption{Simulation truth of cluster means and  weights for Scenario 2 in simulation. Here, cluster 3 is the common cluster shared among all groups.}
\label{tab:multv}
\end{table}



\begin{figure}
    \centering
    \includegraphics[width=0.55\textwidth]{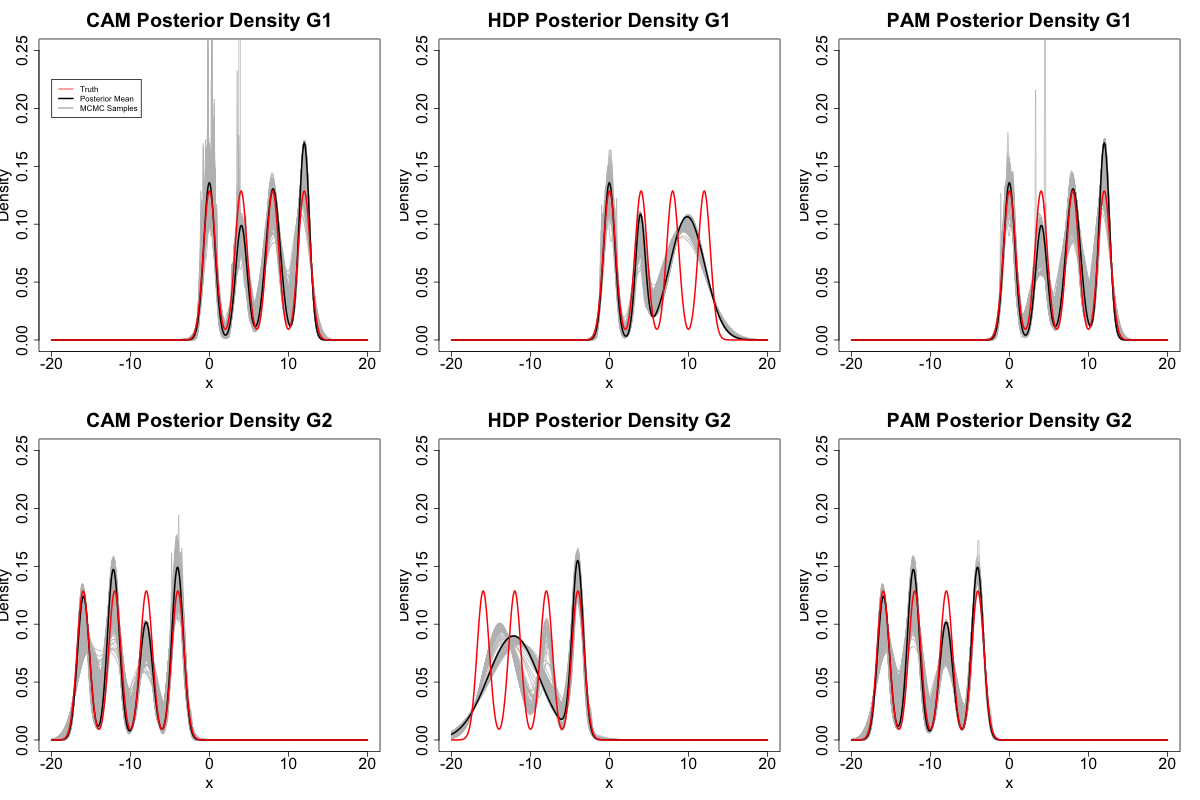}
    \caption{  Posterior density estimation using CAM (first column), HDP (second column), and PAM (third column) for a randomly selected dataset in Case 1 of Scenario 1. Each row corresponds to a specific group. The red lines represent the truth, the grey lines indicate the posterior density estimated in each MCMC iteration, and the black lines represent the point-estimate of the posterior density.  }
    \label{fig:post_density}
\end{figure}

Table \ref{fig:post_density} shows the estiamted posterior density of a randomly selected sample for Case 1 in Scenario 1.

Table \ref{tab:1} shows the performance of CAM, HDP, and PAM over 30 datasets for each sample sizes of Case 3 in Scenario 1.

\begin{table}[H]
    \centering
    \resizebox{\linewidth}{!}{
    \begin{tabular}{l|c|ccc|c}
    \hline
        Sample sizes & Metrics & CAM & HDP & PAM & Truth  \\
    \hline
        \multirow{3}{*}{$n_A = 50$} & Number of clusters in all groups & 4.03 (0.49) &  3.93 (0.53) & 4.93 (0.87) & 6 \\
        & ARI & 0.90 (0.05) &  0.87 (0.05) & 0.87 (0.07) & \\
        & NFD & 0.07 (0.03) &  0.04 (0.02) & 0.06 (0.02) & \\
    \hline
        \multirow{3}{*}{$n_A = 100$} & Number of clusters in all groups & 4.67 (0.61) &  4.00 (0.59) & 5.67 (0.71) & 6 \\
        & ARI & 0.93 (0.04) &  0.87 (0.05) & 0.91 (0.04) & \\
        & NFD & 0.07 (0.02) &  0.04 (0.02) & 0.04 (0.02) & \\
    \hline
        \multirow{3}{*}{$n_A = 150$} & Number of clusters in all groups & 4.97 (0.49) &  4.27 (0.58) & 5.97 (0.62) & 6 \\
        & ARI & 0.95 (0.02) &  0.90 (0.04) & 0.95 (0.03) & \\
        & NFD & 0.07 (0.02) &  0.03 (0.01) & 0.02 (0.01) & \\
    \hline
        \multirow{3}{*}{$n_B = 10$} & Number of clusters in all groups & 4.17 (0.75) &  4.23 (0.50) & 5.77 (0.82) & 6 \\
        & ARI & 0.79 (0.08) &  0.76 (0.08) & 0.73 (0.06) & \\
        & NFD & 0.08 (0.02) &  0.07 (0.03) & 0.09 (0.02) & \\
    \hline
        \multirow{3}{*}{$n_B = 20$} & Number of clusters in all groups & 4.40 (0.56) &  4.30 (0.65) & 6.00 (0.59) & 6 \\
        & ARI & 0.83 (0.08) &  0.78 (0.09) & 0.82 (0.06) & \\
        & NFD & 0.08 (0.02) &  0.06 (0.03) & 0.06 (0.02) & \\
    \hline
        \multirow{3}{*}{$n_B = 40$} & Number of clusters in all groups & 5.43 (0.50) &  4.33 (0.61) & 6.17 (0.38) & 6 \\
        & ARI & 0.93 (0.05) &  0.82 (0.07) & 0.94 (0.05) & \\
        & NFD & 0.05 (0.02) &  0.05 (0.02) & 0.02 (0.01) & \\
    \hline
    \end{tabular}
    }
    \caption{Simulated univariate data in Case 3 of Scenario 1. Clustering performance for CAM, HDP, and PAM are evaluated according to the number of total estimated clusters (truth = 6 clusters), the Adjusted Rand Index (ARI), and the normalized Frobenius distance (NFD). Entries are  Mean (SD) over 30 datasets.}
\label{tab:1}
\end{table}

Table \ref{tab:p_c_u} shows the estimated number of clusters, common clusters, and unique clusters for the sample size of $n_A = 150$ in Case 3 of Scenario 1. For simplicity, except for all groups, the common clusters reported use Group 6 as a reference group, and measures the common clusters between Groups 1 to 5 with Group 6.

\begin{table}[H]
    \centering
    \resizebox{\linewidth}{!}{
    \begin{tabular}{lc|ccc|c}
    \hline
       \multicolumn{2}{c|}{Metrics}  & CAM & HDP & PAM & Truth  \\
    \hline
       \multirow{6}{*}{Number of clusters}  & G1 & 1.00 (0.00) & 1.00 (0.00) & 1.03 (0.18) & 1 \\
         & G2 & 2.00 (0.00) & 2.00 (0.00) & 2.00 (0.00) & 2 \\
         & G3 & 3.30 (0.47) & 3.00 (0.00) & 3.03 (0.18) & 3 \\
         & G4 & 4.00 (0.53) & 3.33 (0.48) & 3.93 (0.25) & 4 \\
         & G5 & 4.43 (0.50) & 3.33 (0.48) & 4.13 (0.57) & 5 \\
         & G6 & 4.60 (0.62) & 3.17 (0.46) & 4.53 (0.68) & 6 \\
    \hline
       \multirow{6}{*}{Common clusters} & All Groups & 1.00 (0.00) & 1.00 (0.00) & 1.00 (0.00) & 1\\
        & G6 and G5 & 4.43 (0.50) & 3.13 (0.35) & 3.47 (0.68) & 5 \\
        & G6 and G4 & 3.70 (0.38) & 3.00 (0.26) & 3.10 (0.66) & 4 \\
        & G6 and G3 & 3.17 (0.38) & 2.67 (0.48) & 2.90 (0.31) & 3 \\
        & G6 and G2 & 2.00 (0.00) &  2.00 (0.00) & 1.96 (0.18) & 2 \\
        & G6 and G1 & 1.00 (0.00) & 1.00 (0.00) & 1.00 (0.00) & 1 \\
    \hline
       \multirow{6}{*}{Uniqe clusters} & G1 & 0.00 (0.00) & 0.00 (0.00) & 0.03 (0.18) & 0\\
        & G2 & 0.00 (0.00) & 0.00 (0.00) & 0.00 (0.00) & 0 \\
        & G3 & 0.00 (0.00) & 0.00 (0.00) & 0.03 (0.18) & 0 \\
        & G4 & 0.17 (0.38) & 0.00 (0.00) & 0.63 (0.56) & 0 \\
        & G5 & 0.00 (0.00) & 0.00 (0.00) & 0.53 (0.51) & 0 \\
        & G6 & 0.13 (0.35) & 0.03 (0.18) & 0.90 (0.40) & 1 \\
    \hline
    \end{tabular}
    }
    \caption{The estimated number of clusters, common, and unique clusters for simulated univariate data in Case 3 of Scenario 1, when the sample size is $n_A = 150$. Note that except all groups, the estimated number of common clusters use Group 6 as a reference. Entries are  Mean (SD) over 30 datasets. }
\label{tab:p_c_u}
\end{table}

Table \ref{tab:multi_30} shows the models' performance on the multivariate data in Scenario 2.

\begin{table}[H]
    \centering
    \resizebox{\linewidth}{!}{
    \begin{tabular}{lc|ccc|c}
    \hline
        Sample sizes & Metrics & CAM & HDP & PAM & Truth  \\
    \hline
        \multirow{3}{*}{$n_j = 50$} & Number of clusters in all groups &  5.40 (1.13) & 5.50 (0.97) & 4.93 (0.64) & 5 \\
         & ARI &  0.90 (0.06) & 0.86 (0.11) & 0.89 (0.08) & \\
         & NFD &  0.05 (0.03) & 0.05 (0.02) & 0.03 (0.03) & \\
    \hline
        \multirow{3}{*}{$n_j = 100$} & Number of clusters in all groups &  5.37 (0.71) & 4.90 (0.76) & 5.07 (0.26) & 5 \\
         & ARI &  0.95 (0.04) & 0.91 (0.07) & 0.96 (0.02) & \\
         & NFD &  0.04 (0.02) & 0.04 (0.02) & 0.01 (0.01) & \\
    \hline
        \multirow{3}{*}{$n_j = 200$} & Number of clusters in all groups &  5.04 (0.19) & 4.93 (0.47) & 5.03 (0.18) & 5 \\
         & ARI &  0.97 (0.01) & 0.96 (0.02) & 0.97 (0.01) & \\
         & NFD &  0.03 (0.02) & 0.01 (0.01) & 0.01 (0.00) & \\
    \hline
    \end{tabular}
    }
    \caption{Simulated multivariate data in Scenario 2. Clustering performance for CAM, HDP, and PAM evaluated according to the number of total estimated clusters (truth = 5 clusters), the Adjusted Rand Index (ARI), and the normalized Frobenius distance (NFD). The entries are Mean (SD) over 30 datasets.}
\label{tab:multi_30}
\end{table}

\begin{figure}
    \centering
    \includegraphics[width=0.8\textwidth]{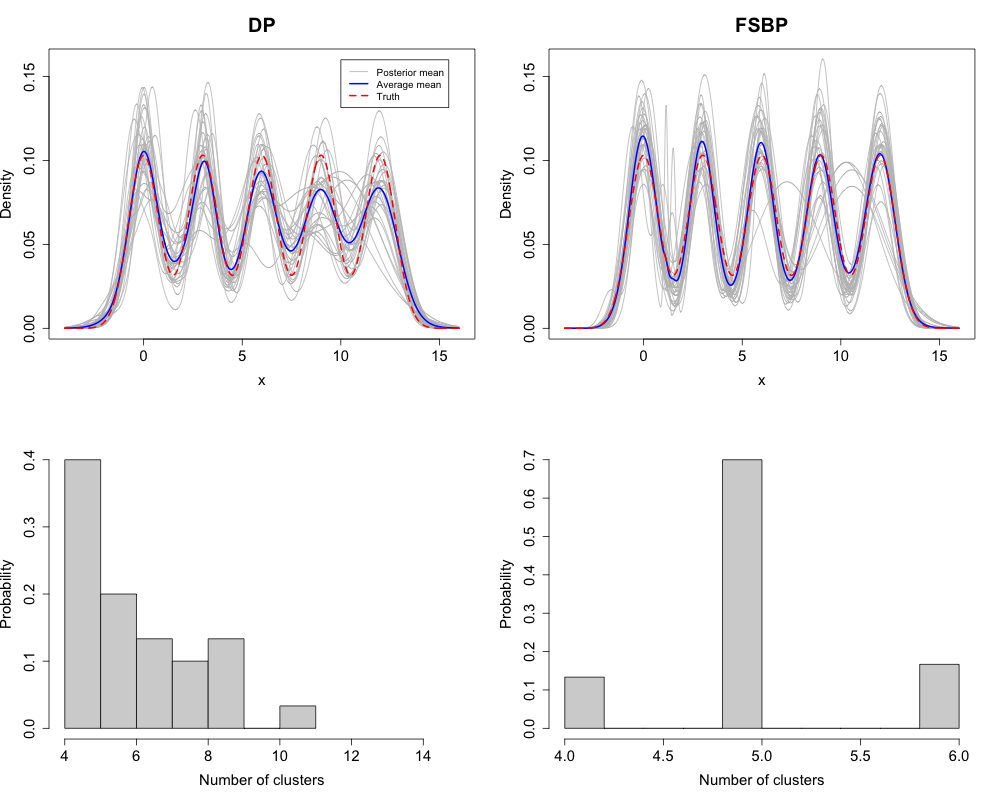}
    \caption{  Estimated posterior density for DP (top-left) and FSBP (top-right), along with histograms depicting the estimated number of clusters (bottom plots). FSBP estimates are based on \cite{wade2018bayesian} using posterior samples. 
    Grey lines represent the posterior mean for each simulated dataset, blue lines show the average of the posterior means across the 30 simulated datasets, and the red dashed lines indicate the truth.  }
    \label{fig:post_FSBP_result}
\end{figure}

Table \ref{fig:post_FSBP_result} shows the performance of FSBP and DP on the univariate data in Scenario 3.

\subsection{Additional Distributions and Results of Microbiome Population in Section 7.1} \label{sec:additional_fig_exp1}

\begin{figure}[H]
\begin{center}
\includegraphics[width=5in]{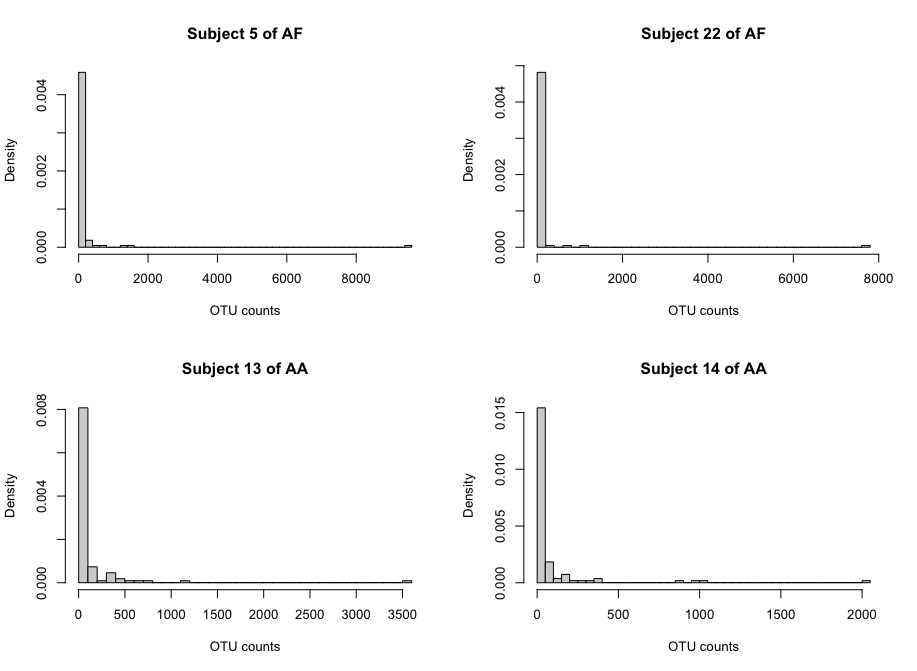}
\end{center}
\caption{Histograms of the microbiome population of the four selected individuals.}
\label{fig:micro_data}
\end{figure}

Figure \ref{fig:micro_data} shows the histogram of OTU counts for the four randomly selected individuals in the analysis of the microbiome dataset.

\begin{figure}
    \centering
    \includegraphics[width=0.8\textwidth]{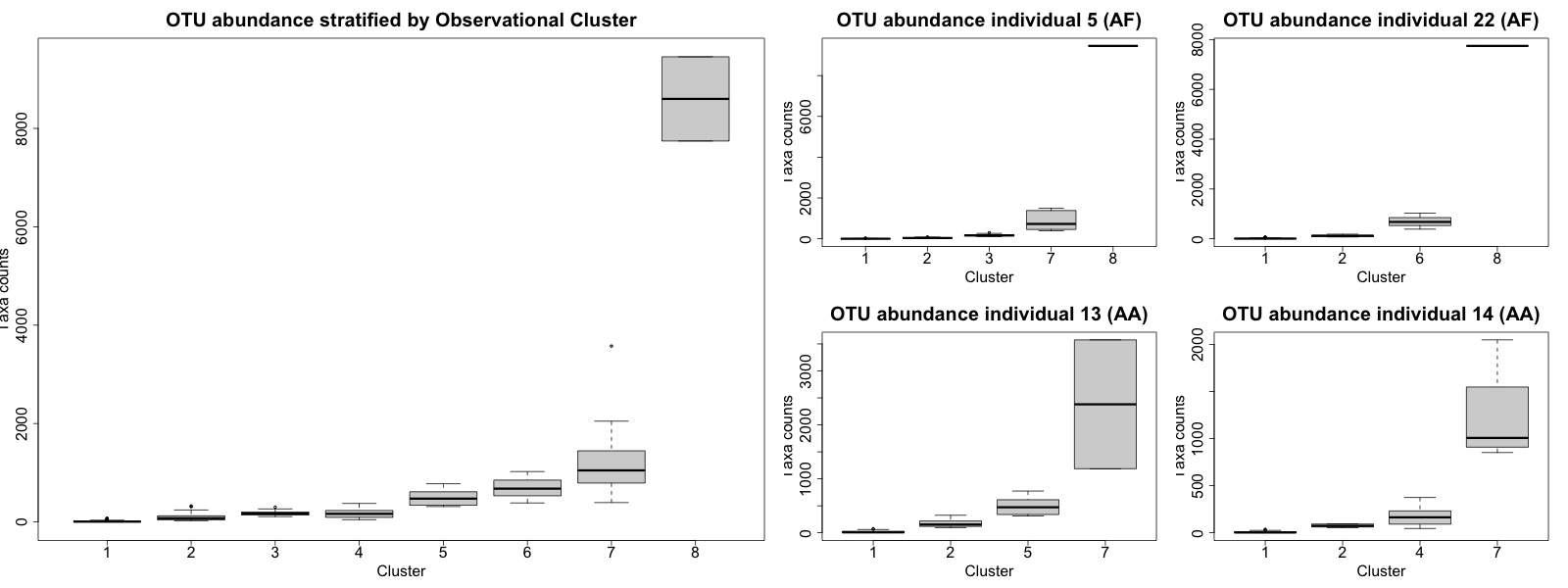}
    \caption{Boxplots of microbiome abundance counts stratified by clusters (Left subplot) and by both clusters and individuals (Four right subplots).}
    \label{fig:2graphs}
\end{figure}

Figure \ref{fig:2graphs} shows barplots of the taxa counts (TC) of OTUs grouped by eight estiamted clusters as well as by both cluster and individuals.

\subsection{Additional Results of Warts Dataset Analysis in Section 7.2}
\label{subsec:warts_appendix}

Figure \ref{fig:warts_sep} below shows the cluster membership of each patient of the warts dataset.

\begin{figure}[H]
\begin{center}
\includegraphics[width=5in]{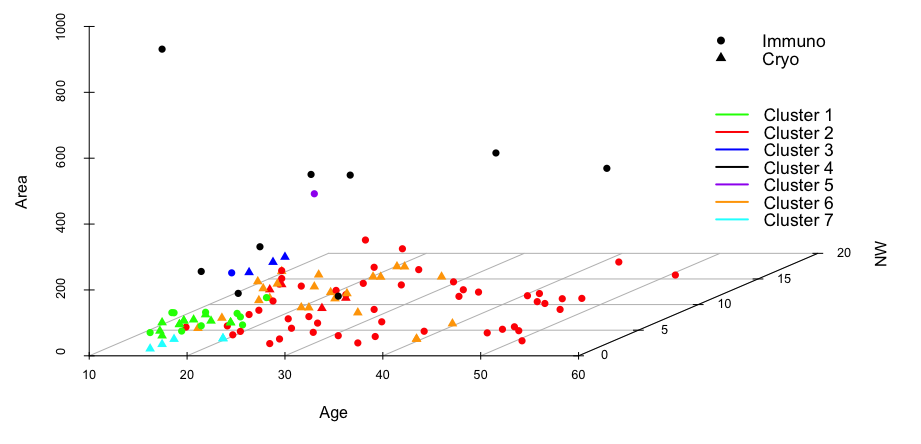}
\end{center}
\caption{Estimated cluster membership of patients in the warts dataset. The cluster labels are shown with different colors, across two groups indicated by the circles and triangles. The clustering result is based on four covariates of area, age, number of warts, and time elaspsed until treatment. We plot three of them: Area, Age, and number of warts (NW).}
\label{fig:warts_sep}
\end{figure}



\end{document}